\documentclass[a4paper,11pt]{article}
\pdfoutput=1 

\usepackage{jheppub, placeins, mathtools,notes2bib,tikz} 
\usepackage{cancel}

\usepackage[T1]{fontenc} 

\title{\boldmath The canonical formulation of E$_{6(6)}$ exceptional\\ field theory}

\author[]{Lars T. Kreutzer}

\affiliation[]{Max Planck Institute for Gravitational Physics (Albert Einstein Institute), \\Am M\"uhlenberg 1, 14476 Potsdam, Germany}
\affiliation[]{Berlin Mathematical School (BMS), Berlin, Germany}

\emailAdd{lars.kreutzer@aei.mpg.de}

\abstract{We investigate the canonical formulation of the (bosonic) E$_{6(6)}$ exceptional field theory. The explicit non-integral (not manifestly gauge invariant) topological term of E$_{6(6)}$ exceptional field theory is constructed and we consider the canonical formulation of a model theory based on the topological two-form kinetic term. Furthermore we construct the canonical momenta and the canonical Hamiltonian of the full bosonic E$_{6(6)}$ exceptional field theory. Most of the canonical gauge transformations and some parts of the canonical constraint algebra are calculated. Moreover we discuss how to translate the results canonically into the generalised vielbein formulation. We comment on the possible existence of generalised Ashtekar variables.}

\begin{document} 
\maketitle
\flushbottom

\section{Introduction}
In 1978 it was discovered that toroidal compactifications of eleven-dimensional supergravity lead to emerging hidden global exceptional E$_{n(n)}(\mathbb{R})$ symmetries in $(11-n)$-dimensional maximal supergravity \cite{Cremmer:1978km,CREMMER197848}. The existence of the exceptional symmetries in maximal supergravity theories continues to be one of their most remarkable features and remains to be fully understood at the quantum level. It is only since 2013 that so-called ``exceptional field theories'' are known, which are fully and manifestly E$_{n(n)}(\mathbb{R})$ covariant and which encompass eleven-dimensional supergravity  \cite{Hohm:2013pua}. See reference \cite{Berman-review-eft} for a recent review of exceptional field theories. These exceptional field theories achieve (local) E$_{n(n)}(\mathbb{R})$ covariance with the use of an extended generalised exceptional geometry and can be reduced to eleven-dimensional supergravity upon the solution of a consistency condition called the section condition. We review the Lagrangian formulation of E$_{6(6)}$ exceptional field theory in section \ref{sec-EFT-lagrangian}\\

So far the canonical formulation of exceptional field theory has not been investigated. In \cite{Naseer:2015fba} the canonical formulation of O$(n,n)$ double field theory has been discussed and very recently E$_{n(n)}$ covariant world-volume theories have been analysed canonically in \cite{osten2021currents}.\\

The complete understanding of the canonical structure of exceptional field theory may shed some light on the role and physical meaning of the section condition. It may furthermore be interesting to use the canonical theory to investigate the local initial value problem for the extended generalised exceptional geometry. In analogy to the canonical double field theory results of \cite{Naseer:2015fba} one could try to generalise the notion of asymptotic flatness and ADM charges for non-compact extended generalised exceptional geometries. From the canonical perspective on exceptional field theory one might moreover hope to identify a suitable notion of a generalised Ashtekar connection, as the results of \cite{Melosch:1997wm} might suggest (we comment on this topic in section \ref{section-conclusios}). Finally the canonical formulation of exceptional field theory may be seen as the starting point for the canonical quantisation procedure. Some loop calculations in exceptional field theory --- for special geometries that are of the form of Minkowski space times a compact torus --- have already been carried out in \cite{Bossard:2017kfv,Bossard:2015foa} and the geometric quantisation of exceptional field theory has recently been commented on in \cite{Alfonsi:2021bot}.\\

In this work we construct the canonical formulation of the (bosonic) E$_{6(6)}$ exceptional field theory \cite{Hohm:2013pua,EFTI-E6} and analyse its canonical structure. The general principles of the canonical formalism for constrained Hamiltonian systems has been described in detail in \cite{dirac-lectures-QM,Henneaux-Teitelboim}. A central result in this work is the calculation of the canonical Hamiltonian of (bosonic) E$_{6(6)}$ exceptional field theory \eqref{can-eq:can-Hamiltonian-EFT-intro}, written here on the primary constraint surface, with $\Pi(X)$ indicating the canonical momenta conjugate to some fields $X$ and $\mathcal{P}^m_M(A)$ being a modified version of the one-form momenta.\footnote{In equation \eqref{can-eq:can-Hamiltonian-EFT-intro} capital indices $K,L,M,\dots$ indicate the (anti-)fundamental $27$ representations of E$_{6(6)}$, $t$ indicates the curved time index, lower case $k,l,m,n,\dots$ indicate the external curved spatial indices while lower case $a,b,\dots$ indicate the external flat spatial indices.} In \eqref{can-eq:can-Hamiltonian-EFT-intro} the secondary (Hamilton, external diffeomorphism, generalised diffeomorphism and tensor gauge) constraints are already apparent. Canonically the generalised diffeomorphisms are generated by the secondary constraints multiplying the Lagrange multipliers $A^M_t$.
\begin{align}\label{can-eq:can-Hamiltonian-EFT-intro}
    \mathcal{H}_{\text{ExFT}}\, = &+ N \cdot \bigg{[} +\frac{1}{4 e} \Pi_{ab}(e)\,  \Pi_{ab}(e) - \frac{1}{12e} \Pi(e)^2 - e\,  \hat{R}  + e\, V_\text{HP}  \nonumber \\
    & \hspace{1.3cm } +\frac{3}{2e} \Pi^{MN}(M) \, \Pi_{MN}(M)  - \frac{e}{24}\, \mathcal{D}_m M_{MN} \, \mathcal{D}^m M^{MN}    \nonumber \\
    & \hspace{1.3cm }  +\frac{e}{4} \, \mathcal{F}^{mn}_M\, \mathcal{F}_{mn}^M       +\frac{1}{2e} \, \mathcal{P}^m_M\,  \mathcal{P}_m^M \bigg{]} \nonumber\\
       & +N^n \cdot \bigg{[}   + 2\, \Pi^m{}_a(e)\, \mathcal{D}_{[n} e_{m]a} - e_{na}\, \mathcal{D}_m \Pi^m{}_a(e)  \nonumber \\
       & \hspace{1.5cm }   +\frac{1}{2} \Pi^{MN}(M)\, \mathcal{D}_n M_{MN}   \nonumber \\ 
        & \hspace{1.5cm } + \mathcal{F}_{nl}^M \mathcal{P}^l_M + \partial_M\left(g_{mn}\, M^{MN}\, \mathcal{P}^m_N \right) \bigg{]}    \nonumber \\
      & +A_t^M \cdot \bigg{[} -\mathcal{D}_l \mathcal{P}^l_M -5 \,d^{NLS} \, d_{MNK}\, A^K_m \partial_S \mathcal{P}^m_L + (\mathcal{H}_{top})_M \nonumber\\
         &\hspace{1.5cm} +\,\Pi^m{}_a(e)\, \partial_M e_{ma}   - \frac{1}{3} \partial_M \Pi(e) \nonumber \\
        & \hspace{1.5cm}+ \frac{1}{2}\,\Pi^{KL}(M) \, \partial_M M_{KL} - 6\, \mathbb{P}^R{}_K{}^S{}_M \, \partial_S\left(\Pi^{KL}(M)\, M_{RL} \right)\bigg{]} \nonumber \\
        & +B_{tlM} \cdot \bigg{[} +10 \,d^{MKL} \partial_K \left( \mathcal{P}^l_L  - \kappa\, \epsilon^{lmnr}\, \mathcal{H}_{mnrL}\right) \bigg{]}
\end{align}

We choose to analyse the E$_{6(6)}$ theory because it is by comparison the simplest of the true exceptional cases and does not involve a pseudo-action with self-dual forms or constrained compensator fields which would further complicate the canonical analysis. For the same reasons this analysis is limited to the bosonic sector of the E$_{6(6)}$ exceptional field theory. The main challenges that are present in the canonical formulation of the bosonic E$_{6(6)}$ exceptional field theory are the complicated topological term, the treatment of the topological two-forms and to some degree the inherent complexity of the underlying extended generalised exceptional geometry.\\
This work may be seen as an extension of the comprehensive canonical analysis of the maximal ungauged E$_{6(6)}$ invariant supergravity theory in \cite{KreutzerSUGRA}.\\

The outline of this work is as follows. In section \ref{sec-EFT-lagrangian} we summarise some of the main findings concerning the Lagrangian formulation of the bosonic E$_{6(6)}$ exceptional field theory based on the references \cite{Hohm:2013pua,EFTI-E6,Baguet:2015xha}. Furthermore we construct the explicit non-integral (not manifestly gauge invariant) $(5+27)$-dimensional form of the topological term of E$_{6(6)}$ exceptional field theory --- which is needed to explicitly carry out the Legendre transformation. In section \ref{can-sec-twoforms-model} we investigate the canonical formulation of a model theory consisting only of the topological two-form kinetic term of E$_{6(6)}$ exceptional field theory. We discuss the canonical constraint algebra of the model and identify some problems regarding the construction of Dirac brackets in the generalised geometry for constraint algebras of this particular form. In section \ref{section-can-eft-chapter} we investigate the canonical formulation of the full bosonic E$_{6(6)}$ exceptional field theory. We calculate the canonical momenta, introduce some redefinitions of the canonical coordinates and carry out the Legendre transformation to arrive at the canonical Hamiltonian of E$_{6(6)}$ exceptional field theory. We calculate most of the gauge transformations generated by the canonical constraints in section \ref{can-eft-gauge-transformations}. In section \ref{sec-canonical-algebra} we calculate parts of the algebra of the canonical constraints and discuss some speculative results based on references \cite{EFTI-E6,Baguet:2015xha,KreutzerSUGRA}. In section \ref{sec-vielbein-usp8} we discuss the USp(8) symmetry and how the results of the previous sections can be translated into the generalised vielbein formulation canonically. Finally we summarise the findings of this work in section \ref{section-conclusios} and comment on the possible existence of a generalised Ashtekar connection and the quantisation of exceptional field theory.

\section{Lagrangian formulation of \texorpdfstring{E$_{6(6)}$}{E6(6)} exceptional field theory}\label{sec-EFT-lagrangian}
It is possible to rewrite eleven-dimensional supergravity in a manifestly E$_{n(n)}(\mathbb{R})$ invariant form, called exceptional field theory (ExFT). ExFT is a Kaluza-Klein-like rewriting of eleven-dimensional supergravity, but without actually truncating any degrees of freedom. ExFT achieves this by making use of an extended generalised exceptional geometry. The E$_{6(6)}$ ExFT was first published in 2013 \cite{Hohm:2013pua,EFTI-E6}, its supersymmetric completion was first published in \cite{Musaev2015} and the theory was later reviewed in \cite{Baguet:2015xha,Berman-review-eft}. The E$_{7(7)}$ and E$_{8(8)}$ ExFTs and their supersymmetric completions have been presented in \cite{Hohm:2013uia,Godazgar:2014nqa} and \cite{Hohm:2014fxa,Baguet:2016jph} respectively. 
In the extended notion of exceptional groups the E$_{2(2)}$ \cite{Berman:2015rcc}, E$_{3(3)}$ \cite{Hohm:2015xna}, E$_{4(4)}$ \cite{Musaev:2015ces} and E$_{5(5)}$ \cite{Abzalov:2015ega} ExFTs have also been constructed. General reviews of exceptional field theories were published in \cite{Hohm:2019bba,Berman-review-eft}.
The structure of exceptional field theory is in some aspects similar to that of gauged supergravity (e.g. tensor hierarchy) \cite{deWit:2002vt,deWit:2003hr,deWit:2004nw,deWit:2007kvg,Wit_2008,Trigiante:2016mnt} and to the Kaluza-Klein rewriting of double field theory (e.g. extended generalised geometry) \cite{Hohm_2013-DFT-extint}.\\

In contrast to the representations of the groups O$(n,n)$ in double field theory, the representations of the exceptional groups E$_{n(n)}$ that occur in ExFT, and therefore also the invariant symbols that they admit, are very different depending on the $n$ chosen. Because of this diversity it is hard to formulate all aspects of exceptional field theory in a way that holds true for all E$_{n(n)}$ simultaneously. In this work we focus on the E$_{6(6)}$ exceptional field theory, which is formulated on an extended $5+27$ dimensional extended geometry \cite{Hohm:2013pua,EFTI-E6}. The E$_{6(6)}$ ExFT is, in a sense, the simplest case of the true exceptional groups $n=6,7,8$, because there are no self-dual forms in the five external dimensions (unlike in the E$_{7(7)}$ ExFT where one has to consider a pseudo-action with an additional self-duality relation) and there are no constrained compensator fields (unlike the E$_{8(8)}$ ExFT). For simplicity we furthermore focus on the bosonic sector of the E$_{6(6)}$ ExFT as described in \cite{Hohm:2013pua,EFTI-E6}.\\

The geometry of E$_{n(n)}$ ($n\leq8$) exceptional field theory is constructed from a Kaluza-Klein-like split \eqref{eft-split-geometry} of the space-time of eleven-dimensional supergravity $ \mathcal{M}_{11}$ into a non-compact external $d$-dimensional ($d:=11-n$) Lorentzian manifold $\mathcal{M}^{\text{ext.}}_{d}$ and an internal $n$-dimensional Riemannian manifold $\mathcal{M}^{\text{int.}}_{n}$. We do not assume any specific topology for the internal manifold $\mathcal{M}^{\text{int.}}_{n}$ and crucially we do not carry out the truncation of any degrees of freedom.
\begin{equation}\label{eft-split-geometry}
    \mathcal{M}_{11} = \mathcal{M}^{\text{ext.}}_{d} \times \mathcal{M}^{\text{int.}}_{n}
\end{equation}
The coordinates of the external geometry $\mathcal{M}^{\text{ext.}}_{d}$ are taken to be $x^\mu$ with $\mu=0,\dots,d-1$ and the coordinates of the internal geometry $\mathcal{M}^{\text{int.}}_{n}$ are $y^m$ with $m=1,\dots,n$. With the eleven-dimensional manifold written in this local factorisation the next step is then to extend the internal manifold and turn it into an extended generalised exceptional geometry. To do so one needs to extend the tangent bundle of the internal geometry while simultaneously adding auxiliary (or dual) coordinates. To construct an extended exceptional geometry we need the internal coordinates to sit in a representation of the duality group E$_{n(n)}$. We therefore add as many auxiliary coordinates to the internal coordinates $y^m$ as are needed to turn them into the generalised internal coordinates $Y^M$. The coordinate index takes the values $M=1,\dots,\dim(R_1($E$_{n(n)}))$ with $R_1$(E$_{n(n)}$) being the coordinate representation, often this is the fundamental representation, as is the case for E$_{6(6)}$ with $R_1$(E$_{6(6)})=27$ \cite{EFTI-E6,Berman-review-eft}. The overall coordinates of the external-internal geometry are then given by $(x^\mu,Y^M)$. The extended internal coordinates $Y^M$ come with associated internal partial derivatives $\partial_M$. Consistency of the extended exceptional geometry requires that the E$_{n(n)}$-covariant projection of certain combinations of internal partial derivatives vanishes, which effectively removes the auxiliary coordinates again. This consistency condition is called the section condition \cite{coimbra2013edd,Berman:2012vc,Bossard:2017aae}.\\

In the section \ref{sec-gen-ext-excep-geom} we discuss the internal E$_{6(6)}$ extended generalised exceptional geometry in more detail and in section \ref{sec-eft-e6-lagrangian} we describe the structure of the E$_{6(6)}$-covariant field theory constructed on this extended geometry.  

\subsection{\texorpdfstring{E$_{6(6)}$}{E6(6)} extended generalised exceptional geometry}\label{sec-gen-ext-excep-geom}
The internal extended 27-dimensional E$_{6(6)}$ generalised exceptional geometry can be thought of as the coordinate-extended version of the generalised exceptional geometry associated to the generalised tangent bundle \eqref{exgegeo-bundle-e6}, where $\text{T}(\mathcal{M}^{\text{int.}}_{6})$ is the ordinary tangent bundle of the unextended internal manifold $\mathcal{M}^{\text{int.}}_{6}$ \cite{Hull_2007,Pacheco_2008,Coimbra:2011nw,Coimbra:2012af}. The extended internal coordinates $Y^M$, with $M=1,\dots,27$, are in the fundamental representation $R_1( $E$_{6(6)})=27$.
\begin{equation}\label{exgegeo-bundle-e6}
\text{E} :=    \text{T}(\mathcal{M}^{\text{int.}}_{6})\oplus\Lambda^2\text{T}^*(\mathcal{M}^{\text{int.}}_{6})\oplus\Lambda^5\text{T}^*(\mathcal{M}^{\text{int.}}_{6})
 \end{equation}
In order for us to be able to explicitly write the objects of E$_{6(6)}$ generalised exceptional geometry we first need to discuss the E$_{6(6)}$-invariant $d$-symbols. The fully symmetric invariant symbols $d_{LMN}$ and $d^{LMN}$ carry three (anti-)fundamental $27$ (or $\overline{27}$) indices \cite{Cremmer,EFTI-E6}. They are the unique invariant symbols of the (anti-)fundamental representation of E$_{6(6)}$ (up to their normalisation) and satisfy the invariance condition \eqref{eft-invariance-d} \cite{EFTI-E6}. We can choose their normalisation to be defined by the condition for them to be inverse in the sense of \eqref{eft-normalisation-d}. The invariant symbols furthermore obey the cubic identities \eqref{eft-cubic-d-1} and \eqref{eft-cubic-d-2}, which are required in some calculations \cite{EFTI-E6}.
\begin{align}
    d^{KLM}\,M_{KN}\,M_{LR}\,M_{MS} & = d_{NRS} \label{eft-invariance-d}\\
    d^{MKL}\,d_{KLN} & = \delta^M_N \label{eft-normalisation-d}\\
    d_{S(MN}\,d_{PQ)T}\,d^{STR} & = \frac{2}{15} \,\delta^R_{(M} \, d_{NPQ)}\label{eft-cubic-d-1} \\
     d_{STR}\,d^{S(MN}\,d^{PQ)T} & = \frac{2}{15} \,\delta^{(M}_R \, d^{NPQ)} \label{eft-cubic-d-2}
\end{align}
Taking $t_\zeta$, with $\zeta=1,\dots,78$, to be the generators of the $\mathfrak{e}_6$ Lie algebra, we can write these generators in the fundamental E$_{6(6)}$ representation as $(t_\zeta)^M{}_N $. The adjoint indices can be raised and lowered with $(t_{\zeta_1})^N{}_M\, (t_{\zeta_2})^M{}_N$, which is proportional to the Cartan-Killing form. The projector $\mathbb{P}^M{}_N{}^K{}_L$ onto the adjoint representation of E$_{6(6)}$ can then be defined by \eqref{eft-def-Proj}, with the normalisation $\mathbb{P}^M{}_N{}^N{}_M = 78 $. We can write the projector \eqref{eft-def-Proj} explicitly in terms of the invariant symbols as in equation \eqref{eft-def-Proj-explicit}.
\begin{align}
\mathbb{P}^M{}_N{}^K{}_L & := (t_\zeta)^M{}_N \, (t^\zeta)^K{}_L \label{eft-def-Proj} \\
\mathbb{P}^M{}_N{}^K{}_L  & = \frac{1}{18} \,\delta^M_N \, \delta^K_L +  \frac{1}{6} \,\delta^M_L \, \delta^K_N  - \frac{5}{3} \,d^{MKR} \, d_{RNL} \label{eft-def-Proj-explicit}
\end{align}
Some aspects of exceptional geometry, across different exceptional groups, can be phrased in terms of an object called the Y-tensor \cite{Berman-review-eft}. The Y-tensor of E$_{6(6)}$, which follows from the projector \eqref{eft-def-Proj-explicit}, is given by \eqref{eft-Y-tensor-e6}. 
\begin{equation}
  Y^{MK}{}_{NL} = 10 \,d^{MKR} \, d_{RNL} \label{eft-Y-tensor-e6}
\end{equation}
In the case of E$_{6(6)}$ exceptional geometry only 6 out of the 27 internal coordinates originate from the physical coordinates of eleven-dimensional supergravity. This fact is encoded in the E$_{6(6)}$-covariant section condition \eqref{eft-e6-section-condition-1}, which implies that at most 6 of the 27 coordinates really exist. As we will see below we can also think of \eqref{eft-e6-section-condition-1} as a consistency condition that arises naturally when considering generalised diffeomorphism.
\begin{equation}\label{eft-e6-section-condition-1}
    d^{KLM}\,\partial_L \,\otimes\, \partial_M  = 0
\end{equation}
Equation \eqref{eft-e6-section-condition-1} is the projection of two internal derivatives with the Y-tensor \eqref{eft-Y-tensor-e6}. We interpret the section condition \eqref{eft-e6-section-condition-1} as the conditions \eqref{eft-e6-section-condition-2} where $\Psi,\,\Phi$ are arbitrary functions (e.g. fields or gauge parameters).
\begin{equation}\label{eft-e6-section-condition-2}
    d^{KLM}\,\partial_L \Phi \, \partial_M \Psi  = 0, \quad   d^{KLM}\,\partial_L  \partial_M  \Phi = 0
\end{equation}
The generalised exceptional Lie derivative $ \mathbb{L}_{\Lambda} V^M$, with generalised vector parameter $\Lambda^M$, of a generalised vector $V^M$ can be written in terms of the projector \eqref{eft-def-Proj-explicit} as \eqref{eft-e6-lie-vec}. The real constant $\lambda(V)$ is the generalised weight of the generalised vector $V^M$. The first term of \eqref{eft-e6-lie-vec} is the transport (or translation) term, the second term can be interpreted as an E$_{6(6)}$ rotation and the last term is a weight term. Equivalently \eqref{eft-e6-lie-vec} can be written as the standard Lie derivative (in terms of internal partial derivatives) with a correction term given by the Y-tensor \eqref{eft-Y-tensor-e6}.  
\begin{equation}\label{eft-e6-lie-vec}
    \mathbb{L}_{\Lambda} V^M := \Lambda^K\,\partial_K V^M -6\, \mathbb{P}^M{}_N{}^K{}_L \, \partial_K \Lambda^L\, V^N + \lambda(V)\,\partial_N \Lambda^N\,V^M 
\end{equation}
Equivalently the generalised Lie derivative of a generalised covector $W_M$ can be written as \eqref{eft-e6-lie-covec}. With the Leibniz rule of the generalised Lie derivative the expressions \eqref{eft-e6-lie-vec} and \eqref{eft-e6-lie-covec} extend in the standard fashion to any generalised tensor.
\begin{equation}\label{eft-e6-lie-covec}
    \mathbb{L}_{\Lambda} W_M := \Lambda^K\,\partial_K W_M +6\, \mathbb{P}^N{}_M{}^K{}_L \, \partial_K \Lambda^L\, W_N + \lambda(W)\,\partial_N \Lambda^N\,W_M 
\end{equation}
Parameters $\Lambda^M$ of the form \eqref{eft-trivial-parameter} are trivial, in the sense that they lead to a vanishing generalised Lie derivative on any other field, when the section condition is applied.
\begin{equation}\label{eft-trivial-parameter}
    \Lambda^M = d^{MNK}\,\partial_N W_K
\end{equation}
It can be verified that the generalised Lie derivative, as defined above, is compatible with the E$_{6(6)}$ invariant $d$-symbols and the relation \eqref{eft-compatible-d} holds.
\begin{equation}\label{eft-compatible-d}
     \mathbb{L}_{\Lambda} d_{MNK} = 0 
\end{equation}
In the extended exceptional geometry the usual Lie bracket is modified by an additional Y-tensor term and the resulting bracket is called the E-bracket. Using the explicit form of the Y-tensor \eqref{eft-Y-tensor-e6} we can write the E$_{6(6)}$ E-bracket of two generalised vectors $\Lambda_1^M , \Lambda_2^N $ as \eqref{eft-e-bracket}.
\begin{equation}\label{eft-e-bracket}
    \left[ \Lambda_1 , \Lambda_2  \right]^M_\text{E} := 2\, \Lambda^K_{[1}\,\partial_K \Lambda^M_{2]}  - 10\,d^{MNP}\,d_{KLP}\, \Lambda^K_{[1}\,\partial_N \Lambda^L_{2]}  
\end{equation}
It can be shown, in a somewhat lengthy calculation, that the commutator of two generalised Lie derivatives is again a generalised Lie derivative, with the parameter of the resulting Lie derivative given by the E-bracket of the original parameters, i.e. the generalised Lie derivative obeys the algebra \eqref{eft-algebra-lie-GD}. The equation \eqref{eft-algebra-lie-GD} is true only up to terms that vanish when the section condition \eqref{eft-e6-section-condition-1} is applied. In this sense we may think of the section condition as a consistency condition that is implied by the closure of the generalised diffeomorphism algebra. The cubic identities \eqref{eft-cubic-d-1} and \eqref{eft-cubic-d-2} also have to be applied repeatedly in the calculation of \eqref{eft-algebra-lie-GD}.
\begin{equation}\label{eft-algebra-lie-GD}
    \left[  \mathbb{L}_{\Lambda_1},   \mathbb{L}_{\Lambda_2}  \right] = \mathbb{L}_{ \left[ \Lambda_1 , \Lambda_2  \right]_\text{E}}
\end{equation}
The E-bracket \eqref{eft-e-bracket} is antisymmetric and satisfies the Leibniz rule and therefore it defines a Leibniz (or Loday) algebra \cite{loday-original,Lavau:2017tvi,Kotov:2018vcz}. However the Jacobi identity of the E-bracket only holds up to a trivial parameter of the form \eqref{eft-trivial-parameter}, equivalently one can say that the Jacobiator \eqref{eft-jacobiator} is of trivial form, where $U^M,V^N,W^K$ are generalised vectors.
\begin{equation}\label{eft-jacobiator}
    J(U,V,W) :=   \left[ \left[ U, V  \right]_\text{E}  , W  \right]_\text{E} +  \left[ \left[  W, U  \right]_\text{E}  , V  \right]_\text{E} +  \left[ \left[ V , W  \right]_\text{E}  , U  \right]_\text{E}
\end{equation}
One can define the Dorfman-like bracket \eqref{eft-dorfman-bracket}, which is helpful in some calculations \cite{EFTI-E6}. 
\begin{equation}\label{eft-dorfman-bracket}
    (V\circ W )^M := \mathbb{L}_V W^M
\end{equation}
The bracket \eqref{eft-dorfman-bracket} does satisfy the Jacobi identity, but it is not antisymmetric and the relation \eqref{eft-rel-dorfman-e} holds, if $\lambda(W)=1/3$.
\begin{equation}\label{eft-rel-dorfman-e}
      (V\circ W )^M =  \left[V, W \right]^M_\text{E} + 5\, d^{MKR}\, \partial_K(d_{RPL} \,V^P\,W^L )
\end{equation}
The symmetric part of \eqref{eft-rel-dorfman-e} is of trivial form and its antisymmetric part is identical to the E-bracket. 

\subsection{\texorpdfstring{E$_{6(6)}$}{E6(6)} exceptional field theory}\label{sec-eft-e6-lagrangian}
We can now discuss the E$_{6(6)}$ exceptional field theory built upon the $5+27$-dimensional extended exceptional geometry. This section reviews some of the results of references \cite{Hohm:2013pua,EFTI-E6,Baguet:2015xha}.\\ The bosonic field content of the E$_{6(6)}$ ExFT is given by \eqref{EFT-field-content}. In general all of the fields \eqref{EFT-field-content} (and all of the gauge parameters) depend on all of the $5+27$ external and internal coordinates $(x^\mu, Y^M)$.
\begin{equation}\label{EFT-field-content}
    \{ E_\mu{}^\alpha, M_{MN}, A_\mu^M, B_{\mu\nu M} \}
\end{equation}
The external vielbein $E_\mu{}^\alpha$ is related to the external metric $G_{\mu\nu}$ by \eqref{eft-vielbein-def}, where $\eta_{\alpha\beta}$ is the external Minkowski metric with signature $(-++++)$. The indices $\alpha,\beta=0,\dots,4$ are flat Lorentz indices and the indices $\mu,\nu=t,1,\dots,4$ are the curved external space-time indices.
\begin{equation}\label{eft-vielbein-def}
    E_\mu{}^\alpha \,  E_\nu{}^\beta \, \eta_{\alpha\beta} = G_{\mu\nu}
\end{equation}
From the perspective of the external geometry the generalised E$_{6(6)}$ metric components $M_{MN}=M_{(MN)}$ are scalar fields and parametrise the coset E$_{6(6)}/$USp(8), which is 42-dimensional. Therefore only 42 of the scalar fields $M_{MN}$ are truly independent. We refer to these relations among the 378 components as the coset constraints of the scalar fields. Among other things the coset constraints imply that $\det(M_{MN})=1$. The inverse scalar fields are defined by $M_{MK}\,M^{KN}=\delta^N_M$. The generalised one- and two-form fields $A_\mu^M$ and $B_{\mu\nu M} $ carry an additional (anti-)fundamental E$_{6(6)}$ index.\\

We can see how the fields of ExFT relate to those of eleven-dimensional supergravity by considering a $5+6$ split of the eleven-dimensional indices. The bosonic field content of eleven-dimensional supergravity is a metric $G_{\hat{\mu}\hat{\nu}}$ and a three-form $C_{\hat{\mu}\hat{\nu}\hat{\rho}}$ \cite{Cremmer:1978km}. Decomposing the eleven-dimensional indices as $\hat{\mu}=(\mu,m)$, with $\mu=t,1,\dots,4$ and $m=1,\dots,6$, we can rearrange the resulting components into the fields of ExFT. The purely external components $G_{\mu\nu}$ of the eleven-dimensional metric become the external metric field of ExFT.
For the other components one has to check whether it is possible to arrive at a differential form of lower degree by Hodge-dualising the field strength of the component, i.e. the purely external components of the three-form $C_{{\mu}{\nu}{\rho}}$ can be Hodge-dualised via their field-strength into one scalar field in five dimensions. Proceeding in this way and dualising all forms to lowest possible degree we find $42=21+1+20 $ scalar fields coming from $G_{mn}$, $C_{{\mu}{\nu}{\rho}}$ and $C_{mnr}$ that constitute the independent components of $M_{MN}$. Furthermore there are $27=6+6+15$ vector fields coming from $G_{\mu n}$, $C_{\mu\nu r}$ and $C_{\mu nr}$ that arrange into the generalised vector fields $A_\mu^M$. In five dimensions the 27 one-forms can equivalently described by 27 two-forms that are arranged into $B_{\mu\nu M}$. Because we keep both the one-forms and the two-forms dual to them we need to impose a duality relation, as otherwise we would add degrees of freedom. This (on-shell) duality relation is given by the equations of motion of the two-forms. The reason for introducing the two-forms lies in the existence of the tensor hierarchy of differential forms, this will become clear when discussing the field-strengths.\\ 
 
The generalised one-forms $A_\mu^M$ are taken to act as the gauge connection for the generalised exceptional diffeomorphisms. Because the generalised diffeomorphism parameters $\Lambda^M(x,Y)$ depend also on the external coordinates we need to introduce the covariant (external) derivative \eqref{eft-cov-deriv}.
\begin{equation}\label{eft-cov-deriv}
    \mathcal{D}_\mu := \partial_\mu - \mathbb{L}_{A_\mu}
\end{equation}
We then require that the one-forms transform as the covariant derivative of the gauge parameter \eqref{eft-cov-GD-A} under a generalised diffeomorphism with parameter $\Lambda^M$. The transformation \eqref{eft-cov-GD-A} can be thought of as the covariantised version of an abelian U$(1)^{27}$ gauge transformation.  
\begin{equation}\label{eft-cov-GD-A}
    \delta_\Lambda A_\mu^M := \mathcal{D}_\mu \Lambda^M
\end{equation}
Naively one can write the field strength \eqref{eft-F-naive}, which replaces the Lie bracket with the E-bracket \eqref{eft-e-bracket}. 
\begin{equation}\label{eft-F-naive}
   F_{\mu\nu}^M := 2\,\partial_{[\mu}A_{\nu]}^M  - [A_\mu ,A_\nu ]_\text{E}^M    
\end{equation}
However just like in the gauged maximal supergravity \cite{deWit:2004nw,Samtleben:2008pe}, the field strength \eqref{eft-F-naive} fails to transform covariantly, due to the non-vanishing Jacobiator of the E-bracket and instead transforms as \eqref{eft-GD-F-naive}.
\begin{equation}\label{eft-GD-F-naive} 
    \delta  F_{\mu\nu}^M =  2\,\mathcal{D}_{[\mu} \,\delta A_{\nu]}^M + 10\,d^{MKR}\,d_{NLR}\,\partial_K\left( A_{[\mu}^N\,\delta A_{\nu]}^L  \right)
\end{equation}
The solution to this problem --- just like in five-dimensional gauged supergravity --- is to introduce the (topological) two-forms $B_{\mu\nu M}$ whose transformation can be defined to absorb the offending term. As we have seen above there are not naturally any two-forms in the field content coming from eleven-dimensional supergravity. Consequently we need to Hodge-dualise the one-forms (via their field strength) with regard to the five external dimensions and require that the resulting two-forms are on-shell dual to the original one-forms in order to not generate any new degrees of freedom. Continuing with the analogy to gauged supergravity one then adds a Stückelberg-type coupling term to the one-form field strength \eqref{eft-F-naive} to arrive at the covariant field strength \eqref{eft-GD-F-covariant-stuckelberg}. 
\begin{align}\label{eft-GD-F-covariant-stuckelberg} 
   \mathcal{F}_{\mu\nu}^M  :=&\,  F_{\mu\nu}^M + 10 \,d^{KLM}\,\partial_K B_{\mu\nu L} \\
    =&\, 2\,\partial_{[\mu}A_{\nu]}^M  - [A_\mu ,A_\nu ]_\text{E}^M    + 10 \,d^{KLM}\,\partial_K B_{\mu\nu L}
\end{align}
The improved field strength \eqref{eft-GD-F-covariant-stuckelberg} transforms covariantly as \eqref{eft-transf-f-cov}.
\begin{equation}\label{eft-transf-f-cov}
    \delta   \mathcal{F}_{\mu\nu}^M =  2\,\mathcal{D}_{[\mu} \,\delta A_{\nu]}^M + 10\,d^{MNK}\,\partial_K \Delta B_{\mu\nu N}
\end{equation}
The modified two-form transformation $\Delta B_{\mu\nu N}$ is defined as \eqref{eft-delta-B} in order to cancel the non-covariant term in \eqref{eft-GD-F-naive}.
\begin{equation}\label{eft-delta-B}
    \Delta B_{\mu\nu N} := \delta B_{\mu\nu N} + d_{NKL}\, A_{[\mu}^K\,\delta A_{\nu]}^L 
\end{equation}
The one-form field strength is generated by the commutator of the covariant derivatives \eqref{eft-comm-cov-derv}. Because the Stückelberg-type coupling term in the covariant field strength is of trivial form (cf. equation \eqref{eft-trivial-parameter}) the commutator does not distinguish between the naive and the covariant field strengths.
\begin{equation}\label{eft-comm-cov-derv}
  \left[  \mathcal{D}_\mu , \mathcal{D}_\nu \right] = -  \mathbb{L}_{F_{\mu\nu}} = - \mathbb{L}_{\mathcal{F}_{\mu\nu}}
\end{equation}
The two-forms $B_{\mu\nu M}$ come with their own covariant field strength $\mathcal{H}_{\rho\sigma\tau N}$ which can be written as \eqref{eft-h-field-strength}, where the ``$\dots$'' indicate terms that vanish under the projection $d^{MNK}\,\partial_K$.
\begin{equation}\label{eft-h-field-strength}
\mathcal{H}_{\rho\sigma\tau N} := 3 \mathcal{D}_{[\rho}B_{\sigma\tau]N}-3 d_{NKL}A^{K}_{[\rho} \bigg(\partial_\sigma A^L_{\tau]}-\frac{1}{3}[A_\sigma,A_{\tau]}]^L_E\bigg) + \dots
\end{equation}
The explicit form \eqref{eft-h-field-strength} of the two-form field strength can be found by solving the Bianchi identity \eqref{eft-bianchi-H}.
\begin{equation}\label{eft-bianchi-H}
    3\,\mathcal{D}_{[\mu} \mathcal{F}^M_{\nu\rho]} = 10\,d^{MNK}\,\partial_K \mathcal{H}_{\mu\nu\rho N}
\end{equation}
In analogy to the tensor hierarchy of gauged supergravity the existence of a three-form is required in order for the two-form field strength \eqref{eft-h-field-strength} to transform covariantly --- just like the two-forms are required for the one-form field strength \eqref{eft-GD-F-covariant-stuckelberg} to transform covariantly. Fortunately in the E$_{6(6)}$ ExFT the three-form terms are contained in the ``$\dots$'' of \eqref{eft-h-field-strength} and do not appear in the Lagrangian because they are being projected out.\\

The action of E$_{6(6)}$ exceptional field theory is given by \eqref{lag-eft-e6-action-integral}. The action \eqref{lag-eft-e6-action-integral} can be thought of as an elegant way of encoding the classical equations of motion. It is not known how the integral over the internal geometry can be carried out explicitly in a meaningful way before the section condition \eqref{eft-e6-section-condition-1} is solved. We treat the integral over the internal coordinates as being symbolic. 
\begin{equation}\label{lag-eft-e6-action-integral}
    S_\text{ExFT} = \int d^5x \int d^{27}Y\, \mathcal{L}_\text{ExFT}
\end{equation}
The E$_{6(6)}$ ExFT Lagrangian consists of the five distinct terms \eqref{eft-lagrangian-terms}. 
\begin{equation}\label{eft-lagrangian-terms}
 \mathcal{L}_\text{ExFT} = \mathcal{L}_\text{EH} + \mathcal{L}_\text{sc} + \mathcal{L}_\text{pot}  + \mathcal{L}_\text{YM} + \mathcal{L}_\text{top}
\end{equation}
The first term is the improved Einstein-Hilbert term \eqref{eft-lagrangian-EH}. This term consists of the Einstein-Hilbert term $E\,R$, where $E$ is the vielbein determinant and $R$ is the $\mathcal{D}_\mu$-covariantised Ricci scalar associated to $E_\mu{}^\alpha$ in which all partial derivatives are replaced by covariant derivatives $\mathcal{D}_\mu$. Additionally there is a one-form dependent improvement term. This improvement term is necessary in order to make the $\mathcal{D}_\mu$-covariantised Riemann-tensor transform tensorially under local Lorentz transformations --- which would otherwise not be the case due to \eqref{eft-comm-cov-derv} \cite{Hohm_2013-DFT-extint}.  
\begin{align}
 \mathcal{L}_\text{EH}   &=  E {\hat R} = ER +E\,\mathcal{F}^M_{\alpha\beta}\,E^{\alpha\rho}\,\partial_M E^\beta_\rho   \label{eft-lagrangian-EH} \\
 \mathcal{L}_\text{sc}   &= \frac{E}{24} \,G^{\mu\nu}\,\mathcal{D}_\mu M_{MN}\,\mathcal{D}_\nu M^{MN}   \label{eft-lagrangian-SC}\\
 \mathcal{L}_\text{pot} &=  -E\,V_\text{pot}(G_{\mu\nu},M_{MN})  \label{eft-lagrangian-pot}\\
 \mathcal{L}_\text{YM}   &= -\frac{E}{4}\, M_{MN}\mathcal{F}_{\mu\nu }^M\,\mathcal{F}^{\mu\nu N}   \label{eft-lagrangian-YM}
\end{align}
The second term in the Lagrangian is the scalar kinetic term \eqref{eft-lagrangian-SC}. The scalar kinetic term can also be seen as an E$_{6(6)}/$USp(8) non-linear coset sigma model with covariantised derivatives. Furthermore there is a scalar potential term \eqref{eft-lagrangian-pot} that only depends on the external metric and the scalar fields. The potential $V_\text{pot}(G_{\mu\nu},M_{MN})$ itself can be written explicitly as \eqref{seft-scalar-pot-expl}, where $G$ is the determinant of the external metric. The name potential is justified because \eqref{seft-scalar-pot-expl} depends only on internal partial derivatives. The generalised Yang-Mills term is given by \eqref{eft-lagrangian-YM}. It is of the standard Yang-Mills form, but written using the improved covariant field strength \eqref{eft-GD-F-covariant-stuckelberg} and with the E$_{6(6)}$ indices contracted by the internal generalised metric $M_{MN}$.
\begin{align}\label{seft-scalar-pot-expl}
    V_\text{pot} & = -\frac{1}{24}\,M^{MN}\,\partial_M M^{KL}\, \partial_N M_{KL}+ \frac{1}{2} \, M^{MN} \, \partial_M M^{KL}\, \partial_L M_{NK} \\
    & - \frac{1}{2}\,G^{-1}\,\partial_M G \,\partial_N M^{MN} - \frac{1}{4}\,M^{MN} \,G^{-1}\,\partial_M G\,G^{-1}\,\partial_N G \nonumber\\
    & - \frac{1}{4}\, M^{MN} \,\partial_M G^{\mu\nu}\,\partial_N G_{\mu\nu} \nonumber
\end{align}
Finally there is the topological term $\mathcal{L}_\text{top}$. In the Lagrangian formulation the explicit non-integral form of the topological term in $5+27$ dimensions, which is not manifestly gauge-invariant, is not needed. Instead the topological term is written as a $6+27$ dimensional integral over an exact six-form \eqref{lag-eft-e6-top-term-int-HS}, where {$\mathcal{F}^M := \frac{1}{2} \mathcal{F}^M_{\mu\nu}\, dx^\mu\wedge dx^\nu$} and {$\mathcal{H}_M := \frac{1}{3!} \mathcal{H}_{\mu\nu\rho M} \, dx^\mu\wedge dx^\nu \wedge dx^\rho$} \cite{EFTI-E6}. We discuss the topological term in more detail in section \ref{can-eft-sec-topl-term}, where we also present its non-integral form. Note that the two-form field strength only appears in the topological term and in this sense the two-forms are topological.
\begin{align}
        \mathcal{S}_\text{top} &= \int d^{27}Y \int d^5x \,\mathcal{L}_\text{top} \\
        &= \kappa  \int d^{27}Y \int_{\mathcal{M}_6} \left(d_{MNK} \,\mathcal{F}^M\wedge\mathcal{F}^N\wedge\mathcal{F}^K - 40\,d^{MNK}\mathcal{H}_M\wedge \partial_N \mathcal{H}_K \right) \label{lag-eft-e6-top-term-int-HS}
\end{align} 
Having discussed the structure of the action \eqref{lag-eft-e6-action-integral} we can now discuss the (infinitesimal) gauge transformations that leave this action invariant.\\
Every term in the Lagrangian \eqref{eft-lagrangian-terms} is individually invariant under generalised diffeomorphisms. With the exception of the $p$-forms, fields transform under generalised diffeomorphisms, with parameter $\Lambda^M$, as the generalised Lie derivative \eqref{eft-e6-lie-vec} acting on them $\delta_\Lambda = \mathbb{L}_{\Lambda}$ --- with the appropriate generalised diffeomorphism weights that are listed in table \ref{eft-diff-weights-lag}. From the perspective of eleven-dimensional supergravity the ExFT generalised diffeomorphisms combine the spatial diffeomorphisms of the six original internal dimensions with three-form gauge transformations that have now become geometrised.
\begin{table} \centering 
    \begin{tabular}{|c|c|}
        \hline
        Weight $\lambda$ & Objects \\ \hline\hline
        $-2/3$           & $G^{\mu\nu},\,\hat{R},\,V_\text{pot}$ \\\hline
        $-1/3$           & $\partial_M  ,\,E_\alpha {}^{\mu}$     \\ \hline
        $0$                         & $\partial_\mu  ,\, \mathbb{L}_{A_\mu}, \, d_{MNK},\,M_{MN},\, \hat{R}_{\mu\nu}{}^{\alpha \beta} $     \\ \hline
        $1/3$            & $A^{M}_\mu , \,  \mathcal{F}^{M}_{\mu\nu} ,\,  \Lambda^{M} ,\,E_\mu {}^{\alpha}$     \\ \hline
        $2/3$         & $B_{\mu\nu M} , \, \Xi_{\mu M} , \, \mathcal{H}_{\mu\nu\rho M} , \,G_{\mu\nu}    $       \\ \hline
        $1$                         & $\mathcal{L}_\text{ExFT} $    \\ \hline
        $5/3$             & $E$      \\\hline
    \end{tabular}
    \caption{\label{eft-diff-weights-lag}The exceptional generalised diffeomorphism weights of the most important objects of E$_{6(6)}$ exceptional field theory.}
\end{table}
Due to the tensor hierarchy the differential forms transform somewhat differently under generalised diffeomorphisms. As was mentioned earlier the transformation of the one-forms combines U$(1)^{27}$ gauge transformations with the generalised Lie derivative to transform as \eqref{eft-GD-A}, where $\Lambda^M$ is the generalised diffeomorphism parameter. The additional transformation in \eqref{eft-GD-A}, with parameter $\Xi_{\mu N}$, is a two-form gauge transformation that is induced in the one-forms by the Stückelberg coupling in the field strength. 
\begin{align}
    \delta A^M_\mu & = \mathcal{D}_\mu \Lambda^M - 10\, d^{MNK}\, \partial_K \Xi_{\mu N} \label{eft-GD-A}\\
    \Delta B_{\mu\nu M} & = 2\,\mathcal{D}_{[\mu} \Xi_{\nu]M} + d_{MKL}\,\Lambda^K \, \mathcal{F}^L_{\mu\nu} + \mathcal{O}_{\mu\nu M} \label{eft-GD-B}
\end{align}
The two-forms transform under two-form gauge transformations and generalised diffeomorphisms as \eqref{eft-GD-B} --- with $\Delta B_{\mu\nu M}$ being the modified transformation defined in \eqref{eft-delta-B}. The two-form gauge transformation with parameter $\Xi_{\nu M}$ is of the standard form, but with the derivative covariantised. The generalised diffeomorphism acts on the two-forms as defined by \eqref{eft-transf-f-cov} and \eqref{eft-delta-B}. Additionally there is a shift transformation with parameter $\mathcal{O}_{\mu\nu M}$ that vanishes under the projection \eqref{eft-transf-shift-B} (cf. Stückelberg coupling \eqref{eft-GD-F-covariant-stuckelberg}).
\begin{equation}\label{eft-transf-shift-B}
d^{MNK} \,\partial_K   \mathcal{O}_{\mu\nu M} = 0
\end{equation}
Besides the internal generalised diffeomorphisms all fields transform under external (covariantised) standard diffeomorphisms. The external diffeomorphism parameter $\xi^\mu(x,Y)$ depends on all coordinates (as does every other gauge parameter), but the external diffeomorphism symmetry is only manifest for parameters that do not depend on the internal coordinates, i.e. ${\partial_M\xi^\mu=0\,\,\forall M}$. For gauge parameters that do depend on the internal coordinates non-trivially, i.e. ${\exists M : \partial_M\xi^\mu\neq0}$, the transformation connects the terms in the Lagrangian \eqref{eft-lagrangian-terms} and thus fixes the relative coefficients. The resulting action \eqref{lag-eft-e6-action-integral} is the unique action that is invariant under both internal and external diffeomorphisms, with gauge parameters that depend on all of the internal and external coordinates \cite{EFTI-E6}.\\

It is remarkable that all relative coefficients in the bosonic Lagrangian are already fixed by the bosonic symmetries, because from the supergravity perspective one would expect that the relative coefficients are fixed by requiring the action to be supersymmetric. This fact may indicate an unknown relation between supersymmetry and the exceptional symmetry. Nonetheless the bosonic action \eqref{lag-eft-e6-action-integral}, with all relative coefficients already determined, admits a supersymmetric completion \cite{Musaev2015}.\\

In general the fields transform under external diffeomorphisms as the standard Lie derivative, but with all the external partial derivatives replaced by covariant derivatives $\mathcal{D}_\mu$ \eqref{eft-cov-deriv}. The transformations of the external vielbein \eqref{eft-diff-e} and the scalar fields \eqref{eft-diff-M} are precisely of this form. 
\begin{align}
    \delta_\xi E_\mu{}^\alpha &= \xi^\nu \, \mathcal{D}_\nu E_\mu{}^\alpha +  \mathcal{D}_\mu\xi^\nu \,  E_\nu{}^\alpha \label{eft-diff-e}\\
    \delta_\xi M_{MN} &= \xi^\mu \, \mathcal{D}_\mu M_{MN}  \label{eft-diff-M}\\
    \Delta_\xi B_{\mu\nu M} &= \frac{1}{16\kappa}\,\xi^\rho \, E\,\epsilon_{\mu\nu\rho\sigma\tau} \,\mathcal{F}^{\sigma\tau N}\, M_{MN}\label{eft-diff-B}\\
       \delta_\xi A^M_\mu &= \xi^\nu \, \mathcal{F}^M_{\nu\mu} + M^{MN}\,g_{\mu\nu}\,\partial_N \xi^\nu \label{eft-diff-A}
\end{align}
The transformation of the differential forms is somewhat modified. In the transformation \eqref{eft-diff-B} of the two-forms the naive transformation, that one would expect of a two-form ${\Delta_\xi B_{\mu\nu M} = \xi^\rho\,\mathcal{H}_{\mu\nu\rho M}}$, has been replaced by the two-form equation of motion \eqref{eft-eom-b}, which is the on-shell duality relation between the one- and two-forms mentioned earlier. This replacement is necessary in order to realise the diffeomorphism symmetry of \eqref{lag-eft-e6-action-integral} off-shell.
\begin{equation}\label{eft-eom-b}
    d^{PML} \,\partial_L \left( E\,M_{MN}\,\mathcal{F}^{\mu\nu N}+\kappa \, \epsilon^{\mu\nu\rho\sigma\tau} \, \mathcal{H}_{\rho\sigma\tau M}   \right) = 0
\end{equation}
The first term in the transformation of the one-forms \eqref{eft-diff-A} is the covariantisation of the expected transformation of a one-form under diffeomorphisms.\\

The second term in \eqref{eft-diff-A} originates from a compensating Lorentz transformation. One should note that this term only exists for diffeomorphism parameters that depend on the internal coordinates. Because this correction term depends also on the vielbein and the scalar fields it leads to the connection of different terms in the Lagrangian.\\
Because this term will be relevant later we need to discuss its origin in some more detail. We need to look at how the ExFT relates to eleven-dimensional supergravity to understand why this term has to exist --- here we follow the calculation presented in \cite{EFTI-E6}.\\
We decompose the eleven-dimensional curved index $\hat{\mu}$ in an $11=5+6$ split as $\hat{\mu}=(\mu,m)$ and the flat Lorentz index as $\hat{\alpha}=(\alpha,a)$. The eleven-dimensional vielbein $  E_{\hat{\mu}}{}^{\hat{\alpha}}$ can be parametrised in a Kaluza-Klein-like decomposition as \eqref{eft-11D-vielbein-KKL}, with $\phi=\det(\phi_m{}^a)$ and $\gamma=-1/3$.\footnote{Note that the nomenclature of the flat indices differs from reference \cite{EFTI-E6} in order to match the notation used in this work.} The $\phi_m{}^a$ can be thought of as an internal vielbein. In order to achieve the upper-triangular form in \eqref{eft-11D-vielbein-KKL} part of the eleven-dimensional Lorentz symmetry has to be gauge-fixed.
\begin{equation}\label{eft-11D-vielbein-KKL}
    E_{\hat{\mu}}{}^{\hat{\alpha}} = 
    \begin{pmatrix}
      \phi^\gamma \, E_\mu{}^\alpha & A_\mu^m\, \phi_m{}^a \\
       0 & \phi_m{}^a\\
    \end{pmatrix}
\end{equation}
The eleven-dimensional vielbein transforms as \eqref{eft-11d-diff} under eleven-dimensional diffeomorphisms and Lorentz transformations.
\begin{equation}\label{eft-11d-diff}
 \delta E_{\hat{\mu}}{}^{\hat{\alpha}} = \xi^{\hat{\nu}} \, \partial_{\hat{\nu}}E_{\hat{\mu}}{}^{\hat{\alpha}} +  \partial_{\hat{\mu}}\xi^{\hat{\nu}} \,  E_{\hat{\nu}}{}^{\hat{\alpha}} + \lambda^{\hat{\alpha}}{}_{\hat{\beta}} \, E_{\hat{\mu}}{}^{\hat{\beta}}
\end{equation}
The condition of the upper-triangular form of \eqref{eft-11D-vielbein-KKL} implies that the vanishing component cannot transform non-trivially under gauge transformations, i.e. $E_{m}{}^\alpha=0 \,\,\Rightarrow\,\, \delta E_{m}{}^\alpha=0$. This leads to the restriction \eqref{eft-lorentz-parameter-fix} on the Lorentz gauge parameters and thereby partially fixes the gauge freedom.
  \begin{equation}\label{eft-lorentz-parameter-fix}
     \lambda^{\alpha}{}_{b} = -\phi^\gamma\,\phi_b{}^m\,\partial_m\xi^\nu \,E_\nu{}^\alpha
  \end{equation}
The Lorentz algebra then implies \eqref{eft-lambda-algebra-relation}, which restricts further parameters. 
\begin{equation}\label{eft-lambda-algebra-relation}
    \lambda^\alpha{}_b = - \delta_{ab}\,\eta^{\alpha\beta}\, \lambda^a{}_\beta
\end{equation}
When combined the relations \eqref{eft-lorentz-parameter-fix} and \eqref{eft-lambda-algebra-relation} for the Lorentz gauge parameters lead to the correction term in the transformation of $A_\mu^m$ \eqref{eft-transf-A-KK-correction} when looking at the transformation of the $E_{\mu}{}^a$ component of the eleven-dimensional vielbein. In \eqref{eft-transf-A-KK-correction} we have defined $\phi^{mn} := \phi_a{}^m\,\phi^{an}$ which is the precursor to the generalised metric $M_{MN}$.
\begin{equation}\label{eft-transf-A-KK-correction}
    \delta_\xi A^m_\mu = \xi^\nu \, F^M_{\nu\mu} + \phi^{2\gamma}\,\phi^{mn}\,g_{\mu\nu}\,\partial_n \xi^\nu
\end{equation}
The transformation \eqref{eft-transf-A-KK-correction} in the Kaluza-Klein rewriting of eleven-dimensional supergravity becomes the transformation \eqref{eft-diff-A} in the extended exceptional geometry. We should note in particular that the correction term in \eqref{eft-transf-A-KK-correction} was a direct consequence of the parametrisation \eqref{eft-11D-vielbein-KKL}. Furthermore the sign of the correction term is completely fixed. We come back to this fact in section \ref{can-sec-External-diffeomorphisms} when we discuss the external diffeomorphisms in the canonical formalism.\\

Finally the external vielbein of ExFT transforms under the five-dimensional external Lorentz transformations as \eqref{eft-vielbein-Lorentz-ext}.
\begin{equation}\label{eft-vielbein-Lorentz-ext}
    \delta_\lambda E_\mu{}^\alpha = \lambda^\alpha{}_\beta\, E_\mu{}^\beta 
\end{equation}
Overall the action \eqref{lag-eft-e6-action-integral} is invariant under external Lorentz transformations, external diffeomorphisms, internal generalised diffeomorphisms, two-form gauge transformations and certain shift transformations.\\

The gauge algebra of the external and internal diffeomorphism transformations is structured as follows \cite{Baguet:2015xha}.
The commutator of the covariantised external diffeomorphisms can be written as \eqref{eft-alg-ee}, it is an external diffeomorphism plus an additional internal diffeomorphism. Furthermore there may appear tensor gauge transformations of higher degree differential forms, from the tensor hierarchy, which are indicated by the dots in \eqref{eft-alg-ee}.
\begin{equation}\label{eft-alg-ee}
    [\delta_{\xi_1} , \delta_{\xi_2} ] = \delta_{\xi_{12}}  + \delta_{\Lambda_{12}} +\dots 
\end{equation}
The effective external diffeomorphism parameter $\xi_{12}$ is given by \eqref{eft-alg-ee-xi}, it is a covariantised version of the expected commutator of the original parameters. The effective internal diffeomorphism parameter $\Lambda_{12}^M$ is given by \eqref{eft-alg-ee-lambda} and its form is essentially the transformation of the one-forms under external diffeomorphisms (cf. equation \eqref{eft-diff-A}).
\begin{align}
    \xi_{12}^\mu &:= \xi^\nu_2 \, \mathcal{D}_\nu \xi^\mu_1 - \xi^\nu_1 \, \mathcal{D}_\nu \xi^\mu_2  \label{eft-alg-ee-xi}\\
    \Lambda_{12}^M &:= \xi^\mu_2 \,\xi^\nu_1 \,\mathcal{F}_{\mu\nu}^M - 2 \, M^{MN}\,g_{\mu\nu} \,\xi^\mu_{[2}\,\partial_N \xi^\nu_{1]}   \label{eft-alg-ee-lambda}
\end{align}
The commutator of an external and an internal diffeomorphism \eqref{eft-alg-ei} is given by an external diffeomorphism and a two-form gauge transformation.
\begin{equation}\label{eft-alg-ei}
 [\delta_{\Lambda} , \delta_{\xi} ] = \delta_{\xi'}  + \delta_{\Xi'}
\end{equation}
The effective gauge parameter $  \xi'^\mu $, of the external diffeomorphism in \eqref{eft-alg-ei}, is the generalised Lie derivative of the original external parameter \eqref{eft-alg-ie-xi}. The effective parameter of the tensor gauge transformation $\Xi'$ is given by \eqref{eft-alg-ie-Xii}, it is a projection of the transformation of the one-forms, under external diffeomorphisms \eqref{eft-diff-A}, which is of the form of a generalised diffeomorphism acting on the two-forms \eqref{eft-GD-B}.
\begin{align}
   \xi'^\mu &:= - \Lambda^M\,\partial_M\xi^\mu \label{eft-alg-ie-xi}\\
   \Xi'_{\mu M} &:=-d_{MNK}\,\left(\xi^\nu\,\mathcal{F}_{\nu\mu}^N+ M^{KL}\, g_{\mu\nu} \,\partial_L \xi^\nu \right)\,\Lambda^K  \label{eft-alg-ie-Xii}
\end{align}
The commutator of two internal generalised diffeomorphisms is given by the commutator of the generalised Lie derivatives \eqref{eft-algebra-lie-GD}, we can write the subalgebra of generalised diffeomorphisms equivalently as \eqref{eft-alg-ii}. 
\begin{equation}\label{eft-alg-ii}
 [\delta_{\Lambda_1} , \delta_{\Lambda_2} ] = \delta_{\Lambda_{12}}
\end{equation}
The effective gauge parameter in \eqref{eft-alg-ii} is then given by the E-bracket \eqref{eft-alg-ii-lambda}.
\begin{equation}\label{eft-alg-ii-lambda}
    \Lambda_{12}^M := [\Lambda_2,\Lambda_1]^M_{\text{E}}
\end{equation}
As was mentioned in section \ref{sec-gen-ext-excep-geom}, the generalised diffeomorphism algebra \eqref{eft-alg-ii} closes only up to terms that vanish upon application of the section condition \eqref{eft-e6-section-condition-1}.\\

In order to go back from the exceptional field theory to eleven-dimensional supergravity one has to solve the section condition \eqref{eft-e6-section-condition-1} for a subset of the internal coordinates. The idea is to reverse the extension of the internal geometry by taking away coordinates, in a way that is consistent with the section condition. Solutions of the section condition contain at most 6 of the 27 coordinates. The solution that leads to eleven-dimensional supergravity is found by embedding the subgroup GL(6) into E$_{6(6)}$ as in \eqref{eft-embed-GL6-in-E66}.
\begin{equation}\label{eft-embed-GL6-in-E66}
    \text{GL}(6) =  \text{SL}(6)\times  \text{GL}(1) \subset \text{E}_{6(6)}
\end{equation}
The fundamental $27$ representation and the adjoint $78$ representation of E$_{6(6)}$ decompose into representations of GL(6) according to equation \eqref{eft-fundamental-decompose} and \eqref{eft-adjoint-decompose} respectively. The index indicates the weight under the GL(1).
\begin{align}
    27 &\rightarrow 6_{+1} \oplus 15_0' \oplus 6_{-1} \label{eft-fundamental-decompose}\\
    78  &\rightarrow 1_{-2} \oplus 20_{-1} \oplus (1\oplus35)_{0} \oplus20_{+1} \oplus 1_{+2}\label{eft-adjoint-decompose}
\end{align}
The internal coordinates $Y^M$ decompose as \eqref{eft-decomp-coords-internal} according to \eqref{eft-fundamental-decompose}, with $y_{mn}=y_{[mn]}$ antisymmetric and the overline indicating the difference of the GL(1) weight.
\begin{equation}\label{eft-decomp-coords-internal}
    Y^M \rightarrow (y^m, y_{mn}, y^{\overline{m}})
\end{equation}
The section condition \eqref{eft-e6-section-condition-1} itself has to be decomposed according to \eqref{eft-fundamental-decompose} too and the only non-vanishing components of the invariant symbol $d^{MNK}$ are \eqref{eft-section-condi-d-symbl}.
\begin{equation}\label{eft-section-condi-d-symbl}
    d^{m\, \overline{n}}{}_{kl} = \frac{1}{\sqrt{5}} \, \delta^m_{[k} \, \delta^{\overline{n}}_{l]}\,\,\,\,,\,\,\,\,d_{mn\,kl\,pq}=\frac{1}{4\sqrt{5}}\,\epsilon_{mnklpq}
\end{equation}
From the decomposed coordinates \eqref{eft-decomp-coords-internal} we can choose the $6_{+1}$ coordinates $y^m$ and drop the other internal coordinates. Keeping only the coordinates $y^m$ and therefore only the internal derivatives $\partial_m$, solves the section condition \eqref{eft-e6-section-condition-1} when inserting the decomposition of the $d$-symbol \eqref{eft-section-condi-d-symbl}. The full set of coordinates that survive are then the $11=5+6$ coordinates $(x^\mu,y^m)$, equivalent to the Kaluza-Klein-like split of the coordinates of eleven-dimensional supergravity. Decomposing all objects according to \eqref{eft-fundamental-decompose} and only keeping the $6_{+1}$ components, one can recover the full structure of eleven-dimensional supergravity in the Kaluza-Klein-like rewriting --- the field strength Hodge dualisations that were described earlier have to be undone in the process \cite{EFTI-E6}.\\

At no point in this process were any degrees of freedom of eleven-dimensional supergravity truncated. In the construction of the ExFT the degrees of freedom of eleven-dimensional supergravity are first rearranged in a Kaluza-Klein-like $5+6$-dimensional split of the coordinates and part of the Lorentz symmetry is gauge fixed, but no truncation is carried out. Then the degrees of freedom are dualised and rearranged on an extended $5+27$-dimensional generalised exceptional space-time, while the section condition \eqref{eft-e6-section-condition-1} is simultaneously imposed. When the section condition is solved, for example by decomposing everything according to \eqref{eft-decomp-coords-internal} and keeping only the $6_{+1}$ parts of the Kaluza-Klein-like split, the eleven-dimensional supergravity is again recovered. One can therefore see exceptional field theory as an E$_{6(6)}$ covariant rewriting of eleven-dimensional supergravity on an extended generalised exceptional geometry.\\

Other non-trivial solutions to the section condition, besides the one described above, exist. For example, one can embed GL(5)$\times$SL(2) $\subset$ E$_{6(6)}$ and only keep the $(5,1)_{+4}$ components of all objects. This is an inequivalent, but consistent, five-dimensional solution to the section condition that leads to the ten-dimensional type IIB supergravity \cite{EFTI-E6}.\\

The section condition can moreover be trivially solved by requiring $\partial_M=0\,\,\,\forall M$. In the case of the trivial solution only the five external coordinates survive. The resulting theory is the ungauged manifestly E$_{6(6)}$ invariant maximal five-dimensional supergravity, which was first described in reference \cite{Cremmer}. This theory can also be obtained by a reduction of eleven-dimensional supergravity on a six-torus, although this does not directly lead to the manifestly E$_{6(6)}$ invariant form. The canonical formulation and analysis of this theory has been investigated in \cite{KreutzerSUGRA}.

\subsection{The explicit non-integral topological term}\label{can-eft-sec-topl-term}
In order to be able to write down the external $1+4$ dimensional ADM decomposition of the ExFT and carry out the Legendre transformation of the Lagrangian --- to find the canonical Hamiltonian --- we need to know all terms of the Lagrangian density explicitly and in a non-integral form. The only term in the Lagrangian \eqref{eft-lagrangian-terms} that is not explicitly stated in its $5+27$-dimensional form is the topological term, which is instead given in its manifestly covariant form as an external six-dimensional boundary integral over an exact six-form \eqref{lag-eft-e6-top-term-int-HS} \cite{EFTI-E6}. Reference \cite{EFTI-E6} does however state the general variation of the topological term explicitly as \eqref{can-top-lag-general-var-EFTI}. The general variation \eqref{can-top-lag-general-var-EFTI} may be sufficient to calculate the canonical momenta but in order to carry out the Legendre transformation of the Lagrangian \eqref{eft-lagrangian-terms} the explicit topological term is needed.
\begin{align}\label{can-top-lag-general-var-EFTI}
\delta\mathcal{L}_\text{top} = \kappa \,\epsilon^{\mu\nu\rho\sigma\tau}(& \frac{3}{4}\, d_{MNK} \,\mathcal{F}_{\mu\nu}^M\,\mathcal{F}_{\rho\sigma}^N\,\delta A^K_\tau + 5\, d^{MNK}\, d_{KQP} \partial_N \mathcal{H}_{\mu\nu\rho M} A_\sigma^P \delta A_\tau ^Q\\
& +5\, d^{MNK}\,\partial_N\mathcal{H}_{\mu\nu\rho M }\delta B_{\sigma\tau K})\nonumber
\end{align}
The structure of ExFT and gauged supergravity is in general very similar and reference \cite{EFTI-E6} mentions the similarity between the topological terms explicitly. 
We can find the explicit non-integral form of the topological term by considering a general ansatz, inspired by the topological term of five-dimensional maximal gauged supergravity (see equation (3.11) of reference \cite{deWit:2004nw}) and then comparing its general variation to \eqref{can-top-lag-general-var-EFTI} in order to fix the relative coefficients of all terms.\\

In five-dimensional gauged supergravity the generators of the group of the gauging are called $X_M$ \cite{deWit:2004nw,Samtleben:2008pe}. The generators obey an algebra \eqref{eft-top-gauged-sugra-alg} with ``structure constants'' $X_{MN}{}^P$ that appear in the topological term of gauged supergravity.
\begin{equation}\label{eft-top-gauged-sugra-alg}
    [X_M, X_N] = -X_{MN}{}^P \, X_P
\end{equation}
The structure constants $X_{MN}{}^P$ are however not antisymmetric and we can split them into a symmetric and antisymmetric part as \eqref{eft-top-gauged-sugra-split-sym} (cf. equation (3.15) in \cite{Samtleben:2008pe}). 
\begin{equation}\label{eft-top-gauged-sugra-split-sym}
    X_{MN}{}^P = X_{(MN)}{}^P + X_{[MN]}{}^P  
\end{equation}
The split \eqref{eft-top-gauged-sugra-split-sym} can be compared to the symmetric and antisymmetric parts of the Dorfman bracket \eqref{eft-rel-dorfman-e}. Similarly we can compare the improved one-form field strength of gauged supergravity (see equations (3.2) and (3.4) of reference \cite{deWit:2004nw}) to that of ExFT \eqref{eft-GD-F-covariant-stuckelberg}. We find that the E-bracket takes the place of the antisymmetric part of the structure constants $X_{[MN]}{}^K\,A^M_\mu\,A^N_\nu  \sim \, [A_\mu, A_\nu]^K_E$ and the projector $Z^{KL}\sim d^{KLM}\,\partial_M$ (cf. equation (3.4) of \cite{deWit:2004nw}) corresponds roughly to the symmetric part of the structure constants. The precise relations are not relevant as we should introduce general coefficients in the ansatz. Inserting these relations in the topological term of gauged supergravity (see equation (3.11) of reference \cite{deWit:2004nw}) we arrive at a suitable ansatz. Comparing the general variation of the ansatz to the general variation \eqref{can-top-lag-general-var-EFTI} we can fix all coefficients. \\

We find that the topological term of the E$_{6(6)}$ exceptional field theory can take the form of equation \eqref{topological-term-E-six} and that this expression yields the same general variation $\delta\mathcal{L}_\text{top}$ as given in equation \eqref{can-top-lag-general-var-EFTI}.
\begin{align}
\label{topological-term-E-six}
\mathcal{L}_{\text{top}} = &-\frac{5\kappa}{2} \epsilon^{\mu\nu\rho\sigma\tau} \enspace d^{MNR}\enspace\partial_R B_{\mu\nu M}\\
&\times\left[3 D_{\rho}B_{\sigma\tau N}-6d_{NKL}\,A^K_{\rho}\left(\partial_\sigma A^L_{\tau}-\frac{1}{3}[A_\sigma,A_{\tau}]^L_\text{E}\right) \right] \nonumber\\
            &+ \kappa\,\, \epsilon^{\mu\nu\rho\sigma\tau}d_{MNP}\enspace A^N_\mu \enspace\partial_\nu A^M_\rho \enspace\partial_\sigma A^P_\tau\nonumber\\
            &-  \frac{3\kappa}{4}\epsilon^{\mu\nu\rho\sigma\tau}d_{MNP} \enspace A^N_\mu\enspace\left[A_\nu,A_\rho\right]^M_E\enspace\partial_\sigma A^P_\tau\nonumber\\
            &+  \frac{3\kappa}{20}\epsilon^{\mu\nu\rho\sigma\tau}d_{MNP}\enspace A^N_\mu\enspace \left[A_\nu,A_\rho\right]^M_E \enspace\left[A_\sigma,A_\tau\right]^P_E\nonumber
\end{align}
We can write \eqref{topological-term-E-six} in a slightly more covariant form by making use of the definition of two-form field strength \eqref{eft-h-field-strength} to find equation \eqref{topological-term-E-six-v2}. In the following we will only need the more explicit form \eqref{topological-term-E-six}.
\begin{align}
\label{topological-term-E-six-v2}
\mathcal{L}_{\text{top}} = &-\frac{5\kappa}{2} \epsilon^{\mu\nu\rho\sigma\tau} \enspace d^{MNR}\enspace\partial_R B_{\mu\nu M}\\
&\times\left[\mathcal{H}_{\rho\sigma\tau N}-3d_{NKL}\,A^K_{\rho}\left(\partial_\sigma A^L_{\tau}-\frac{1}{3}[A_\sigma,A_{\tau}]^L_\text{E}\right) \right] \nonumber\\
            &+ \kappa\,\, \epsilon^{\mu\nu\rho\sigma\tau}d_{MNP}\enspace A^N_\mu \enspace\partial_\nu A^M_\rho \enspace\partial_\sigma A^P_\tau\nonumber\\
            &- \frac{3\kappa}{4}\epsilon^{\mu\nu\rho\sigma\tau}d_{MNP} \enspace A^N_\mu\enspace\left[A_\nu,A_\rho\right]^M_E\enspace\partial_\sigma A^P_\tau\nonumber\\
            &+ \frac{3\kappa}{20}\epsilon^{\mu\nu\rho\sigma\tau}d_{MNP}\enspace A^N_\mu\enspace \left[A_\nu,A_\rho\right]^M_E \enspace\left[A_\sigma,A_\tau\right]^P_E\nonumber
\end{align}
The numerical value of the overall coefficient of the topological term in the Lagrangian, as used in this work, is $\kappa=+\sqrt{10}/6$ \cite{EFTI-E6, Baguet:2015xha, Musaev2015, Hohm:2019bba}. The bosonic theory does not fix the sign of this constant, only its modulus and therefore its sign is conventional \cite{Musaev2015}.\footnote{There is some confusion in the literature about this constant that originates from a hidden renaming of the constant in reference \cite{EFTI-E6} --- the calculations are nonetheless all correct, if we are aware of this renaming and do not mix up the different values. If we call $\kappa_1=\sqrt{10}/6$ and $\kappa_2=\frac{3}{4}\kappa_1=\sqrt{10}/8$, then $\kappa_1$ is used in the equations (3.7), (3.8) and (3.9) of \cite{EFTI-E6}. However starting from equation (3.29) of \cite{EFTI-E6}, when the value of this constant is determined and in particular in the equation (3.31) which states $\kappa^2=\frac{5}{32}$ the rescaled constant $\kappa_2$ is used. The additional factor $\frac{3}{4}$ comes from taking the variation of the term, as can be seen in equation (3.8) of \cite{EFTI-E6}. References \cite{Musaev2015, Hohm:2019bba} continue to use $\kappa_2$ consistently, while reference \cite{Baguet:2015xha} uses $\kappa_1$ consistently.}\\

The coefficient of the topological term of ungauged five-dimensional maximal supergravity, as described in \cite{KreutzerSUGRA} --- we can call the coefficient $\kappa_{5}$ --- is related to the that of ExFT by $\kappa_5 = \frac{1}{4}\kappa$. This is due to the use of the ungauged abelian field strength in the writing of the topological term.

\section{Canonical topological 2-forms in \texorpdfstring{$5+27$}{5+27} dimensions}\label{can-sec-twoforms-model}
As a preparation to the canonical formulation of the full E$_{6(6)}$ ExFT we investigate the canonical formulation of only the topological kinetic term of the two-forms $B_{\mu\nu M}$ (cf. equation \eqref{topological-term-E-six}) in this section. For this model theory to be a good analogy to the two-forms of ExFT we need to consider the theory on the same $5+27$-dimensional extended geometry as ExFT itself.\\

We begin by calculating the canonical momenta of the two-forms and constructing the canonical Hamiltonian of the two-forms in section \ref{ham-can-B-model-canmom-canham-sec}. In section \ref{ham-can-B-sec-constr-consist} we find the complete set of canonical constraints. We then compute all canonical transformations and the full constraint algebra in section \ref{ham-can-sec-gauge-dof}. In doing so we can confirm that there are no propagating degrees of freedom in this theory. Moreover we find that external diffeomorphisms are not canonically generated because of the topological nature of the two-forms in this model. Finally, in section \ref{ham-can-sec-dirac-brackets}, we identify some obstacles related to the construction of Dirac brackets in exceptional generalised geometry for constraint algebras of a certain form.\\

For the model we want to consider an action of the form \eqref{ham-can-B-model-action} with five external coordinates $x^\mu$ and 27 internal coordinates $Y^M$. The integral over the internal coordinates of the generalised exceptional geometry in \eqref{ham-can-B-model-action} is taken to be symbolic, since we do not know how to carry out this integral explicitly while observing the section condition \eqref{eft-e6-section-condition-1}.
\begin{equation}\label{ham-can-B-model-action}
    S_B = \int \,d^5x\,\int\,d^{27}Y\,\mathcal{L}_B
\end{equation}
The Langrangian we want to consider can be written as in equation \eqref{ham-can-B-model-lagrangian1}, with $\varrho$ being the overall constant. Compared to the two-form kinetic term in ExFT \eqref{topological-term-E-six} we have dropped the covariantisation of the external derivative in \eqref{ham-can-B-model-lagrangian1} in order to make the model as simple as possible --- the theory is nonetheless interesting enough to be useful. As a consequence thereof the internal generalised diffeomorphisms are not generated canonically in this model theory.
\begin{equation}\label{ham-can-B-model-lagrangian1}
   \mathcal{L}_\text{B}=-\varrho \,\epsilon^{\mu\nu\rho\sigma\tau}\, d^{MNR}\, \partial_R B_{\mu\nu M}\, \partial_\rho B_{\sigma\tau N} 
\end{equation}
Alternatively we can make use of the naive two-form field strength $H_M := \text{d}B_M$ ({$ H_{\rho\sigma\tau N} := 3 \, \partial_{[\rho} B_{\sigma\tau] N} $}) to write the Lagrangian as in \eqref{ham-can-B-model-lagrangian2} or \eqref{ham-can-B-model-lagrangian3}. 
It is important to note here that this topological Lagrangian is linear in the field strength and hence in the time derivative. It is this fact that leads to the peculiar canonical structure that we find in the following analysis. 
\begin{align}
   \mathcal{L}_\text{B}= &-\frac{\varrho}{30} \, d^{MNR}\, \partial_R B_{M}\wedge H_{N} \label{ham-can-B-model-lagrangian2}\\
  = &-\frac{\varrho}{3} \,\epsilon^{\mu\nu\rho\sigma\tau}\, d^{MNR}\, \partial_R B_{\mu\nu M}\, H_{\rho\sigma\tau N} \label{ham-can-B-model-lagrangian3}
\end{align}
In analogy to the topological term of ExFT we can furthermore write the action \eqref{ham-can-B-model-action} equivalently as a boundary term in a $6+27$ dimensional geometry \eqref{ham-can-B-model-action-6dimensions}. Because the two-form field strength is closed $\text{d}H_M=0$ we can write the action also as \eqref{ham-can-B-model-action-6dimensions2}.
\begin{align}
    S_B & = -\frac{\varrho}{30} \int \,d^6\tilde{x}\,d^{27}Y\,\text{d}\left(  d^{MNR}\,\partial_R B_{M}\wedge H_{N} \right) \label{ham-can-B-model-action-6dimensions} \\
    & = -\frac{\varrho}{30} \int \,d^6\tilde{x}\,d^{27}Y\, d^{MNR}\, \partial_R H_{M}\wedge H_{N}  \label{ham-can-B-model-action-6dimensions2}
\end{align}
We can decompose the external five-dimensional indices in a $1+4$-dimensional space-time split as $\mu=(t,m)$, where $t$ indicates the curved time index and $m=1,\dots,4$ the curved spatial index.
The space-time split of the Lagrangian can then be written as \eqref{ham-can-B-model-lagrangian-ADM}.
\begin{align}
   \mathcal{L}_\text{B}=&-\varrho \,\epsilon^{\mu\nu\rho\sigma\tau}\, d^{MNR}\, \partial_R B_{\mu\nu M}\, \partial_\rho B_{\sigma\tau N} \\
    =& -\varrho  \,\epsilon^{tnrsl}\, d^{MNR}\, \partial_R B_{nr M}\, \partial_t B_{sl N} \label{ham-can-B-model-lagrangian-ADM}\\
    & -2\varrho\,\epsilon^{tnrsl}\, d^{MNR}\, \partial_R B_{tn M}\, \partial_r B_{sl N} \nonumber\\
    & +2\varrho \,\epsilon^{tnrsl}\, d^{MNR}\, \partial_R B_{nr M}\, \partial_s B_{tl N} \nonumber
\end{align}
\subsection{Canonical momenta and canonical Hamiltonian}\label{ham-can-B-model-canmom-canham-sec}
From the space-time split of the Lagrangian \eqref{ham-can-B-model-lagrangian-ADM} we can read off the canonical momenta of the time and spatial components of the two-forms, which we call $\Pi^{tlN}(B),\,\Pi^{slN}(B)$, the canonical momenta can be stated as \eqref{hamilton-2forms-canonical-momentum-B1} and \eqref{hamilton-2forms-canonical-momentum-B2}. Due to the linearity of the Lagrangian in the time derivative $\partial_t$ we find that the canonical momenta do not contain any time derivatives themselves --- therefore implying that they all lead to primary constraints, which we name $\mathcal{H}_\text{P1}$ and $\mathcal{H}_\text{P2}$. 
\begin{align}
(\mathcal{H}_\text{P1})^{lN}:=\Pi^{tlN}(B) & =0  
\label{hamilton-2forms-canonical-momentum-B1}\\
(\mathcal{H}_\text{P2})^{slN}:=\Pi^{slN}(B) + 2\varrho\, \epsilon^{tmnsl} \,d^{MNR} \, \partial_R B_{mnM} & =0 
\label{hamilton-2forms-canonical-momentum-B2}
\end{align}
The Legendre transformation of the Lagrangian \eqref{ham-can-B-model-lagrangian-ADM} is given by \eqref{hamilton-2forms-canonical-legendre}, the factor of $1/2$ needs to be inserted in the second term in order to avoid overcounting. Inserting the space-time decomposition of the Lagrangian and using the definition of the primary constraints we can write the canonical Hamiltonian $\mathcal{H}_\text{B}$ as \eqref{hamilton-2forms-canonical-hamiltonian}.
\begin{align}
    \mathcal{H}_\text{B} &=  \dot{B}_{tlN}\cdot\, \Pi^{tlN}(B) +\frac{1}{2} \, \dot B_{mnN}\cdot \Pi^{mnN}(B) - \mathcal{L}_\text{B}\label{hamilton-2forms-canonical-legendre} \\
    &=  \dot{B}_{tlN}\cdot\, (\mathcal{H}_\text{P1})^{lN} + B_{tn M}\cdot (\mathcal{H}_\text{S1})^{nM}\label{hamilton-2forms-canonical-hamiltonian}
\end{align}
We have introduced the object $\mathcal{H}_\text{S1}$, as defined by equation \eqref{hamilton-2forms-canonical-constr-S1}, here --- already indicating that this term will yield a secondary constraint.
\begin{equation}\label{hamilton-2forms-canonical-constr-S1}
  (\mathcal{H}_\text{S1})^{nM} :=  -4\varrho\,\epsilon^{tnrsl}\,d^{MNR} \, \partial_R\partial_r  B_{sl N}
\end{equation}
Note that we can rewrite \eqref{hamilton-2forms-canonical-constr-S1}, using the two-form field strength, as in \eqref{hamilton-2forms-canonical-constr-S1-H}.
\begin{equation}\label{hamilton-2forms-canonical-constr-S1-H}
  (\mathcal{H}_\text{S1})^{nM} =  -\frac{4\varrho}{3}\,\epsilon^{tnrsl}\,d^{MNR} \, \partial_R H_{rsl N}
\end{equation}
To verify the consistency of the primary constraints \eqref{hamilton-2forms-canonical-momentum-B1} and \eqref{hamilton-2forms-canonical-momentum-B2} we need to make use of the total Hamiltonian $  \mathcal{H}_\text{B-Total}$ \eqref{hamilton-2forms-total-ham} which is defined as the canonical Hamiltonian plus a linear phase space sum over the primary constraints with general coefficients.
\begin{equation}\label{hamilton-2forms-total-ham}
      \mathcal{H}_\text{B-Total} :=   \mathcal{H}_\text{B}+ (u_1)_{lN}\cdot  (\mathcal{H}_\text{P1})^{lN} + (u_2)_{slN}\cdot (\mathcal{H}_\text{P2})^{slN}
\end{equation}
\subsection{Constraints and consistency}\label{ham-can-B-sec-constr-consist}
Before we begin with checking the consistency of the primary constraints it is useful to determine their algebra using the fundamental equal time Poisson brackets given in equations \eqref{hamilton-2forms-canonical-poisson1} and \eqref{hamilton-2forms-canonical-poisson2}. Here we use the notation that $X_1:=(x_1,Y_1)$ denotes both the spatial external and internal coordinates, $X_1-X_2=(x_1-x_2, Y_1-Y_2)$ and $\delta^{(4+27)}(X_1-X_2)$ is the $(4+27)$-dimensional Dirac delta distribution.
\begin{align}
      \{ B_{tlR}(X_1), \Pi^{tnS}(B)(X_2)\}  & = \delta^n_l \delta^S_R \,\delta^{(4+27)}(X_1-X_2) \label{hamilton-2forms-canonical-poisson1} \\
       \{ B_{klR}(X_1), \Pi^{mnS}(B)(X_2)\}  & = \left(\delta^m_k \delta^n_l - \delta^m_l \delta^n_k\right)  \delta^S_R \,\delta^{(4+27)}(X_1-X_2) \label{hamilton-2forms-canonical-poisson2}
\end{align}
What we find is that the primary constraints all Poisson-commute amongst each other. This implies that their total time evolution, as generated by the total Hamiltonian, is equivalent to the time evolution generated by the canonical Hamiltonian \eqref{hamilton-2forms-canonical-hamiltonian}.
\begin{align}
    \{(\mathcal{H}_\text{P1})^{kK} ,(\mathcal{H}_\text{P1})^{lL}  \} &= 0 \\
    \{(\mathcal{H}_\text{P1})^{kK} ,(\mathcal{H}_\text{P2})^{mnM} \} &= 0\\
    \{(\mathcal{H}_\text{P2})^{klK}  ,(\mathcal{H}_\text{P2})^{mnM} \} & =0
\end{align}
Consistency of the primary constraints is equivalent to the requirement that their total time evolution is vanishing, thus preserving the constraints in time \cite{Henneaux-Teitelboim}. Requiring the time evolution of the primary constraint $\mathcal{H}_\text{P1}$ to vanish \eqref{ham-can-B-P1-consistency-S1} confirms that the expression $(\mathcal{H}_\text{S1})$ is indeed a secondary constraint.
\begin{equation}\label{ham-can-B-P1-consistency-S1}
    \{(\mathcal{H}_\text{P1})^{kK}  , \mathcal{H}_\text{B-Total}\} = - (\mathcal{H}_\text{S1})^{nM} =0
\end{equation}
Before we go on to check the consistency of the primary constraint $\mathcal{H}_\text{P2}$ it is convenient to define the smeared (or integrated) secondary constraint $\mathcal{H}_\text{S1}[\Phi]$ as in equation \eqref{hamilton-2forms-canonical-S1-smeared}. The smeared constraint allows us to avoid having to write derivatives of Dirac delta distributions (see e.g. references \cite{Blair_2014,KreutzerSUGRA} for the use of smeared constraints). The smearing function $\Phi_{nM}(x,Y)$ can be thought of as a tensor of test functions which allows us to integrate the constraint over the full spatial geometry and makes it possible to move derivatives onto the smearing function.
\begin{equation}\label{hamilton-2forms-canonical-S1-smeared}
  \mathcal{H}_\text{S1}[\Phi] :=  \int d^4x \,\int d^{27}Y\,
\Phi_{n M}\cdot (\mathcal{H}_\text{S1})^{nM}
\end{equation}
Calculating the Poisson brackets of the smeared constraint with the primary constraints we find that $\mathcal{H}_\text{P1}$ Poisson-commutes with the new secondary constraint, whereas $\mathcal{H}_\text{P2}$ does not. It follows that $\mathcal{H}_\text{P2}$ and $\mathcal{H}_\text{S1}$ are second class constraints.
\begin{align}
\{(\mathcal{H}_\text{P1})^{mM}  ,   \mathcal{H}_\text{S1}[\Phi] \} & = 0 \\
   \{(\mathcal{H}_\text{P2})^{mnM}  ,   \mathcal{H}_\text{S1}[\Phi] \} & = 8\varrho\,\epsilon^{tklmn}\,d^{KLM} \, \partial_K\partial_l \Phi_{kL} \label{hamilton-2forms-canonical-S1-P2-Poisson}
\end{align}
With the explicit expression for the Poisson bracket of equation \eqref{hamilton-2forms-canonical-S1-P2-Poisson} it is straightforward to compute the total time evolution of the primary constraint $\mathcal{H}_\text{P2}$. We find that it is given by the same expression as \eqref{hamilton-2forms-canonical-S1-P2-Poisson} but with the field $B_{tkL}$ taking the place of the smearing function. Requiring this time evolution to vanish leads to another secondary constraint $\mathcal{H}_\text{S2}$ as defined in \eqref{hamilton-2forms-canonical-S2}.
\begin{equation}\label{hamilton-2forms-canonical-S2}
    \{(\mathcal{H}_\text{P2})^{mnM}  , \mathcal{H}_\text{B-Total}\} =: (\mathcal{H}_\text{S2})^{mnM} = 8\varrho\,\epsilon^{tklmn}\,d^{KLM} \, \partial_K\partial_l B_{tkL} = 0
\end{equation}
The consistency of the secondary constraints has to be verified similarly, but no new constraints can be found by requiring their time evolution to vanish --- we do however find that the coefficient functions in the total Hamiltonian should be vanishing $u_1\equiv0, \, u_2\equiv0$. The consistency and constraint finding procedure thus terminates and the complete set of constraints is listed below.
\begin{align}
 (\mathcal{H}_\text{P1})^{lN} & = \Pi^{tlN}(B)  \\
(\mathcal{H}_\text{P2})^{slN} & =\Pi^{slN}(B) + 2\varrho\, \epsilon^{tmnsl} \,d^{MNR} \, \partial_R B_{mnM} \\
 (\mathcal{H}_\text{S1})^{nM} & =  -4\varrho\,\epsilon^{tnrsl}\,d^{KLM} \, \partial_K\partial_r  B_{sl L}  \\
 (\mathcal{H}_\text{S2})^{mnM}& = +8\varrho\,\epsilon^{tklmn}\,d^{KLM} \, \partial_K\partial_l B_{tkL} 
\end{align}
\subsection{Canonical transformations, algebra and degrees of freedom}\label{ham-can-sec-gauge-dof}
Using the smeared version of all constraints we can write all non-trivial transformations generated by the constraints as follows. In the context of the canonical transformations the smearing functions are interpreted as the (gauge) parameters of the transformations.
\begin{align}
    \{ B_{tnN}  , \mathcal{H}_\text{P1}[\chi_1] \}  & = 2\,(\chi_1)_{nN}   \label{hamilton-2forms-canonical-gauge-transf-shift1}  \\
    \{ B_{mnN}  , \mathcal{H}_\text{P2}[\chi_2] \}  & =2\,(\chi_2)_{mnN}    \label{hamilton-2forms-canonical-gauge-transf-shift2}   \\
      \{ \Pi^{mnN}(B)  ,  \mathcal{H}_\text{P2}[\chi_2]  \}  & =    + 4\varrho\, \epsilon^{tmnsl} \,d^{MNR} \, \partial_R (\chi_2)_{slM}  \label{ham-two-forms-gauge-P2-on-PiB} \\
    \{ \Pi^{mnM}(B)  ,  \mathcal{H}_\text{S1}[\Phi_1]  \}  & = +8\varrho\,\epsilon^{tlkmn}\,d^{KLM} \, \partial_K\partial_k  (\Phi_1)_{l L}   \label{ham-two-forms-gauge-S1-on-PiB}  \\
     \{ \Pi^{tnM}(B)  , \mathcal{H}_\text{S2}[\Phi_2] \}  & =  +16\varrho\,\epsilon^{tnklm}\,d^{KLM} \, \partial_K\partial_k (\Phi_2)_{lmL}  \label{ham-two-forms-gauge-S2-on-PiB}
\end{align}
The shift transformations \eqref{hamilton-2forms-canonical-gauge-transf-shift1} and \eqref{hamilton-2forms-canonical-gauge-transf-shift2} of the two-forms are generic and expected for fields that appear with only a single time derivative in the Lagrangian. Because the primary constraints directly relate the fields to their canonical momenta these shift transformations appear. It should be noted that this in particular includes transformations where the parameter is a derivative, e.g. $(\chi_2)_{mnN} =: \partial_{[m} \lambda_{n]N} $ leading to the perhaps more familiar transformation $  \delta_{\mathcal{H}_\text{P2}[\lambda]}B_{mnN}$ as in equation \eqref{hamilton-2forms-canonical-gauge-transf-shift2-example}. 
\begin{equation}\label{hamilton-2forms-canonical-gauge-transf-shift2-example}
    \delta_{\mathcal{H}_\text{P2}[\lambda]}B_{mnN}=   \{ B_{mnN}  , \mathcal{H}_\text{P2}[\lambda] \}   = 2\,\partial_{[m} \lambda_{n]N}
\end{equation}
In chapter 19 of reference \cite{Henneaux-Teitelboim} it is shown that in the case of free Maxwell theory the shift transformations can be made into the usual form by way of the parameters of the extended Hamiltonian. In our case there are however no first class constraints (as is determined below) and therefore the extended Hamiltonian agrees with the already determined total Hamiltonian, it is thus unclear how an analogous procedure would work in this case. Nonetheless it seems probable that a way of fully rearranging the shift transformations in the usual form should exist and it may be instructive to study the case of three-dimensional Chern-Simons theory $\mathcal{L}=A\wedge F$, whose canonical constraints are structured in a similar way as those of our model theory.\\

Since we can write the action in terms of differential forms as a boundary integral \eqref{ham-can-B-model-action-6dimensions} we should expect that (external) diffeomorphisms are a symmetry of this action. Canonically we do however not find any constraint that leads to diffeomorphism gauge transformations. Fundamentally this is expected in a topological theory because we would normally expect the canonical diffeomorphism constraint to arise from the consistency requirement (i.e. secondary constraint) of the primary constraint that tells us that the canonical momentum of the shift vector is vanishing $\Pi_n(N^n)=0$. But in this model theory there is no metric field and even if there was one it would not appear in the topological term. It is therefore impossible to see the (external) diffeomorphism symmetry of purely topological fields in the canonical formalism and this seems to be a general fact. The existence of Lagrangian symmetries in the canonical formalism has been discussed in \cite{Gracia:1992af}.\\ 

Having computed all non-vanishing canonical transformations we can now determine the full algebra formed by the constraints. There are only two non-vanishing Poisson brackets among the constraints. The first relation \eqref{hamilton-2forms-canonical-S1-P2-Poisson-alg} is equivalent to what we have already seen in equation \eqref{hamilton-2forms-canonical-S1-P2-Poisson}, the other relation is given by equation \eqref{hamilton-2forms-canonical-S2-P1-Poisson-alg}.
\begin{align}
    \{ \mathcal{H}_\text{P1}[\chi_1],  \mathcal{H}_\text{S2}[\Phi_2] \}  & =  +16\varrho\,\epsilon^{tklmn}\,d^{KLM} \, (\chi_1)_{kL}\,\partial_K\partial_l (\Phi_2)_{mnM} \label{hamilton-2forms-canonical-S1-P2-Poisson-alg} \\
    \{ \mathcal{H}_\text{P2}[\chi_2],  \mathcal{H}_\text{S1}[\Phi_1] \}  & =+8\varrho\,\epsilon^{tklmn}\,d^{KLM} \, (\chi_2)_{mnM}\,\partial_K\partial_l  (\Phi_1)_{k L}    \label{hamilton-2forms-canonical-S2-P1-Poisson-alg}
\end{align}
Because all constraints are involved in these two relations we can conclude that all canonical constraints in this model are second class constraints. 
\FloatBarrier
\begin{table}[ht!]\centering
    \begin{tabular}{ccccc}
        \# & Fields & Momenta & Primary & Secondary \\ \hline
         108 & $B_{tnN}$ & $\Pi^{tnN}$ & $(\mathcal{H}_\text{P1})^{nN}$ &  $(\mathcal{H}_\text{S1})^{nN}$ \\ 
         162 & $B_{mnN}$  & $\Pi^{mnN}$ & $(\mathcal{H}_\text{P2})^{mnN}$  & $(\mathcal{H}_\text{S2})^{mnN}$  \\  
    \end{tabular}
\caption{\label{hamilton-2forms-canonical-table}This table lists the number and names of the independent components of all the fields, canonical momenta as well as of the primary and secondary constraints.}
\end{table}
\FloatBarrier
The number of physical degrees of freedom of the theory described by the Lagrangian \eqref{ham-can-B-model-lagrangian1} is zero and hence there are no propagating degrees of freedom --- as is expected in a purely topological theory. Canonically this is because the number of independent components of all the second class constraints $\mathcal{H}_\text{P1}$, $\mathcal{H}_\text{P2}$, $\mathcal{H}_\text{S1}$ and $\mathcal{H}_\text{S2}$ taken together exactly cancels the independent phase space variables of the theory --- as can be seen in table \ref{hamilton-2forms-canonical-table}.
\subsection{Dirac brackets in extended exceptional generalised geometry}\label{ham-can-sec-dirac-brackets}
Because of the existence of second class constraints the next step in the canonical analysis of the theory should be the construction of a Dirac bracket $\{.,.\}_\text{DB}$. We can define symbolic indices $a,b \in \{P1, P2, S1, S2\}$ to label the canonical constraints and define the matrix $ M_{ab}$ as in equation \eqref{ham-can-B-model-M-dirac}.
\begin{equation}
    M_{ab}(x_1,x_2,Y_1,Y_2) := \{\mathcal{H}_a(x_1,Y_1) , \mathcal{H}_b(x_2,Y_2) \} \label{ham-can-B-model-M-dirac}
\end{equation}
The components of the matrix $ M_{ab}$ are given by the constraint algebra relations but with the smearing parameters now replaced by Dirac delta distributions and thus dependent on all coordinates. The indices that were contracted into parameters are now open, but we will understand them to be covered by the symbolic indices $a,b$ too.\\

One may now try to define the Dirac bracket for this theory as in equation \eqref{ham-can-B-model-dirac-bracket}, however there are several difficulties and potential problems with this definition.
\begin{align}
   \{f, g \}_\text{DB} := \{f, g \} &-\underset{a,b}{\sum} \int d^4x_1\int d^4x_2\int d^{27}Y_1\int d^{27}Y_2 \nonumber\\
     &\cdot\left(\{f , \mathcal{H}_a(x_1, Y_1) \} \, M^{ab}(x_1,x_2,Y_1,Y_2) \, \{ \mathcal{H}_b(x_2,Y_2), g  \} \right) \label{ham-can-B-model-dirac-bracket}
\end{align}
The first difficulty is the question of what the inverse matrix $M^{ab}$ actually is. Since its components depend on Dirac delta distributions and have open indices the inversion should be defined by a condition such as equation \eqref{ham-can-B-model-dirac-bracket-M-inverse}.
\begin{align}
  \underset{b}{\sum}\int d^4x_2\int d^{27}Y_2 \,&   M_{ab}(x_1,x_2,Y_1,Y_2)\,  M^{bc}(x_2,x_3,Y_2,Y_3) \nonumber\\
  &= \delta_a{}^c\,\delta^{(4+27)}(x_1-x_3,Y_1-Y_3)\label{ham-can-B-model-dirac-bracket-M-inverse}
 \end{align}
 However due to the form of the components of $M_{ab}$, or equivalently due to the form of the algebra relations \eqref{hamilton-2forms-canonical-S1-P2-Poisson-alg} and \eqref{hamilton-2forms-canonical-S2-P1-Poisson-alg}, solving equation \eqref{ham-can-B-model-dirac-bracket-M-inverse} requires us to find distributions $\Psi$, as components of $M^{ab}$, that satisfy equations of the type \eqref{ham-can-stammfunktion}, with mixed derivatives of $\Psi$ yielding the $4+27$-dimensional Dirac delta distribution.
 \begin{equation}\label{ham-can-stammfunktion}
     (\dots)^{mM}\,\partial_M\partial_m\Psi(x_1-x_3,Y_1-Y_3)=\delta^{(4+27)}(x_1-x_3,Y_1-Y_3)
 \end{equation}
Solving equations of the form of \eqref{ham-can-B-model-dirac-bracket-M-inverse} to determine the inverse $M^{ab}$ hence turns out to be a rather difficult problem as we need to identify a primitive function of the $4+27$-dimensional Dirac delta distribution. This is a general problem that arises when the constraint algebra is of a form that includes (mixed) derivative terms. If we could identify such distributions then we could solve \eqref{ham-can-B-model-dirac-bracket-M-inverse} because the $d$-symbol and the Levi-Civita symbol are invertible.\\

Furthermore the internal integrals in equations \eqref{ham-can-B-model-dirac-bracket} and  \eqref{ham-can-B-model-dirac-bracket-M-inverse} have to be carried out while observing the section condition and it is hence not obvious that these objects are well defined or how the internal integration should be carried out explicitly. \\
In reference \cite{osten2021currents} Dirac brackets have recently been used in the context of exceptional world volume theories with a definition somewhat similar to \eqref{ham-can-B-model-dirac-bracket} but in a very different set up.\\ 
 
In principle we should be able to circumvent the introduction of the Dirac bracket entirely by ``unfixing'' the (gauge) conditions that make the constraints of this model second class \cite{Henneaux-Teitelboim,Gomis:1989vy}. However this procedure of introducing a new set of constraints that are first class, together with additional gauge fixing conditions, is not unique and it is not immediately clear how one should proceed in this model. It may be worth pointing out again that the canonical structure of the constraints in this model is similar to the structure of three-dimensional Chern-Simons theory and it may be possible to identify a solution there --- although there is of course no analogue to the generalised geometry used here.

\section{Canonical formulation of \texorpdfstring{E$_{6(6)}$}{E6(6)} exceptional field theory}\label{section-can-eft-chapter}
In this section we construct the canonical formulation of the (bosonic) E$_{6(6)}$ ExFT. In section \ref{can-eft-sec-notation} we clarify the notation and list some of the conventions used in the following sections. We then go on to compute the external ADM decomposition of the full ExFT Lagrangian in section \ref{can-eft-adm-sec}. In section \ref{can-eft-can-momenta-sec} we compute all canonical momenta and some field redefinitions are introduced. In section \ref{can-eft-sec-primary-constraints} we identify all primary constraints. The Legendre transformation of the ExFT Lagrangian is carried out sector by sector in section \ref{can-eft-sec-legendre} and the resulting canonical Hamiltonian is presented in section \ref{can-eft-sec-can-ham-disc}. The fundamental Poisson brackets are defined in section \ref{can-eft-sec-fundamental-poisson}. In section \ref{can-eft-can-constr-sec} we go through the consistency algorithm of the canonical constraints. Some of the secondary constraints, that follow from the primary constraints associated to the two-form momenta, are found to be of an unusual form and this discussion is continued in section \ref{can-eft-gauge-transformations} where we discuss the canonical gauge transformations.
\subsection{Notation and conventions}\label{can-eft-sec-notation}
In the following sections we need to make use of a large number of different indices, which we list in table \ref{table-notation-EFT}. The index $t$ is reserved for the curved time coordinate and $0$ for the flat time coordinate. The external curved five-dimensional index decomposes in the ADM split as $\mu=(t,m)$ and the flat five-dimensional index decomposes as $\alpha=(0,a)$.
\begin{table}[h]
\centering
\begin{tabular}{|l|l|l|}
\hline
Type of index & Dimension (real) & Letters used \\ \hline\hline
Fundamental rep. of $E_{6(6)}$ & 27  & $K,L,M,N,\dots,X,Y,Z$\\ \hline
Fundamental rep. of USp(8) & 8 & $A,B,C,D,E,\dots,J$\\ \hline
Curved (external) & 5 & $\mu,\nu,\rho,\sigma,\tau,\dots$  \\ \hline
Curved (time) & 1 & $t$\\\hline
Curved (external spatial) & 4 & $k,l,m,n,o,p,q,r,s,u,\dots$ \\ \hline
Flat (external) & 5 & $\alpha,\beta,\gamma,\delta,\dots$ \\ \hline
Flat (time)  & 1 & $0$ \\\hline
Flat (external spatial) & 4 & $a,b,c,d,e,f,g,h,\dots$ \\ \hline
\end{tabular}
\caption{
\label{table-notation-EFT}Conventions of the indices used, their dimensions and descriptions of the types of indices.}
\end{table}
We may occasionally use the convention $\epsilon^{klmn} := \epsilon^{tklmn}$ for the spatial components of the five-dimensional Levi-Civita symbol, which does not contain any vielbein factors. For the external geometry we use the Minkowski signature $(-++++)$. A dot $\dot{X}$ on a variable $X$ indicates a curved time derivative. For the canonical momenta we employ the notation that $\Pi(X)$, with appropriate indices, denotes the canonical momenta conjugate to any field $X$. The notation used in this work generally agrees with the notation of reference \cite{KreutzerSUGRA}.\\

The scalar fields $M_{MN}$ are E$_{6(6)}/$USp(8) coset representatives and only have 42 independent components. We call the relations that connect the 378 components of the symmetric matrix $M_{MN}=M_{(MN)}$ the coset constraints. In the canonical formalism they appear as canonical constraints if they are added explicitly to the Lagrangian. Alternatively they can be considered implicitly, in which case one can treat $M_{MN}$ as a generic symmetric matrix of fields, which greatly simplifies the canonical analysis. In the implicit case one cannot apply the coset constraints before all Poisson brackets are fully evaluated. The explicit and the implicit treatment of coset constraints has been discussed in \cite{KreutzerSUGRA} and both formalisms have been described in detail for the case of SL$(n)$ in appendix D of \cite{KreutzerSUGRA}. In the following we will be working in the implicit formalism in order to simplify the analysis. For an alternative and more explicit vielbein formalism approach to the canonical formulation of coset space sigma models see reference \cite{Matschull:1994vi}\\

The section condition \eqref{eft-e6-section-condition-1} cannot be considered as a canonical constraint because it is a condition on the internal coordinate derivatives and not on the canonical variables. If we wanted to add the section condition explicitly to the Lagrangian we would have to add infinitely many constraints because \eqref{eft-e6-section-condition-1} applies to all fields and gauge parameters.\\  

The generalised Lie derivative $\mathbb{L}_\Lambda$ is always understood to include a weight term with the generalised diffeomorphism weight determined according to table \ref{can-eft-table-gen-diff-weights}.
\begin{table} \centering 
    \begin{tabular}{|c|c|}
        \hline
        Weight $\lambda$ & Objects \\ \hline\hline
        $-2/3$           & $G^{\mu\nu},\, g^{\mu\nu},\,\hat{R},\,V_\text{pot}$ \\\hline
        $-1/3$           & $\partial_M  ,\,E_\alpha {}^{\mu}, \, e_a{}^m $     \\ \hline
        $0$                         & $\partial_\mu  ,\, \mathbb{L}_{A_\mu}, \, d_{MNK},\,M_{MN},\,\mathcal{V}_{M}{}^{AB},\, \hat{R}_{\mu\nu}{}^{\alpha \beta} ,\,N^n$       \\ \hline
        $1/3$            & $A^{M}_\mu , \,  \mathcal{F}^{M}_{\mu\nu} ,\,  \Lambda^{M} , \,  N  , \, N_a ,\,E_\mu {}^{\alpha}, \, e_m{}^a , \, \Pi^{stN}(B)$     \\ \hline
        $2/3$         & $B_{\mu\nu M} , \, \Xi_{\mu M} , \, \mathcal{H}_{\mu\nu\rho M} , \, \Pi_M^m(A) , \,G_{\mu\nu},\, g_{mn} , \,N_n, \, \Pi(e)^{m}_a     $       \\ \hline
        $1$                         & $\mathcal{L}_\text{ExFT} , \,  \Pi^{MN}(M) , \, \Pi^M{}_{AB}(\mathcal{V})$     \\ \hline
        $4/3$             & $e$      \\ \hline
        $5/3$             & $E$      \\\hline
    \end{tabular}
    \caption{\label{can-eft-table-gen-diff-weights}The generalised exceptional diffeomorphism weights of the most important objects of canonical E$_{6(6)}$ exceptional field theory.}
\end{table}

\subsection{ADM decomposition of the Lagrangian}\label{can-eft-adm-sec}
We can now compute the ADM decomposition of all terms in the E$_{6(6)}$ ExFT Lagrangian \eqref{eft-lagrangian-terms}. We will be able to make use of the results of this section in the computation of the canonical momenta and in carrying out the Legendre transformation of the Lagrangian. Explicitly seeing the ADM decomposition of the Lagrangian also gives us some intuition as to what the various terms are doing. Because we are working on expressions that are part of the Lagrangian in this section, we sometimes drop true total derivative terms (that include the Lagrange multipliers), in analogy to the procedure in \cite{KreutzerSUGRA}, as they are not relevant to the results of this work. 
Useful relations regarding the ADM decomposition, in the same notation as here, can be found in \cite{KreutzerSUGRA}. General information on the ADM decomposition can be found in \cite{PhysRev.116.1322,1962ADM,MTW,Nicolai:1992xx,Wald:1984,Kiefer:2004gr}

\subsubsection*{ADM decomposition of the improved Einstein-Hilbert term}
The improved Einstein-Hilbert term is given by \eqref{can-eft-EH-term}. 
\begin{equation}\label{can-eft-EH-term}
\mathcal{L}_{EH} = E\,\hat{R} =E\,R  + E\,\mathcal{F}^M_{\mu\nu} \,E^{\alpha\rho}\,\partial_M E_\rho{}^\beta\,E_\alpha{}^\mu \,E_\beta{}^\nu 
\end{equation}
We start by looking at the covariantised Einstein-Hilbert term $E\,R$. The Ricci scalar is defined with the covariantised coefficients of anholomonomy as defined in \eqref{can-eft-anholonomy-def}, where $\mathcal{D}_\mu$ is the covariant derivative \eqref{eft-cov-deriv}.
It is this dependence on the one-forms $A^M_\mu$, through the covariant derivative, in the coefficients of anholomonomy that leads to the vielbein transforming under generalised diffeomorphisms.
\begin{equation}\label{can-eft-anholonomy-def}
    \Omega_{\alpha\beta\gamma} := 2\, E_{[\alpha}{}^\mu E_{\beta]}{}^\nu\, \mathcal{D}_\mu E_{\nu\gamma}.
\end{equation}
We choose to fix part of the Lorentz symmetry and parametrise the external vielbein in the ADM split as \eqref{can-adm-param-vielbein-adm-split}, where $N$ is the lapse function, $N^a$ the shift vector and $e_m{}^a$ the spatial vielbein. We flatten or unflatten spatial indices with the spatial vielbein. 
\begin{equation}\label{can-adm-param-vielbein-adm-split}
E_\mu{}^\alpha =: \begin{pmatrix} 
                N & N^a \\
                0 & e_{m}{}^a\\
                \end{pmatrix}
\end{equation}
In the ADM decomposition the components of the coefficients of anholonomy take the standard form \eqref{can-eft-anholonomy-components1}, \eqref{can-eft-anholonomy-components2} and \eqref{can-eft-anholonomy-components3}, but with all the derivatives covariantised. 
\begin{align}
    \Omega_{abc} &= 2 \,e_{[a}{}^m e_{b]}{}^n\,\mathcal{D}_me_{nc} \label{can-eft-anholonomy-components1}\\
    \Omega_{ab0} &= 0 \label{can-eft-anholonomy-components2}\\
    \Omega_{0b0} & = - e_b {}^nN^{-1}\, \mathcal{D}_n N\label{can-eft-anholonomy-components3}
\end{align}
As in general relativity there is only a single component of the coefficients of anholonomy \eqref{can-EFT-coeff-anholonomy-zeroab} that contains a time derivative.
\begin{equation}
\label{can-EFT-coeff-anholonomy-zeroab}
\Omega_{0bc} = \frac{1}{N} \bigg(e_b{}^n\,(\mathcal{D}_0-N^m \mathcal{D}_m)\,e_{nc}-e_b{}^m\,e_{nc}\,\mathcal{D}_m N^n \bigg)
\end{equation}
We can invert the relation \eqref{can-EFT-coeff-anholonomy-zeroab} to express the time derivative of the spatial vielbein as in equation \eqref{can-EFT-coeff-vielbein-derivative}.
\begin{equation}\label{can-EFT-coeff-vielbein-derivative}
\partial_0 e_{kc} = N e_k{}^b \,\Omega_{0bc} + (\mathbb{L}_{A_0} + N^m \mathcal{D}_m)\, e_{kc} + e_{nc}\, \mathcal{D}_k N^n
\end{equation}
Using the ADM decomposition of the coefficients of anholonomy we can write the ADM decomposition of the Einstein-Hilbert term as the standard formula \eqref{can-EFT-compact-ricci-ADM}, where $R_d$ is the Ricci scalar in $d$-dimensions \cite{Wald:1984,Nicolai:1992xx}.
\begin{equation}
\label{can-EFT-compact-ricci-ADM}
   E\,R_5 = e\,N\, (\Omega_{0(ab)}\Omega_{0(ab)}-\Omega_{0aa}\Omega_{0bb}+R_4)
\end{equation}
The ADM decomposition of the Einstein-Hilbert improvement term is given in \eqref{can-eft-einstein-improve-adm}. 
\begin{align}\label{can-eft-einstein-improve-adm}
    + E\,\mathcal{F}^M_{\mu\nu} \,E^{\alpha\rho}\,\partial_M E_\rho{}^\beta\,E_\alpha{}^\mu \,E_\beta{}^\nu  =&+\frac{e}{N}\,\mathcal{F}^M_{tn}   \,\partial_M N^n \nonumber\\
    &-\frac{e}{N}\,\mathcal{F}^M_{mn} \, N^m   \,\partial_M N^n \nonumber\\
    &+e\,N\,\mathcal{F}^M_{mn} \, e^{ar}\,\partial_M e_r{}^b\, e_a{}^m \, e_b{}^n  
\end{align}
The first term in \eqref{can-eft-einstein-improve-adm} will contribute to the canonical momenta of the one-forms. Note in particular the sign of the second term, this point is further discussed in section \ref{sec-can-EFT-legendre-1forms}. The last line is the spatial improvement term and will join the spatial Ricci scalar $R_4$ of equation \eqref{can-EFT-compact-ricci-ADM}. 
\subsubsection*{ADM decomposition of the Yang-Mills term}
The ADM decomposition of the generalised Yang-Mills term takes the form \eqref{can-eft-adm-ym}. There is one term that is quadratic and one that is linear in the time components of the one-form field strength. The third term is the spatial Yang-Mills term. The fourth term will drop out in the Legendre transformation.
\begin{align}\label{can-eft-adm-ym}
   -\frac{E}{4}\,M_{MN} \, \mathcal{F}^{\mu\nu M} \,\mathcal{F}^N_{\mu\nu}    = & + \frac{e}{2\,N}\, M_{MN} \,   \mathcal{F}^{M}_{ts} \,\mathcal{F}^N_{tn}\,g^{sn}     \nonumber\\
     & -  \frac{e}{N}\,M_{MN} \,   \mathcal{F}^{M}_{ts} \,\mathcal{F}^N_{mn}\,g^{sn}\,N^m     \nonumber\\
     & -  \frac{e\,N}{4}\,M_{MN} \, \mathcal{F}^{M}_{rs} \,\mathcal{F}^N_{mn}\,g^{rm}\,g^{sn}     \nonumber\\
       & +  \frac{e}{2\,N}\,M_{MN} \, \mathcal{F}^{M}_{rs} \,\mathcal{F}^N_{mn}\,N^r\,N^m\,g^{sn}    
\end{align}

\subsubsection*{ADM decomposition of the scalar kinetic term}
The ADM decomposition of the scalar kinetic term can be written in the form of \eqref{can-eft-adm-scalar}. The first term of the last line is the spatial kinetic term. The structure of the scalar terms will become clearer in the Legendre transformation.
\begin{align}\label{can-eft-adm-scalar}
    &+ \frac{1}{24} \, E\,g^{\mu\nu} \, \mathcal{D}_\mu M_{MN} \, \mathcal{D}_\nu M^{MN} \\
    =&-\frac{e}{24\,N} {\bigg(}- \dot{M}_{MN}\,\dot{M}_{RS}\,M^{RM}\,M^{SN}-\dot{M}_{MN}\,\mathbb{L}_{A_t}M^{MN} \nonumber\\
    &\hspace{1.7cm}+\mathbb{L}_{A_t}M_{MN}\,\dot{M}_{RS}\,M^{RM}\,M^{SN}+\mathbb{L}_{A_t}M_{MN}\,\mathbb{L}_{A_t}M^{MN}{\bigg )}  \nonumber\\
    &+\frac{e}{24\,N}\,N^l {\bigg(}+ \dot{M}_{MN}\,\mathcal{D}_l M^{MN} -\dot{M}_{RS}\,M^{RM}\,M^{SN}\,\mathcal{D}_l M_{MN}  \nonumber\\
    &\hspace{2.2cm}-\mathbb{L}_{A_t}M_{MN}\,\mathcal{D}_l M^{MN} - \mathcal{D}_l M_{MN}  \,\mathbb{L}_{A_t}M^{MN}{\bigg )} \nonumber\\
    &+\frac{1}{24}\,{\Big(} e\, N\, g^{kl} - \frac{e}{N} \, N^k\,N^l {\Big)}\,\mathcal{D}_k M_{MN} \,\mathcal{D}_l M^{MN} \nonumber
\end{align}

\subsubsection*{ADM decomposition of the topological term}
We split the topological term \eqref{topological-term-E-six} into the individual terms, in order to make the expression more manageable and then compute the ADM decompositions. We start with the kinetic term of the B-field \eqref{can-eft-adm-top-term-b-kin}. This kinetic term \eqref{can-eft-adm-top-term-b-kin} is the $\mathcal{D}_\mu$-covariantised version of the model Lagrangian in section \ref{can-sec-twoforms-model}. The last term in \eqref{can-eft-adm-top-term-b-kin} is the only term in the ExFT Lagrangian with a time derivative on the B-field. 
\begin{align}\label{can-eft-adm-top-term-b-kin}
    -\frac{15\kappa}{2} \,\epsilon^{\mu\nu\rho\sigma\tau}\, d^{MNR}\, \partial_R B_{\mu\nu M}\, \mathcal{D}_\rho B_{\sigma\tau N} =& -15\kappa \,\epsilon^{tnrsl}\, d^{MNR}\, \partial_R B_{tn M}\, \mathcal{D}_r B_{sl N} \nonumber\\
    & +15\kappa \,\epsilon^{tnrsl}\, d^{MNR}\, \partial_R B_{nr M}\, \mathcal{D}_s B_{tl N} \nonumber\\
    & +\frac{15\kappa}{2}  \,\epsilon^{tnrsl}\, d^{MNR}\, \partial_R B_{nr M}\, \mathbb{L}_{A_t} B_{sl N} \nonumber\\
    & -\frac{15\kappa}{2}  \,\epsilon^{tnrsl}\, d^{MNR}\, \partial_R B_{nr M}\, \partial_t B_{sl N} 
\end{align}
The ADM decomposition of the two other B-field dependent terms are given in \eqref{can-eft-adm-top-term-2} and \eqref{can-eft-adm-top-term-3}. 
\begin{align}\label{can-eft-adm-top-term-2}
    &+15\kappa \epsilon^{\mu\nu\rho\sigma\tau} \enspace d^{MNR}\enspace d_{NKL}\,\partial_R B_{\mu\nu M} \, A^K_{\rho} \partial_\sigma A^L_{\tau}\\
  = &+30\kappa \epsilon^{tnrsl} \, d^{MNR}\, d_{NKL}\,\partial_R B_{tn M} \, A^K_{r} \,\partial_s A^L_{l} \nonumber\\
  &-15\kappa \epsilon^{tnrsl} \, d^{MNR}\, d_{NKL}\,\partial_R B_{nr M} \, A^K_{s} \,\dot{A}^l_{l} \nonumber\\
  &-15\kappa \epsilon^{tnrsl} \, d^{MNR}\, d_{NKL}\,\partial_l\partial_R B_{nr M} \, A^K_{s} \, A^L_{t} \nonumber\\
  &+30\kappa \epsilon^{tnrsl} \, d^{MNR}\, d_{NKL}\,\partial_R B_{nr M} \, A^K_{t} \,\partial_s A^L_{l} \nonumber
\end{align}
There is a time derivative on the one-form in the third line of \eqref{can-eft-adm-top-term-2}.
\begin{align}\label{can-eft-adm-top-term-3}
    & -5\kappa\,\epsilon^{\mu\nu\rho\sigma\tau} \, d^{MNR}\, d_{NKL}\,\partial_R B_{\mu\nu M} \,A^K_{\rho}\,[A_\sigma,A_{\tau}]^L_\text{E}\\
  =  & -10\kappa\,\epsilon^{tnrsl} \, d^{MNR}\, d_{NKL}\,\partial_R B_{tn M} \,A^K_{r}\,[A_s,A_l]^L_\text{E} \nonumber\\
  & -5\kappa\,\epsilon^{tnrsl} \, d^{MNR}\, d_{NKL}\,\partial_R B_{nr M} \,A^K_t\,[A_s,A_l]^L_\text{E} \nonumber\\
  & +10\kappa\,\epsilon^{tnrsl} \, d^{MNR}\, d_{NKL}\,\partial_R B_{nr M} \,A^K_{s}\,[A_t,A_l]^L_\text{E} \nonumber
\end{align}
The term \eqref{can-eft-adm-top-term-4} is the analogue of the topological term in five-dimensional ungauged maximal E$_{6(6)}$ invariant supergravity (cf. reference \cite{KreutzerSUGRA}), because this is the only part of the topological term \eqref{topological-term-E-six} that does not depend on any internal derivative and therefore survives in the trivial solution of the section condition.
\begin{align}\label{can-eft-adm-top-term-4}
    &+ \kappa\,\, \epsilon^{\mu\nu\rho\sigma\tau}d_{MNP}\enspace A^N_\mu \enspace\partial_\nu A^M_\rho \enspace\partial_\sigma A^P_\tau\\
    =& + \kappa\,\, \epsilon^{tnrsl}\,d_{MNP}\enspace A^N_t \enspace\partial_n A^M_r \enspace\partial_s A^P_l \nonumber \\
    & -2\, \kappa\,\, \epsilon^{tnrsl}\,d_{MNP}\enspace A^N_n \enspace\partial_t A^M_r \enspace\partial_s A^P_l \nonumber \\
    & +2\, \kappa\,\, \epsilon^{tnrsl}\,d_{MNP}\enspace A^N_n \enspace\partial_r A^M_t \enspace\partial_s A^P_l \nonumber 
\end{align}
In the term \eqref{can-eft-adm-top-term-5} there is a time derivative on the one-form in the fourth line.
\begin{align}\label{can-eft-adm-top-term-5}
     &- \frac{3\kappa}{4}\epsilon^{\mu\nu\rho\sigma\tau}d_{MNP} \enspace A^N_\mu\enspace\left[A_\nu,A_\rho\right]^M_E\enspace\partial_\sigma A^P_\tau \\
    =&- \frac{3\kappa}{4}\epsilon^{tnrsl}\,d_{MNP}{\bigg(}+ A^N_t\,\left[A_n,A_r\right]^M_E\,\partial_s\, A^P_l \nonumber \\
    & \hspace{3.2cm} -2  A^N_n\,\left[A_t,A_r\right]^M_E\,\partial_s\, A^P_l \nonumber \\
    & \hspace{3.2cm} -  A^N_n\,\left[A_r,A_s\right]^M_E\,\dot{A}^P_l \nonumber \\
    & \hspace{3.2cm} +  A^N_n\,\left[A_r,A_s\right]^M_E\,\partial_l\, A^P_t {\bigg)} \nonumber 
\end{align}
The terms \eqref{can-eft-adm-top-term-6} and \eqref{can-eft-adm-top-term-2} are the only parts of the topological term that are purely internal in derivatives. For this reason they cannot contribute to any of the canonical momenta.
\begin{align}\label{can-eft-adm-top-term-6}
    &+ \frac{3\kappa}{20}\epsilon^{\mu\nu\rho\sigma\tau}d_{MNP}\enspace A^N_\mu\enspace \left[A_\nu,A_\rho\right]^M_E \enspace\left[A_\sigma,A_\tau\right]^P_E \\
    = &+ \frac{3\kappa}{20}\epsilon^{tnrsl}\,d_{MNP}{\bigg (} +A^N_ t\, \left[A_n,A_r\right]^M_E \,\left[A_s,A_l\right]^P_E \nonumber\\
     &\hspace{3.2cm} -4 \,A^N_n\, \left[A_t ,A_r\right]^M_E \,\left[A_s,A_l\right]^P_E {\bigg )} \nonumber
\end{align}

\subsubsection*{ADM decomposition of the scalar potential term}
The ADM decomposition of the scalar potential is given in \eqref{can-eft-adm-pot}. Note that we have already written the potential with the sign with which it will appear in the Hamiltonian. The first term in \eqref{can-eft-adm-pot} depends on several Lagrange multipliers and we cannot interpret this term canonically --- fortunately we find that this term cancels in the Legendre transformation. The remaining terms of \eqref{can-eft-adm-pot} form the scalar potential of the Hamiltonian, which will be part of the Hamilton constraint.
\begin{align}\label{can-eft-adm-pot}
    + E \,V_\text{pot} = & - \frac{e}{2N} g_{mn} M^{MN} \partial_M N^n \,\partial_N N^m  \\
            & - \frac{N \,e}{4}  M^{MN} \partial_M g^{mn}\, \partial_N g_{mn}  \nonumber \\
            & - \frac{N}{e}  M^{MN} \partial_M e \, \partial_N e  \nonumber \\
             & - \frac{N\,e}{24}  M^{MN} \partial_M M^{KL} \, \partial_N M_{KL} \nonumber \\
             & + \frac{N\,e}{2}  M^{MN} \partial_M M^{KL} \, \partial_L M_{NK} \nonumber \\
             & + N   \, \partial_M \partial_N M^{MN} \,  e \nonumber \\     
             & + N  \, 2  \,  M^{MN} \, \partial_M \partial_N e \nonumber \\
             & + N   \,2   \, \partial_M M^{MN} \, \partial_N e \nonumber 
\end{align}

\subsection{Canonical momenta}\label{can-eft-can-momenta-sec}
Having computed the ADM decomposition of all terms of the ExFT Lagrangian in section \ref{can-eft-adm-sec} we can now compute the canonical momenta. In this section we also discuss some important redefinitions of the canonical coordinates that simplify the later calculations and the form of the canonical Hamiltonian.

\subsubsection*{The canonical momenta of the one-form fields}\label{can-eft-momenta-oneform-sec}
Because there are no time derivatives on the time component of the one-forms we find that the momenta \eqref{can-EFT-original-canonical-momentum-A-t-comp} vanish, as is expected and they double as primary constraints.
\begin{align}
\Pi^{t}_T(A) = &\,0 \label{can-EFT-original-canonical-momentum-A-t-comp} \\
\Pi^{l}_T(A) =  & \frac{e}{N} g^{ln}\;M_{TN}\;\bigg( \mathcal{F}^{N}_{tn}+ N^k \mathcal{F}^{N}_{nk}  \bigg) \label{can-EFT-original-canonical-momentum-A}\\
  & -\frac{3\kappa}{4} \epsilon^{lmnr}d_{MNT}A^N_m\;[A_n,A_r]^M_E \nonumber \\
  & +2\kappa\,\,\epsilon^{lmnr}d_{MNT}\;A^M_m\;\partial_n A^N_r \nonumber \\
  &+15 \kappa\,\, \epsilon^{lmnr} d^{MNR} d_{NKT} \; \partial_R B_{mnM}\; A^K_r \nonumber \\
  &+ \frac{e}{N} \partial_T N^l \nonumber
\end{align}
The canonical momenta of the spatial components of the one-forms are given by equation \eqref{can-EFT-original-canonical-momentum-A}. The first line of \eqref{can-EFT-original-canonical-momentum-A} is the covariantised version of the expression expected in Yang-Mills theory coupled to general relativity. The next three terms originate from the three time derivatives on the one-forms in the topological term. The $\frac{e}{N} \partial_T N^l$ contribution comes from the Einstein-Hilbert improvement term.\\

In reference \cite{KreutzerSUGRA} it was found in five-dimensional E$_{6(6)}$ invariant supergravity that if we define modified canonical momenta, where all topological contributions to the momenta $\Pi^{l}_T(A)$ are subtracted, this greatly simplifies the canonical Hamiltonian and has nice transformation properties under the canonical constraints. In direct analogy we define the modified momenta-like variables \eqref{can-EFT-original-canonical-momentum-A-redef} by subtracting all topological contributions from the canonical momenta. Explicitly we can thus also write the modified momenta $\mathcal{P}^{l}_T(A)$ as in \eqref{can-EFT-original-canonical-momentum-A-redef-expl}. 
\begin{align}
  \mathcal{P}^{l}_T(A) := & +\Pi^{l}_T(A) \label{can-EFT-original-canonical-momentum-A-redef} \\
  & +\frac{3\kappa}{4} \epsilon^{lmnr}d_{MNT}A^N_m\;[A_n,A_r]^M_E \nonumber \\
  & -2\kappa\,\,\epsilon^{lmnr}d_{MNT}\;A^M_m\;\partial_n A^N_r \nonumber \\
  &-15 \kappa\,\, \epsilon^{lmnr} d^{MNR} d_{NKT} \; \partial_R B_{mnM}\; A^K_r \nonumber \\
  = & + \frac{e}{N} g^{ln}\;M_{TN}\;\bigg( \mathcal{F}^{N}_{tn}+ N^k \mathcal{F}^{N}_{nk}  \bigg)  + \frac{e}{N} \partial_T N^l  \label{can-EFT-original-canonical-momentum-A-redef-expl}
\end{align}
We find that $\mathcal{P}^{l}_T(A)$ are indeed the right variables to use, because they lead to the simplest canonical Hamiltonian and hide a large number of topological contributions that would otherwise clutter the Hamiltonian. Without this redefinition the Legendre transformation is itself also very messy due to the large number of terms that are produced. In section \ref{can-eft-sec-legendre} we find that the momenta $\mathcal{P}^{l}_T(A)$ allow us to make the Legendre transformation of the one-forms comparatively simple.\\

Just as in reference \cite{KreutzerSUGRA} this redefinition of the momenta leads to complications when evaluating Poisson brackets, because the redefinition \eqref{can-EFT-original-canonical-momentum-A-redef} is not a canonical transformation and the new variables do not Poisson-commute with themselves $\{\mathcal{P}^n_N(A),\mathcal{P}^m_M(A)\} \neq 0$. This fact combined with the complicated topological term \eqref{topological-term-E-six} presents some of the greatest difficulties in the canonical analysis of E$_{6(6)}$ exceptional field theory. Not using the redefined momenta $\mathcal{P}^{l}_T(A)$ does not circumvent these issues, as in this case the complications are just displaced and already apparent at the level of the Hamiltonian.\\

By definition we find that $\Pi^n_N(A) = \mathcal{P}^n_N(A)$ if we set the coefficient of the topological term to zero $\kappa=0$. This fact allows us to work in orders of the topological coefficient $\kappa$ to break up the calculations in more manageable parts. For some very difficult calculations we only present the calculation at $\kappa=0$, but often we can already see the main structure of the full result at this level. A notable exception to this are all calculations that concern the B-fields as their dynamics are entirely topological. We should stress the fact that the case $\kappa=0$ is only considered as a \textit{computational tool} --- because it removes one of the main difficulties --- and this case does very likely not correspond to any physically meaningful theory upon solution of the section condition.  

\subsubsection*{The canonical momenta of the two-form fields}
The canonical momenta of the two-forms are given by equations \eqref{can-EFT-canonical-momentum-B-primary-1} and \eqref{can-EFT-canonical-momentum-B-primary-2}.
Because the only time derivative on the B-fields comes from the topological kinetic term of equation 
\eqref{can-eft-adm-top-term-b-kin} the resulting canonical momenta are identical to the canonical momenta of the model theory that we studied in section \ref{ham-can-B-model-canmom-canham-sec}. 
\begin{align}
\Pi^{tl\,N}(B) = &\,0 \label{can-EFT-canonical-momentum-B-primary-1}\\
\Pi^{sl\,N}(B) = &-15\kappa\,\,\, \epsilon^{mnsl} d^{MNR} \, \partial_R B_{mnM} \label{can-EFT-canonical-momentum-B-primary-2}
\end{align}
We can see that both \eqref{can-EFT-canonical-momentum-B-primary-1} and \eqref{can-EFT-canonical-momentum-B-primary-2} are primary canonical constraints, because there are no time derivatives on the variables.

\subsubsection*{The canonical momenta of the scalar fields}
There is a slight subtlety in the calculation of the canonical momenta of the scalar fields concerning the scaling of the diagonal components of $M_{MN}$. This issue is purely formal but we want to briefly explain the issue here since it plays a role in the Legendre transformation that we will carry out in section \ref{can-eft-sec-legendre}.\\

Because we want both $\frac{\partial \dot M_{11}}{\partial \dot M_{11}} = 1$ and $\frac{\partial \dot M_{12}}{\partial \dot M_{12}} = 1$ to be true we have to subtract a Kronecker delta term in the general derivative \eqref{can-eft-momenta-M-derivative} in order to get the correct result for the diagonal components.
\begin{equation}\label{can-eft-momenta-M-derivative}
    \frac{\partial \dot M_{QP}}{\partial \dot M_{MN}} = \delta^M_Q \delta^N_P + \delta^M_P \delta^N_Q -(\delta^\text{\tiny Kronecker}_{MN})\delta^M_Q \delta^N_P 
\end{equation}
The canonical momenta of the scalar fields are then found to be given by \eqref{can-EFT-canonical-momentum-M} (there is no sum on $R,\,S$ in this expression).
\begin{equation}
\label{can-EFT-canonical-momentum-M}
\Pi^{RS}(M) = \frac{e}{12\,N}(2-\delta^\text{\tiny Kronecker}_{RS})\,  \bigg{[}+\dot{M}_{QP}\; M^{QR}M^{PS} +N^n\;\mathcal{D}_nM^{RS} +\mathbb{L}_{A_t}M^{RS}\bigg{]} 
\end{equation}
Because the Kronecker delta term in \eqref{can-EFT-canonical-momentum-M} is somewhat unappealing, we can choose to remove it by rescaling the diagonal component of the canonical momenta as in \eqref{can-eft-momenta-scalar-rescaling}.
\begin{equation}\label{can-eft-momenta-scalar-rescaling}
\tilde{\Pi}^{RS}(M) := \begin{cases}
                       2\cdot\Pi^{RS}(M), \quad\text{if}\quad R=S  \\
                       \,\,\quad\Pi^{RS}(M), \quad\text{if}\quad R\neq S
                       \end{cases}
\end{equation}
The rescaled canonical momenta of the scalars are then given by equation \eqref{can-EFT-canonical-momentum-M-rescaled}.
\begin{equation}
\label{can-EFT-canonical-momentum-M-rescaled}
\tilde{\Pi}^{RS}(M) = \frac{e}{6\,N}\enspace \bigg{[}+\dot{M}_{QP}\; M^{QR}M^{PS} +N^n\;\mathcal{D}_nM^{RS} +\mathbb{L}_{A_t}M^{RS}\bigg{]} 
\end{equation}
During the Legendre transformation in section \ref{can-eft-sec-legendre} we have to pay attention to this rescaling to get the prefactor of the scalar terms right. After the Legendre transformation we will simply write $\Pi^{RS}(M)=\tilde{\Pi}^{RS}(M)$ as this distinction is no longer necessary.

\subsubsection*{The canonical momenta of the external metric components}
Only the $\Omega_{0bc}$ components of the coefficients of anholonomy \eqref{can-EFT-coeff-anholonomy-zeroab} contain a time derivative and that is on the spatial vielbein. Therefore the Einstein-Hilbert term \eqref{can-EFT-compact-ricci-ADM} leads to the vanishing of the canonical momenta of the lapse function \eqref{can-EFT-canonical-momentum-lapse} and the shift vector \eqref{can-EFT-canonical-momentum-shift}, which become primary constraints. Only the canonical momenta of the spatial vielbein are non-vanishing and given by equation \eqref{can-EFT-canonical-momentum-vierbein}. While the momenta \eqref{can-EFT-canonical-momentum-vierbein} look exactly like in general relativity the coefficients of anholonomy are nonetheless written in terms of the covariant derivatives.
\begin{align}
    \Pi(N) & = 0 \label{can-EFT-canonical-momentum-lapse}\\
    \Pi^{a}(N_a) & = 0  \label{can-EFT-canonical-momentum-shift}\\
    \Pi^{m}_a(e) &  = 2\,e\,e^m_b (\Omega_{0(ab)}-\delta_{ab}\,\Omega_{0cc}) \label{can-EFT-canonical-momentum-vierbein}
\end{align}
We furthermore define the contractions \eqref{can-eft-contr-vielbein-1} and \eqref{can-eft-contr-vielbein-2} of the vielbein momenta. 
\begin{align}
    \Pi_{ab}(e) := &+ e_{m(a} \,\Pi^m{}_{b)}(e) \label{can-eft-contr-vielbein-1}\\
        \Pi(e) := &+e_{m}{}^a \, \Pi^m{}_a(e)\label{can-eft-contr-vielbein-2}
\end{align}

\subsection{Primary constraints}\label{can-eft-sec-primary-constraints}
Having computed all canonical momenta in section \ref{can-eft-can-momenta-sec} we can identify the following complete set of primary constraints.
The primary constraints \eqref{can-EFT-primary-canonical-constraint-lapse}, \eqref{can-EFT-primary-canonical-constraint-shift}, \eqref{can-EFT-primary-canonical-constraint-A} and \eqref{can-EFT-primary-canonical-constraint-B} are of shift type, meaning that they will only generate shifts in their conjugate canonical variables and nothing else. We will see that the consistency requirement of each of them generates a corresponding secondary constraint.
\begin{align}
   \Pi(N) &= 0       \label{can-EFT-primary-canonical-constraint-lapse}\\
    \Pi^{a}(N_a) &= 0 \label{can-EFT-primary-canonical-constraint-shift}\\
    \Pi_M(A^M_t) &= 0 \label{can-EFT-primary-canonical-constraint-A}
\end{align}
We furthermore introduce the names $\mathcal{H}_\text{P1}$ and $\mathcal{H}_\text{P2}$ for the primary constraints \eqref{can-EFT-primary-canonical-constraint-B} and \eqref{can-EFT-primary-canonical-constraint-B-nontrivial} coming from the B-fields. These two-form constraints are in direct analogy to the primary constraints of the model theory in section \ref{can-sec-twoforms-model}.
\begin{align}
(\mathcal{H}_\text{P1})^{mM} &:= \Pi^{tm\,M}(B) = 0 \label{can-EFT-primary-canonical-constraint-B} \\
(\mathcal{H}_\text{P2})^{slN} &:= \bigg(\Pi^{sl\,N}(B) + 15\,\kappa\, \epsilon^{tmnsl} d^{MNR} \, \partial_R B_{mnM} \bigg)= 0 \label{can-EFT-primary-canonical-constraint-B-nontrivial}
\end{align}
Finally there are the primary Lorentz constraints \eqref{can-eft-Lorentz-constraints}, which behave exactly like in five-dimensional supergravity and whose canonical properties have been discussed in that case in detail in \cite{KreutzerSUGRA}.
\begin{equation}\label{can-eft-Lorentz-constraints}
 L_{ab} := e_{m[a} \Pi(e)^m{}_{b]} = 0 
\end{equation}
Overall we count a total of $1+4+27+108+162+6=308$ primary constraints.

\subsection{Legendre transformation}\label{can-eft-sec-legendre}
In this section we go through the Legendre transformation of the (bosonic) Lagrangian of E$_{6(6)}$ exceptional field theory. We begin by clarifying how we want to split up the computation, in order to manage the large number of terms and then proceed to compute the partial results using the ADM decompositions presented in section \ref{can-eft-adm-sec}.\\

The Legendre transformation of the ExFT Lagrangian \eqref{eft-lagrangian-terms} is formally given by equation \eqref{can-eft-legendre-1}. 
\begin{align}\label{can-eft-legendre-1}
\mathcal{H}_{\text{ExFT}} =\quad&+\dot N \cdot \Pi(N) + \sum_{\substack{a=1,\dots,4}} \dot N_a\cdot \Pi^a(N_a)+ \sum_{\substack{a=1,\dots,4\\m=1,\dots,4}}\dot e_{ma}\cdot \Pi^{ma}(e_{ma})\\
&+ \sum_{\substack{N=1,\dots,27}} \dot A_{t}^N\cdot \Pi_N(A_t^N) +  \sum_{\substack{N=1,\dots,27\\n=1,\dots,4}} \dot A_{n}^N\cdot \Pi_N^n(A_n^N)   \nonumber\\
&+ \sum_{\substack{N=1,\dots,27\\n=1,\dots,4}} \dot B_{tnN}\cdot \Pi^{nN}(B_{tnN}) + \frac{1}{2} \sum_{\substack{N=1,\dots,27\\m,n=1,\dots,4}} \dot B_{mnN}\cdot \Pi^{mnN}(B_{mnN})   \nonumber\\
&+\sum_{\substack{R,S=1,\dots,27\\R\leq S}} \dot M_{RS}\cdot \Pi^{RS}(M_{RS}) - \mathcal{L}_{\text{ExFT}}\nonumber
\end{align}
We explicitly write out the summation here in order to avoid overcounting. In order to only sum over the independent field components we introduce a factor $1/2$ in the B-Field transformation term and restrict the sum on the scalar fields. We cannot sum over all components of the scalar fields by inserting a factor because that would give the wrong prefactor for the diagonal components. Following the calculation starting from \eqref{can-EFT-legendre-rescaling-momenta-scalar} we find that we can indeed write it as the unrestricted sum \eqref{can-EFT-legendre-rescaling-momenta-scalar-final} if we use the rescaled scalar momenta instead. We can treat the scalar fields as a generic symmetric matrix here because we use the implicit formalism of the coset constraints as explained above. This treatment is identical to the treatment of the scalar fields in \cite{KreutzerSUGRA}.
\begin{align}
\label{can-EFT-legendre-rescaling-momenta-scalar}
\mathcal{H}_{\text{ExFT}} &= \sum_{R \leq S} \Pi^{RS}(M) \dot M_{RS} + \dots \\
& = \sum_{R < S} \Pi^{RS}(M) \dot M_{RS} + \Pi^{RS}(M) \dot M_{RS} \bigg\rvert_{R=S} + \dots \\
& = \sum_{R < S} \tilde{\Pi}^{RS}(M) \dot M_{RS} + \frac{1}{2}\tilde{\Pi}^{RS}(M) \dot M_{RS} \bigg\rvert_{R=S} + \dots \\
& = \frac{1}{2} \sum_{R \neq S} \tilde{\Pi}^{RS}(M) \dot M_{RS} + \frac{1}{2}\tilde{\Pi}^{RS}(M) \dot M_{RS} \bigg\rvert_{R=S} + \dots \\
& = \frac{1}{2} \sum_{R,S} \tilde{\Pi}^{RS}(M) \dot M_{RS} + \dots \label{can-EFT-legendre-rescaling-momenta-scalar-final}
\end{align}
Expanding the Lagrangian we arrive at \eqref{can-EFT-legendre-collection-terms-hamiltonian}, which is written in a way that already suggests how we can split up the computation according to the various sectors. 
\begin{align}
\label{can-EFT-legendre-collection-terms-hamiltonian}
\mathcal{H}_{\text{ExFT}} =\quad&+  \sum_{\substack{a=1,\dots,4\\m=1,\dots,4}}\dot e_{ma}\cdot \Pi^{ma}(e_{ma}) -E\,R_5\quad\\
& +  \sum_{\substack{N=1,\dots,27\\n=1,\dots,4}}\dot A_{n}^N\cdot \Pi_N^n(A_n^N)  -\mathcal{L}_\text{YM} -  E\;\mathcal{F}^M_{\alpha\beta}E^{\alpha\rho}\partial_M E^\beta_\rho   \nonumber\\
& +  \frac{1}{2} \sum_{\substack{N=1,\dots,27\\m,n=1,\dots,4}}\dot B_{mnN}\cdot \Pi^{mnN}(B_{mnN}) -\mathcal{L}_\text{top.}   \nonumber\\
&+ \frac{1}{2}\sum_{\substack{R,S=1,\dots,27}} \dot M_{RS}\cdot \tilde{\Pi}^{RS}(M_{RS})- \mathcal{L}_\text{sc}-\mathcal{L}_\text{pot} \nonumber\\
&  +\dot N \cdot \Pi(N) + \sum_{\substack{a=1,\dots,4}} \dot N_a\cdot \Pi^a(N_a)  \nonumber\\ 
& + \sum_{\substack{N=1,\dots,27}} \dot A_{t}^N\cdot \Pi_N(A_t^N)  + \sum_{\substack{N=1,\dots,27\\n=1,\dots,4}} \dot B_{tnN}\cdot \Pi^{nN}(B_{tnN})   \nonumber
\end{align}
In the following sections we will look at the computation for each sector of the theory individually. The main difficulty lies in the transformation of the terms concerning primarily the one-forms, which appear in the generalised Yang-Mills term, the Einstein-Hilbert improvement term and in the topological term.

\subsubsection[Einstein-Hilbert term]{Legendre transformation of the Einstein-Hilbert term}
For the Legendre transformation of the pure Einstein-Hilbert term (without the improvement) regarding the spatial vielbein time derivative we find that the contributions to the Hamiltonian are given by \eqref{can-eft-legendre-adm-sector}. 
\begin{align}
\Pi^{m}_a(e) \,{\dot e}_{ma} - E\,R_5 = & + N \cdot\bigg(\frac{1}{4e}\,\Pi_{ab}(e)\,\Pi_{ab}(e) -\frac{1}{12e} \Pi^2(e)-e \,R_4  \bigg) \label{can-eft-legendre-adm-sector}\\
&+ N^n \cdot  \bigg(2\,\Pi^m{}_a(e)\, \mathcal{D}_{[n}e_{m]a} - e_{na}\,\mathcal{D}_m \Pi^m{}_a(e)\bigg) \nonumber\\
&+ A^K_t\cdot \bigg( \Pi^m{}_a(e) \,\partial_K e_{ma} -\frac{1}{3}\partial_K \Pi(e) \bigg) \nonumber
\end{align}
The first two lines of \eqref{can-eft-legendre-adm-sector} are the covariantised version of what we expect in pure general relativity \cite{dewittCanGR,Nicolai:1992xx,KreutzerSUGRA}. Because of the gauging of the derivatives with the one-forms we find the additional $A^K_t$ contribution and it is these terms that canonically generate generalised diffeomorphisms on the vielbein and on its momenta.

\subsubsection[Scalar kinetic term]{Legendre transformation of the scalar kinetic term}
For the Legendre transformation of the scalar kinetic term we find \eqref{can-eft-scalar-legendre}. The contributions to the Hamilton and (external) diffeomorphism constraints are the covariantised version of the terms that are found in the Hamiltonian of five-dimensional E$_{6(6)}$ invariant supergravity \cite{KreutzerSUGRA}. The additional $A^K_t$ contributions will turn out to be the correct terms to generate generalised diffeomorphisms on the scalar fields and their canonical momenta. The explicit projector $\mathbb{P}^R{}_L{}^S{}_K$ in this expression might appear strange, however it only appears here because we need to factor out the Lagrange multiplier $A^K_t$ from the generalised Lie derivatives $\mathbb{L}_{A_t}$ in the ADM decomposition \eqref{can-eft-adm-scalar}. Projectors will appear in this context in general for fields that are not scalars under E$_{6(6)}$ --- e.g. there are no projectors in \eqref{can-eft-legendre-adm-sector} for the external vielbein because it does not carry any E$_{6(6)}$ index.
\begin{align}\label{can-eft-scalar-legendre}
&\frac{1}{2}\sum_{\substack{R,S=1,\dots,27}} \dot M_{RS}\cdot \tilde{\Pi}^{RS}(M_{RS})- \mathcal{L}_\text{sc} \\ 
=&+N\cdot\bigg(\frac{3}{2\,e}\; \tilde{\Pi}^{MN}(M) \;\tilde{\Pi}^{RS}(M)\;M_{MR}\;M_{NS} - \,\frac{e}{24} \; g^{kl} \; \mathcal{D}_k M_{MN} \; \mathcal{D}_l M^{MN}\bigg)\nonumber\\
&+N^n\cdot  \bigg( \frac{1}{2} \;\tilde{\Pi}^{MN}(M)\, \mathcal{D}_n M_{MN}  \bigg) \nonumber\\
&+ A^K_t \cdot \bigg( \frac{1}{2}\,\tilde{\Pi}^{MN}(M)\,\partial_K M_{MN} -6\,\mathbb{P}^R{}_L{}^S{}_K \, \partial_S \left( \tilde{\Pi}^{LN}(M)\, M_{RN} \right) \bigg) \nonumber
\end{align}

\subsubsection[Two-form kinetic term]{Legendre transformation of the two-form kinetic term}
\label{sec-can-legendre-2form-EFT}
We want to treat the Legendre transformations regarding the one-forms and the two-forms separately and therefore we need to single out the covariantised B-field kinetic term $ \mathcal{L}_\text{BK}$, as defined in \eqref{can-eft-B-kinetic-langrangian}, from the topological term. The remaining part of the topological term $\mathcal{L}_\text{top.}-\mathcal{L}_\text{BK}$ is considered in the transformation of the remaining one-form terms in section \ref{sec-can-EFT-legendre-1forms}.
\begin{equation}\label{can-eft-B-kinetic-langrangian}
 \mathcal{L}_\text{BK} :=   -\frac{15\kappa}{2} \epsilon^{\mu\nu\rho\sigma\tau} d^{MNR}\, \partial_R B_{\mu\nu M}\, D_\rho B_{\sigma\tau N} 
\end{equation}
With the ADM split \eqref{can-eft-adm-top-term-b-kin} we find that the Legendre transformation of $ \mathcal{L}_\text{BK}$ is given by \eqref{can-eft-legendre-B}. 
\begin{align}\label{can-eft-legendre-B}
    & \frac{1}{2} \, \dot B_{mnN}\cdot \Pi^{mnN}(B_{mnN}) - \mathcal{L}_\text{BK} \\
    =&-\frac{15\kappa}{2}\,\epsilon^{nrsl}\,d^{MNR}\, \partial_R B_{nr M} \,\mathbb{L}_{A_t}B_{sl N} -30\kappa\,\epsilon^{nrsl}\,d^{MNR} \, B_{tn M} \,\mathcal{D}_r \partial_R B_{sl N} \nonumber
\end{align}
Because $\mathcal{L}_\text{BK}$ is the covariantised version of the Lagrangian of the model theory of section \ref{can-sec-twoforms-model} we find the covariantised version of the $\mathcal{H}_{S1}$ constraint from section \ref{can-sec-twoforms-model} plus an extra $A_t$ dependent term. In section \ref{can-sec-twoforms-model} we could rewrite the $ B_{tn M}$ term in terms of the naive two-form field strength as \eqref{hamilton-2forms-canonical-constr-S1-H}. In \eqref{can-eft-legendre-B} we can commute the covariant derivative with the internal derivative because of the contraction into the $d$-symbol $d^{MNR} \partial_R B_{slN}$ by both the derivative and the B-field (cf. identity (2.36) of \cite{EFTI-E6}). However in \eqref{can-eft-legendre-B} we are missing the remaining one-form dependent terms that appear in the covariantised field strength $\mathcal{H}_{lmnL}$ --- we will see that these terms appear in the Legendre transformation of the one-forms.\\

Taking a closer look at the $A_t$ dependent term in \eqref{can-eft-legendre-B} we can see that we can make use of the definition of the primary constraint \eqref{can-EFT-primary-canonical-constraint-B-nontrivial} to write this term as \eqref{can-eft-legendre-b-at-term-v1}. The first term in \eqref{can-eft-legendre-b-at-term-v1} already appears to be the correct term to generate generalised diffeomorphisms on the B-fields, however we are not allowed to set $\mathcal{H}_{P2}=0$ and go to the primary constraint surface at this point. We will come back to this point when discussing the canonical gauge transformations. Because we need to factor out the Lagrange multipliers we arrive at the constraint contribution \eqref{can-eft-legendre-b-at-term-v1-2} and similarly to \eqref{can-eft-scalar-legendre} the projector $\mathbb{P}^R{}_K{}^S{}_M$ is explicitly visible in the constraint due to the integration by parts. 
\begin{align}
    &-\frac{15\kappa}{2}\,\epsilon^{nrsl}\,d^{MNR}\, \partial_R B_{nr M} \,\mathbb{L}_{A_t}B_{sl N}  \nonumber\\
  =&\,+\frac{1}{2}\,\left(\Pi^{slN}(B)-(\mathcal{H}_{P2})^{slN} \right)    \,\mathbb{L}_{A_t}B_{sl N} \label{can-eft-legendre-b-at-term-v1} \\     
  =&\,  A^M_t\cdot\bigg(  + \frac{1}{2}\,\Pi^{lnN}(B) \, \partial_M B_{lnN} - 3\, \mathbb{P}^R{}_K{}^S{}_M \, \partial_S\left(\Pi^{lnK}(B)\, B_{lnR} \right)  \nonumber\\ 
     & - \frac{1}{2}\,(\mathcal{H}_\text{P2})^{lnN} \, \partial_M B_{lnN} + 3\, \mathbb{P}^R{}_K{}^S{}_M \, \partial_S\left((\mathcal{H}_\text{P2})^{lnK}\, B_{lnR} \right)  \nonumber\\ 
    &-\frac{1}{3}\,\partial_M \left(B_{mnN} \Pi^{mnN}(B) \right) +\frac{1}{3}\,\partial_M \left(B_{mnN} (\mathcal{H}_\text{P2})^{mnN} \right) \bigg)\label{can-eft-legendre-b-at-term-v1-2}
\end{align}
Alternatively we can factor out the Lagrange multiplier $A^M_t$ without using the primary constraint to arrive at the simpler form of the contribution \eqref{can-eft-legendre-b-at-term-v2}. In the calculation of \eqref{can-eft-legendre-b-at-term-v2} one has to make use of the section condition \eqref{eft-e6-section-condition-1}. This is notable insofar as it is the only instance where this becomes necessary in the calculation of the canonical Hamiltonian. The use of the section condition can however be avoided by the use of the form \eqref{can-eft-legendre-b-at-term-v1-2}.
\begin{align}
    &-\frac{15\kappa}{2}\,\epsilon^{nrsl}\,d^{MNR}\, \partial_R B_{nr M} \,\mathbb{L}_{A_t}B_{sl N}  \nonumber\\
  &=  A^M_t\cdot\bigg( -75\,\kappa\,\epsilon^{nrsl}\,d^{QNR}\,d^{LST}\,d_{NMT}\,\partial_R B_{nrQ}\,\partial_S B_{slL} \bigg) \label{can-eft-legendre-b-at-term-v2}
\end{align}
While \eqref{can-eft-legendre-b-at-term-v2} is of simpler form than \eqref{can-eft-legendre-b-at-term-v1-2} it makes it harder to see that it may lead to generalised diffeomorphisms. In the following we will need to make use of both forms of this term depending on the situation.

\subsubsection[Remaining one-form terms]{Legendre transformation of the Yang-Mills, Einstein-Hilbert improvement and topological terms}
\label{sec-can-EFT-legendre-1forms}
In this section we finally calculate the Legendre transformation of the terms that contribute to the dynamics of the one-forms \eqref{can-eft-legendre-one-forms-formal} --- namely the generalised Yang-Mills term, the Einstein-Hilbert improvement term and the remaining part $(\mathcal{L}_\text{top.}-\mathcal{L}_\text{BK})$ of the topological term. 
\begin{equation}\label{can-eft-legendre-one-forms-formal}
\dot A_{n}^N\cdot \Pi_N^n(A_n^N)  -\mathcal{L}_\text{YM} -  E\;\mathcal{F}^M_{\alpha\beta}E^{\alpha\rho}\partial_M E^\beta_\rho   -(\mathcal{L}_\text{top.}-\mathcal{L}_\text{BK} )
\end{equation}
The one-form sector is by far the most complicated part of the Legendre transformation of ExFT. To simplify the calculation we transiently introduce the expression $ \Upsilon^M_s$ as defined by \eqref{can-eft-legendre-upsilon-def} but we only make use of it in this calculation. The use of $\Upsilon^M_s$, but more importantly the use of the modified momenta $\mathcal{P}^{l}_T(A)$ is what allows us to carry out this calculation in a relatively simple form.
\begin{align}
    \Upsilon^M_s &:=\mathcal{F}^M_{ts} - \dot{A}^M_s \label{can-eft-legendre-upsilon-def}\\
    &= -\partial_s A_t^M - \left[ A_t, A_s \right]^M_\text{E} + 10 \, d^{MNK}\,\partial_N B_{tsK}
\end{align}
We begin with the computation of the terms that involve a time derivative, i.e. either terms with a $\dot{A}^N_n$ or $\mathcal{F}^M_{tl}$. In the following equations the dots '$\dots$' always indicate the same collection of terms that do not have a time derivative in the ADM decomposition --- we will write these terms explicitly once we have dealt with the time derivatives. We start from the Legendre transformation \eqref{can-eft-legendre-a-1} and then write out the time derivative terms. We can then insert \eqref{can-eft-legendre-upsilon-def} and arrive at \eqref{can-eft-legendre-a-2} after some rearrangements of the terms.
\begin{align}
&\dot A_{n}^N\cdot \Pi_N^n(A_n^N)  -\mathcal{L}_\text{YM} -  E\;\mathcal{F}^M_{\alpha\beta}E^{\alpha\rho}\partial_M E^\beta_\rho   -(\mathcal{L}_\text{Top.}-\mathcal{L}_\text{BK} )  \label{can-eft-legendre-a-1}\\
=& \,\dot A_{n}^N\cdot \Pi_N^n(A_n^N)  - \frac{e}{2\,N}\, M_{MN} \,   \mathcal{F}^{M}_{ts} \,\mathcal{F}^N_{tn}\,g^{sn}  +  \frac{e}{N}\,M_{MN} \,   \mathcal{F}^{M}_{ts} \,\mathcal{F}^N_{mn}\,g^{sn}\,N^m  \nonumber\\
& -\frac{e}{N}\,\mathcal{F}^M_{tn}   \,\partial_M N^n +15\kappa\, \epsilon^{tnrsl} \, d^{MNR}\, d_{NKL}\,\partial_R B_{nr M} \, A^K_{s} \,\dot{A}^l_{l} \nonumber\\
& +2\, \kappa\,\, \epsilon^{tnrsl}\,d_{MNP}\enspace A^N_n \enspace\partial_t A^M_r \enspace\partial_s A^P_l -\frac{3\kappa}{4}\epsilon^{tnrsl}\,d_{MNP}  A^N_n\,\left[A_r,A_s\right]^M_E\,\dot{A}^P_l +\dots \nonumber \\
=& \,\dot A_{n}^N\cdot \Pi_N^n(A_n^N) \label{can-eft-legendre-a-2} \\
& - \frac{e}{N}\, M_{MN} \,   \mathcal{F}^{M}_{ts} \,\dot{A}^N_n\,g^{sn} + \frac{e}{2\,N}\, M_{MN} \, (\dot{A}^M_s + \Upsilon^M_s) \,(\dot{A}^N_n  - \Upsilon^N_n)\,g^{sn} \nonumber \\
&+  \frac{e}{N}\,M_{MN} \,  (\dot{A}^M_s + \Upsilon^M_s)\,\mathcal{F}^N_{mn}\,g^{sn}\,N^m  \nonumber\\
& -\frac{e}{N}\,(\dot{A}^M_n + \Upsilon^M_n)  \,\partial_M N^n +15\kappa\, \epsilon^{tnrsl} \, d^{MNR}\, d_{NKL}\,\partial_R B_{nr M} \, A^K_{s} \,\dot{A}^l_{l} \nonumber\\
& +2\, \kappa\,\, \epsilon^{tnrsl}\,d_{MNP}\enspace A^N_n \enspace\partial_t A^M_r \enspace\partial_s A^P_l -\frac{3\kappa}{4}\epsilon^{tnrsl}\,d_{MNP}  A^N_n\,\left[A_r,A_s\right]^M_E\,\dot{A}^P_l +\dots \nonumber
\end{align}
Comparing \eqref{can-eft-legendre-a-2} to the explicit form of the canonical momenta \eqref{can-EFT-original-canonical-momentum-A} we can identify all terms that are needed to cancel the $\dot A_{n}^N\cdot \Pi_N^n(A_n^N)$ term of the Legendre transformation. There is then only one quadratic term with time derivatives left over and we arrive at \eqref{can-eft-legendre-a-3}.
\begin{align}\label{can-eft-legendre-a-3}
&\dot A_{n}^N\cdot \Pi_N^n(A_n^N)  -\mathcal{L}_\text{YM} -  E\;\mathcal{F}^M_{\alpha\beta}E^{\alpha\rho}\partial_M E^\beta_\rho   -(\mathcal{L}_\text{Top.}-\mathcal{L}_\text{BK} )  \\
=&+ \frac{e}{2\,N}\, M_{MN} \, (\dot{A}^M_s + \Upsilon^M_s) \,(\dot{A}^N_n  - \Upsilon^N_n)\,g^{sn} \nonumber \\
&+  \frac{e}{N}\,M_{MN} \, \Upsilon^M_s\,\mathcal{F}^N_{mn}\,g^{sn}\,N^m  -\frac{e}{N}\,\Upsilon^M_n \,\partial_M N^n +\dots\nonumber
\end{align}
To replace the remaining time derivatives of the one-forms in \eqref{can-eft-legendre-a-3} we need to write the modified canonical momenta $\mathcal{P}^{l}_T(A)$ as in \eqref{can-eft-legendre-a-4-P} using \eqref{can-eft-legendre-upsilon-def}. We can now invert \eqref{can-eft-legendre-a-4-P} to arrive at the expression \eqref{can-eft-legendre-a-5-P-inverse} for $\dot{A}^N_n $.
\begin{align}
 \mathcal{P}^{l}_T(A) &= \frac{e}{N} g^{ln}\;M_{TN}\;\bigg( \dot{A}^N_n + \Upsilon^N_n + N^k \mathcal{F}^{N}_{nk}  \bigg)  + \frac{e}{N} \partial_T N^l \label{can-eft-legendre-a-4-P} \\
 \Rightarrow \dot{A}^N_n   &= \frac{N}{e}\,g_{ln}\,M^{TN} \bigg(\mathcal{P}^{l}_T(A) -\frac{e}{N} \partial_T N^l \bigg)  -\Upsilon^N_n - N^k \mathcal{F}^{N}_{nk} \label{can-eft-legendre-a-5-P-inverse}
\end{align}
Inserting the time derivative \eqref{can-eft-legendre-a-5-P-inverse} into \eqref{can-eft-legendre-a-3} we arrive at the expression \eqref{can-eft-legendre-a-6} which does not contain any time derivatives.
\begin{align}\label{can-eft-legendre-a-6}
&\dot A_{n}^N\cdot \Pi_N^n(A_n^N)  -\mathcal{L}_\text{YM} -  E\;\mathcal{F}^M_{\alpha\beta}E^{\alpha\rho}\partial_M E^\beta_\rho   -(\mathcal{L}_\text{Top.}-\mathcal{L}_\text{BK} )    \nonumber\\
=&+\frac{N}{2\,e}\, g_{lm}\, M^{KL} \, \mathcal{P}^l_L(A)\,\mathcal{P}^m_K(A)+\frac{e}{2\,N}\, g_{lm}\, M^{KL} \partial_L N^l \, \partial_K N^m \nonumber\\
&-g_{lm}\, M^{KL} \mathcal{P}_L^l(A) \, \partial_K N^m - \mathcal{P}^n_N(A)\, \Upsilon_n^N + \frac{e}{N}\,N^n\,\mathcal{F}^M_{mn}\,\partial_M N^m\nonumber\\
&+ N^n \, \mathcal{F}^M_{nl} \, \mathcal{P}^l_M(A)  +\frac{e}{2\,N}\,g^{mn}\,M_{MN}\,N^k\,N^l\,\mathcal{F}^M_{mk}\,\mathcal{F}^N_{nl} +\dots
\end{align}
We can now take a closer look at the terms we have found. The origin of the term $+\frac{e}{2\,N}\, g_{lm}\, M^{KL} \partial_L N^l \, \partial_K N^m$ is the interplay of the improvement term of the Einstein-Hilbert term and the generalised Yang-Mills term. The improvement term leads to the $\frac{e}{N}\partial_N N^n$ contribution in $\mathcal{P}_N^n(A)$ and the Yang-Mills term creates the $\dot{A}^2$ from which the term in question originates via \eqref{can-eft-legendre-a-5-P-inverse}. This is the only term that cancels against a scalar potential contribution. Note that the sign of the improvement term of the Einstein-Hilbert term is irrelevant for this cancellation.\\
The term $-g_{lm}\, M^{KL} \mathcal{P}_L^l(A) \, \partial_K N^m$ exists due to the same interplay of terms and we discuss it in more detail in section \ref{can-sec-External-diffeomorphisms} when discussing the external diffeomorphism transformations. \\

Taking a closer look at the $- \mathcal{P}^n_N(A)\, \Upsilon_n^N$ term and opening up the $\Upsilon_n^N$ leads us to the relation \eqref{can-eft-legendre-relation-gen-diff-gen1}.
\begin{equation}\label{can-eft-legendre-relation-gen-diff-gen1}
     - \mathcal{P}^n_N(A)\, \Upsilon_n^N = \partial_n A^M_t \, \mathcal{P}^n_M(A) + \mathcal{P}^n_M(A) \, \left[A_t, A_n \right]^M_\text{E} -10\,d^{MKL}\,\partial_K B_{tnL}\,\mathcal{P}^n_M(A)
\end{equation}
Furthermore we find that we can rewrite the first two terms of \eqref{can-eft-legendre-relation-gen-diff-gen1} as \eqref{can-eft-legendre-relation-gen-diff-gen2}. We can think of equation \eqref{can-eft-legendre-relation-gen-diff-gen2} as the covariantised version of the Gau\ss{} constraint from \cite{KreutzerSUGRA}, which generates U$(1)^{27}$ gauge transformations, plus an extra momentum term.
\begin{align}
    & \partial_n A^M_t \, \mathcal{P}^n_M(A) + \mathcal{P}^n_M(A) \, \left[A_t, A_n \right]^M_\text{E} \nonumber\\
    =&  A^M_t \cdot\left( -\mathcal{D}_n \mathcal{P}^n_M(A) - 5\, d^{KLR}\,d_{MRS}\,A^S_n\,\partial_L \mathcal{P}^n_K(A)\right) \label{can-eft-legendre-relation-gen-diff-gen2}
\end{align}
This extra momentum term $A^M_t \left(- 5\, d^{KLR}\,d_{MRS}\,A^S_n\,\partial_L \mathcal{P}^n_K(A)\right)$ is a direct consequence of the E-bracket in the covariantised field strength of the one-forms $\mathcal{F}^M_{mn}$. However without the E-bracket term in $\mathcal{F}^M_{mn}$ we would also not be able to write down the covariantised $-\mathcal{D}_n \mathcal{P}^n_M(A)$ term. When we compute the canonical gauge transformations we will see that this additional momentum term can be thought of as being related to the tensor gauge symmetry of the B-field.\\

We can now write out the '$\dots$' in \eqref{can-eft-legendre-a-6}, which is a rather large number of terms. We find that there are some cancellations with the explicit terms of \eqref{can-eft-legendre-a-6} and we arrive at equation \eqref{can-eft-legendre-a-final1}.
\begin{align}
&\dot A_{n}^N\cdot \Pi_N^n(A_n^N)  -\mathcal{L}_\text{YM} -  E\;\mathcal{F}^M_{\alpha\beta}E^{\alpha\rho}\partial_M E^\beta_\rho   -(\mathcal{L}_\text{Top.}-\mathcal{L}_\text{BK} )  \nonumber \\
=&+\frac{N}{2\,e}\, g_{lm}\, M^{KL} \, \mathcal{P}^l_L(A)\,\mathcal{P}^m_K(A)+\frac{e}{2\,N}\, g_{lm}\, M^{KL} \partial_L N^l \, \partial_K N^m \label{can-eft-legendre-a-final1}\\
&-g_{lm}\, M^{KL} \mathcal{P}_L^l(A) \, \partial_K N^m -10\,d^{MKL}\,\partial_K B_{tnL}\,\mathcal{P}^n_M(A) + N^n \, \mathcal{F}^M_{nl} \, \mathcal{P}^l_M(A)  \nonumber \\
& -  A^M_t \mathcal{D}_n \mathcal{P}^n_M(A) - 5\, A^M_t\, d^{KLR}\,d_{MRS}\,A^S_n\,\partial_L \mathcal{P}^n_K(A)  \nonumber \\
    & -e\,N\,\mathcal{F}^M_{mn} \, e^{ar}\,\partial_M e_r{}^b\, e_a{}^m \, e_b{}^n +  \frac{e\,N}{4}\,M_{MN} \, \mathcal{F}^{M}_{rs} \,\mathcal{F}^N_{mn}\,g^{rm}\,g^{sn}     \nonumber\\
&-30\kappa \epsilon^{tnrsl} \, d^{MNR}\, d_{NKL}\,\partial_R B_{tn M} \, A^K_{r} \,\partial_s A^L_{l} \nonumber\\
  &+15\kappa \epsilon^{tnrsl} \, d^{MNR}\, d_{NKL}\,\partial_l\partial_R B_{nr M} \, A^K_{s} \, A^L_{t} \nonumber\\
  &-30\kappa \epsilon^{tnrsl} \, d^{MNR}\, d_{NKL}\,\partial_R B_{nr M} \, A^K_{t} \,\partial_s A^L_{l} \nonumber\\
  & +10\kappa\,\epsilon^{tnrsl} \, d^{MNR}\, d_{NKL}\,\partial_R B_{tn M} \,A^K_{r}\,[A_s,A_l]^L_\text{E} \nonumber\\
  & +5\kappa\,\epsilon^{tnrsl} \, d^{MNR}\, d_{NKL}\,\partial_R B_{nr M} \,A^K_t\,[A_s,A_l]^L_\text{E} \nonumber\\
  & -10\kappa\,\epsilon^{tnrsl} \, d^{MNR}\, d_{NKL}\,\partial_R B_{nr M} \,A^K_{s}\,[A_t,A_l]^L_\text{E} \nonumber\\
  & -3 \kappa\,\, \epsilon^{tnrsl}\,d_{MNP}\enspace A^N_t \enspace\partial_n A^M_r \enspace\partial_s A^P_l \nonumber \\
    &+ \frac{3\kappa}{4}\epsilon^{tnrsl}\,d_{MNP}\, A^N_t\,\left[A_n,A_r\right]^M_E\,\partial_s\, A^P_l \nonumber \\
    &- \frac{3\kappa}{2}\epsilon^{tnrsl}\,d_{MNP} \, A^N_n\,\left[A_t,A_r\right]^M_E\,\partial_s\, A^P_l \nonumber \\
    & + \frac{3\kappa}{4}\epsilon^{tnrsl}\,d_{MNP}\, A^N_n\,\left[A_r,A_s\right]^M_E\,\partial_l\, A^P_t  \nonumber \\
     &- \frac{3\kappa}{20}\epsilon^{tnrsl}\,d_{MNP}\,A^N_ t\, \left[A_n,A_r\right]^M_E \,\left[A_s,A_l\right]^P_E \nonumber\\
     &+ \frac{3\kappa}{5}\epsilon^{tnrsl}\,d_{MNP}\,A^N_n\, \left[A_t ,A_r\right]^M_E \,\left[A_s,A_l\right]^P_E \nonumber
\end{align}
In order to rewrite \eqref{can-eft-legendre-a-final1} in a more useful form we can factor out the Lagrange multipliers in the non-topological terms and organise the terms accordingly to arrive at \eqref{can-eft-legendre-a-final2}.
\begin{align}
&\dot A_{n}^N\cdot \Pi_N^n(A_n^N)  -\mathcal{L}_\text{YM} -  E\;\mathcal{F}^M_{\alpha\beta}E^{\alpha\rho}\partial_M E^\beta_\rho   -(\mathcal{L}_\text{Top.}-\mathcal{L}_\text{BK} )   \nonumber\\
=&+\frac{e}{2\,N}\, g_{lm}\, M^{KL} \partial_L N^l \, \partial_K N^m \label{can-eft-legendre-a-final2}\\
&+ N \cdot \bigg( +\frac{N}{2\,e}\, g_{lm}\, M^{KL} \, \mathcal{P}^l_L(A)\,\mathcal{P}^m_K(A) +  \frac{e}{4}\,M_{MN} \, \mathcal{F}^{M}_{rs} \,\mathcal{F}^N_{mn}\,g^{rm}\,g^{sn}       \nonumber \\
 &\hspace{1.4cm} -e\,\mathcal{F}^M_{mn} \, e^{ar}\,\partial_M e_r{}^b\, e_a{}^m \, e_b{}^n \bigg)\nonumber\\
&+ N^n\cdot\bigg( + \mathcal{F}^M_{nl} \, \mathcal{P}^l_M(A) + \partial_K\left( g_{mn}\, M^{KL} \mathcal{P}_L^m(A)\right) \bigg)  \nonumber \\
& + A^M_t \cdot\bigg( -  \mathcal{D}_n \mathcal{P}^n_M(A) - 5\, d^{KLR}\,d_{MRS}\,A^S_n\,\partial_L \mathcal{P}^n_K(A) \bigg) \nonumber \\
  &+15\kappa \epsilon^{tnrsl} \, d^{MNR}\, d_{NKL}\,\partial_l\partial_R B_{nr M} \, A^K_{s} \,A^L_{t} \nonumber\\
  &-30\kappa \epsilon^{tnrsl} \, d^{MNR}\, d_{NKL}\,\partial_R B_{nr M} \, A^K_{t} \,\partial_s A^L_{l} \nonumber\\
  & +5\kappa\,\epsilon^{tnrsl} \, d^{MNR}\, d_{NKL}\,\partial_R B_{nr M} \,A^K_t\,[A_s,A_l]^L_\text{E} \nonumber\\
  & -10\kappa\,\epsilon^{tnrsl} \, d^{MNR}\, d_{NKL}\,\partial_R B_{nr M} \,A^K_{s}\,[A_t,A_l]^L_\text{E} \nonumber\\
  & -3 \kappa\,\, \epsilon^{tnrsl}\,d_{MNP}\enspace A^N_t \enspace\partial_n A^M_r \enspace\partial_s A^P_l \nonumber \\
    &+ \frac{3\kappa}{4}\epsilon^{tnrsl}\,d_{MNP}\, A^N_t\,\left[A_n,A_r\right]^M_E\,\partial_s\, A^P_l \nonumber \\
    &- \frac{3\kappa}{2}\epsilon^{tnrsl}\,d_{MNP} \, A^N_n\,\left[A_t,A_r\right]^M_E\,\partial_s\, A^P_l \nonumber \\
    & + \frac{3\kappa}{4}\epsilon^{tnrsl}\,d_{MNP}\, A^N_n\,\left[A_r,A_s\right]^M_E\,\partial_l\, A^P_t  \nonumber \\
     &- \frac{3\kappa}{20}\epsilon^{tnrsl}\,d_{MNP}\,A^N_ t\, \left[A_n,A_r\right]^M_E \,\left[A_s,A_l\right]^P_E \nonumber\\
     &+ \frac{3\kappa}{5}\epsilon^{tnrsl}\,d_{MNP}\,A^N_n\, \left[A_t ,A_r\right]^M_E \,\left[A_s,A_l\right]^P_E \nonumber\\
     &+10\,d^{MKL}\,B_{tnL}\,\partial_K  \mathcal{P}^n_M(A) \nonumber\\
     &+30\kappa \epsilon^{tnrsl} \, d^{MNR}\, d_{NKL}\, B_{tn M} \,\partial_R( A^K_{r} \,\partial_s A^L_{l} )\nonumber\\
      & -10\kappa\,\epsilon^{tnrsl} \, d^{MNR}\, d_{NKL}\, B_{tn M} \,\partial_R( A^K_{r}\,[A_s,A_l]^L_\text{E}) \nonumber
\end{align}
In \eqref{can-eft-legendre-a-final2} we find the expected quadratic one-form field strength $\mathcal{F}^2$ and momenta terms $\mathcal{P}^2$ of the Hamilton constraint. We can furthermore see the spatial Einstein-Hilbert improvement term. We also find the expected one-form diffeomorphism constraint term $\mathcal{F}\mathcal{P}$ plus the additional term $+ \partial_K\left( g_{mn}\, M^{KL} \mathcal{P}_L^m(A)\right)$ that was already mentioned earlier.\\

Combining the last three terms of \eqref{can-eft-legendre-a-final2} with the $B_{tnM}$ term from section \ref{sec-can-legendre-2form-EFT} we can construct the expression \eqref{can-eft-legendre-a-S1-constraint}. Here we define $\mathcal{H}_\text{S1}$ in analogy to the secondary constraint from section \ref{can-sec-twoforms-model} that carries the same name. We will find that it is indeed a secondary canonical constraint and it generates part of the tensor gauge transformations.
\begin{equation}\label{can-eft-legendre-a-S1-constraint}
    +B_{tlM} \cdot \bigg{[}10 \,d^{MKL} \partial_K \left( \mathcal{P}^l_L  - \kappa\, \epsilon^{lmnr}\, \mathcal{H}_{mnrL}\right) \bigg{]} =:  +B_{tlM} \cdot (\mathcal{H}_\text{S1})^{lM}
\end{equation}
Lastly we should factor out the Lagrange multipliers of the remaining topological terms. We only state them in this form in the final result \eqref{can-eft-Htop} as this expression is not interesting and makes the Hamiltonian much more complicated. This is in particular due to the terms with a time index in the E-bracket $\left[A_t ,A_r\right]^M_E $.

\subsection{Canonical Hamiltonian}\label{can-eft-sec-can-ham-disc}
Combining the partial results for the computation of the Legendre transformation \eqref{can-EFT-legendre-collection-terms-hamiltonian} from section \ref{can-eft-sec-legendre} we finally arrive at the complete canonical bosonic E$_{6(6)}$ exceptional field theory Hamiltonian $\mathcal{H}_{\text{ExFT}}$ \eqref{can-eq:can-Hamiltonian-EFT}. In order to make the secondary constraints apparent we have written \eqref{can-eq:can-Hamiltonian-EFT} in the form where the Lagrange multipliers have been factored out.
\begin{align}\label{can-eq:can-Hamiltonian-EFT}
    \mathcal{H}_{\text{ExFT}}\,\, = &+ N \cdot \bigg{[} +\frac{1}{4 e} \Pi_{ab}(e)\,\,  \Pi_{ab}(e) - \frac{1}{12e} \Pi(e)^2 - e\,\,  \hat{R}  + e\, V_\text{HP} \nonumber\\
                              & \hspace{1.3cm } +\frac{3}{2e} \Pi^{MN}(M) \,\, \Pi^{RS}(M)\,\,  M_{MR}\,\,  M_{NS}   - \frac{e}{24} g^{kl} \,\, \mathcal{D}_k M_{MN} \,\, \mathcal{D}_l M^{MN}      \nonumber \\
                              &\hspace{1.3cm }+\frac{e}{4} M_{MN}\,\,  g^{rm}\,\,  g^{sn} \,\, \mathcal{F}_{rs}^M\,\, \mathcal{F}_{mn}^N       +\frac{1}{2e} g_{lm}\,\,  M^{KL} \,\, \mathcal{P}^l_L\,\,  \mathcal{P}^m_K \bigg{]} \nonumber \\
                          & +N^n \cdot \bigg{[}   + 2\,\, \Pi^m{}_a(e)\,\, \mathcal{D}_{[n} e_{m]a} - e_{na}\,\, \mathcal{D}_m \Pi^m{}_a(e)     \nonumber \\
                              &\hspace{1.5cm } +\frac{1}{2} \Pi^{MN}(M)\,\, \mathcal{D}_n M_{MN}     \nonumber \\
                              & \hspace{1.5cm } + \mathcal{F}_{nl}^M \mathcal{P}^l_M + \partial_M\left(g_{mn}\, M^{MN}\, \mathcal{P}^m_N \right) \bigg{]}    \nonumber \\
                          & +A_t^M \cdot \bigg{[} -\mathcal{D}_l \mathcal{P}^l_M -5 \,d^{NLS} \, d_{MNK}\, A^K_m \partial_S \mathcal{P}^m_L + (\mathcal{H}_{top})_M \nonumber\\
                           &\hspace{1.5cm} +\,\Pi^m{}_a(e)\, \partial_M e_{ma}   - \frac{1}{3} \partial_M \Pi(e) \nonumber \\
                          & \hspace{1.5cm}+ \frac{1}{2}\,\Pi^{KL}(M) \, \partial_M M_{KL} - 6\, \mathbb{P}^R{}_K{}^S{}_M \, \partial_S\left(\Pi^{KL}(M)\, M_{RL} \right)\bigg{]} \nonumber \\
                            & +B_{tlM} \cdot \bigg{[} +10 \,d^{MKL} \partial_K \left( \mathcal{P}^l_L  - \kappa\, \epsilon^{lmnr}\, \mathcal{H}_{mnrL}\right) \bigg{]} \nonumber \\
                            &+ \dot{N} \cdot \Pi(N)  + \dot{N}_a\cdot \Pi^a(N_a) + \dot{A}_t^M \cdot\Pi_M(A_t) +\dot{B}_{tnN}\cdot \Pi^{nN}(B_{tnN})  
\end{align}
In the Hamiltonian \eqref{can-eq:can-Hamiltonian-EFT} $\hat{R}$ is the improved spatial Ricci scalar \eqref{can-eft-impr-spat-ricci} (cf. equation \eqref{eft-lagrangian-EH}).
\begin{equation}\label{can-eft-impr-spat-ricci}
    - e\,  \hat{R} =  - e\,  R_4  - e \,\mathcal{F}^M_{mn}\, e_a{}^m e_b{}^n \,(e^{ra} \partial_M e_r{}^b)
\end{equation}
The scalar potential of the Hamiltonian $V_\text{HP}$ is closely related to the scalar potential of the Lagrangian $V_\text{pot}$ by \eqref{can-eft-potantial-relation}. We add to the Lagrangian potential the contribution from the Einstein-Hilbert improvement term, which we found in section \ref{sec-can-EFT-legendre-1forms}, thus cancelling this term overall.
\begin{equation}\label{can-eft-potantial-relation}
    N\,e\,V_\text{HP}  = N\, e\,V_\text{pot}  + \frac{e}{2N} g_{mn} M^{MN} \partial_M N^n \,\partial_N N^m 
\end{equation}
The Hamiltonian scalar potential takes the explicit form of equation \eqref{can-Hamiltonian-eq:EFT-potential}.
\begin{align}\label{can-Hamiltonian-eq:EFT-potential}
    + e\,V_\text{HP} :=  & - \frac{e}{4}  M^{MN} \partial_M g^{mn}\, \partial_N g_{mn}  - \frac{1}{e}  M^{MN} \partial_M e \, \partial_N e \\
             & - \frac{e}{24}  M^{MN} \partial_M M^{KL} \, \partial_N M_{KL} + \frac{e}{2}  M^{MN} \partial_M M^{KL} \, \partial_L M_{NK}\nonumber \\
             & +  \partial_M \partial_N M^{MN} \,  e+  2  \,  M^{MN} \, \partial_M \partial_N e + 2   \, \partial_M M^{MN} \, \partial_N e\nonumber 
\end{align}
The Hamiltonian topological term $\mathcal{H}_{top}$ collects all terms that originate from $\mathcal{L}_{top}$ \eqref{topological-term-E-six}, with the exception of the terms that go into the modified momenta $\mathcal{P}^n_N$ or into the two-form field strength $\mathcal{H}_{klmN}$. $\mathcal{H}_{top}$ is explicitly given by \eqref{can-eft-Htop}. $\mathcal{H}_{top}$ is the ExFT analogue of the Hamiltonian topological $F^2$ $\theta$-term in the Gau\ss{} constraint of five-dimensional ungauged maximal E$_{6(6)}$ invariant supergravity, cf. reference \cite{KreutzerSUGRA}. We can also see this term in the third line of \eqref{can-eft-Htop}, it is the only part of the topological term that does not depend on any internal derivatives. The great complexity of $\mathcal{H}_{top}$ is a consequence of the complexity of the topological term in the Lagrangian $\mathcal{L}_{top}$ and is made slightly worse by the need to factor out the Lagrange multiplier $A^M_t$. Note that $\mathcal{H}_{top}$ is by definition linear in the coefficient of the Lagrangian topological term $\kappa$. This factor is hidden for the two-form kinetic term in \eqref{can-eft-Htop} because it is written in terms of the primary constraint $\mathcal{H}_\text{P2}$ (cf. discussion in section \ref{sec-can-legendre-2form-EFT}).
\begin{align}
& (\mathcal{H}_{top})_M = \label{can-eft-Htop} \\
     &+ \frac{1}{2}\,\Pi^{lnN}(B) \, \partial_M B_{lnN} - 3\, \mathbb{P}^R{}_K{}^S{}_M \, \partial_S\left(\Pi^{lnK}(B)\, B_{lnR} \right) -\frac{1}{3}\,\partial_M \left(B_{mnN} \Pi^{mnN}(B) \right) \nonumber\\  
      & - \frac{1}{2}\,(\mathcal{H}_\text{P2})^{lnN} \, \partial_M B_{lnN} + 3\, \mathbb{P}^R{}_K{}^S{}_M \, \partial_S\left((\mathcal{H}_\text{P2})^{lnK}\, B_{lnR} \right) +\frac{1}{3}\,\partial_M \left(B_{mnN} (\mathcal{H}_\text{P2})^{mnN} \right) \nonumber\\
      & - 3\kappa\, \epsilon^{tlmnr}\,\, d_{MNP}\,\, \partial_l A^N_{m} \,\, \partial_n A^P_{r} \nonumber \\
     &   - 15\kappa\, \epsilon^{tlmnr}\,\, d^{SRN}\,\, d_{MNK}\,\partial_l \partial_SB_{mnR}\,A^K_r   - 30\kappa\, \epsilon^{tlmnr}\,\, d^{SRN}\,\, d_{MNK}\,\partial_SB_{mnR}\,\partial_l\,A^K_r \nonumber\\
&  +5\kappa \,\epsilon^{tlmnr} \,d^{SRN} \, d_{MNK}\, [A_l,A_m]^K_E \,\partial_S B_{nrR}  -20\kappa \,\epsilon^{tlmnr} \,d^{SRN} \, d_{QNK}\, A^K_l\, \partial_M A^Q_m  \,\partial_S B_{nrR} \nonumber \\
     &  +100\kappa \,\epsilon^{tlmnr} \,d^{NKT}\,d^{QRS}\, d_{MNL}\, d_{TPQ}\, A^P_l \, \partial_K A^L_m \,\partial_S B_{nrR} \nonumber \\
 &  +\frac{3}{2}\kappa \,\epsilon^{tlmnr} \,d_{MNK}\,\, \partial_l A_m^N \,\, [A_n,A_r]^K_E  -3\kappa \,\epsilon^{tlmnr} \,d_{QNK}\,\,A^Q_l \, \partial_m A_n^N \,\, \partial_M A_r^K \nonumber \\
 & +15\kappa \,\epsilon^{tlmnr} \,d^{NRS}\,d_{MNK}\,d_{RLP}\, A_l^L \,\,\partial_SA_m^K \, \partial_n  A^P_r \nonumber \\
  & -\frac{3}{2}\kappa \,\epsilon^{tlmnr} \,d_{MNP}\,\,A^N_l \, \partial_m A_n^S \,\, \partial_S A_r^P   -\frac{3}{2}\kappa \,\epsilon^{tlmnr} \,d_{MNP}\,\,A^N_l \,  A_n^S \,\, \partial_m\partial_S A_r^P \nonumber \\
 & +\frac{15}{2}\kappa \,\epsilon^{tlmnr} \,d^{PXS}\,d_{MNP}\,d_{XYZ}\, A_l^N \,\,\partial_m A_n^Y \, \partial_S  A^Z_r \nonumber \\
 &+\frac{15}{2}\kappa \,\epsilon^{tlmnr} \,d^{PXS}\,d_{MNP}\,d_{XYZ}\, A_l^N \,\, A_n^Y \, \partial_m\partial_S  A^Z_r \nonumber \\
  &  -\frac{3}{20}\kappa \,\epsilon^{tlmnr} \,d_{MNK}\, [A_l,A_m]^N_E\,[A_n,A_r]^K_E +\frac{6}{5}\kappa \,\epsilon^{tlmnr} \,d_{QNK}\, A^Q_l \,\partial_M A^N_m \,[A_n,A_r]^K_E\nonumber \\
 &  -6\kappa \,\epsilon^{tlmnr} \,d^{NRS}\,d_{MNK}\,d_{RLQ}\, A^L_l \,\partial_S A^K_m \,[A_n,A_r]^Q_E\nonumber 
\end{align}
We can further compare the E$_{6(6)}$ ExFT Hamiltonian \eqref{can-eq:can-Hamiltonian-EFT} to the Hamiltonian $\mathcal{H}_\text{5D}$ of five-dimensional ungauged maximal E$_{6(6)}$ invariant supergravity \cite{KreutzerSUGRA}. As is expected we find that upon applying the trivial solution of the section condition ($\partial_M = 0\,\,\forall M$) to the ExFT Hamiltonian \eqref{can-eq:can-Hamiltonian-EFT} it reduces to the five-dimensional supergravity Hamiltonian $\mathcal{H}_\text{5D}$. 
The ExFT Hamiltonian contains all terms that are in form identical to those of $\mathcal{H}_\text{5D}$, but with the derivatives, one-form field strength and Ricci scalar replaced by the covariantised expressions $\mathcal{D}_\mu,\, \mathcal{F}_{\mu\nu}^M$ and $\hat{R}$.
In addition to the terms found in $\mathcal{H}_\text{5D}$ there are internal derivative terms in ExFT that vanish completely in the trivial solution of the section condition. This includes the scalar potential $V_\text{HP}$, the term $+ N^n\, \partial_M\left(g_{mn}\, M^{MN}\, \mathcal{P}^m_N \right)$ --- which is further discussed in section \ref{can-sec-External-diffeomorphisms}, as well as all of the $B_{tlM}$ dependent and most of the $A^M_t$ dependent terms. In $\mathcal{H}_\text{5D}$ the $A^M_t$ dependent terms form the Gau\ss{} constraint which generates U(1)$^{27}$ gauge transformations. In ExFT the analogue expression is much more complicated because the one-forms act as a connection in the covariant derivative $\mathcal{D}_\mu$ and because the Lagrangian topological term of ExFT is much larger than that of five-dimensional supergravity. In section \ref{can-eft-GD-sec} we will see that the $A^M_t$ dependent terms form the constraint that generates generalised exceptional diffeomorphisms.
The $B_{tlM}$ dependent terms do not have any analogue in $\mathcal{H}_\text{5D}$ because there is no two-form in the five-dimensional theory. In section \ref{can-eft-tensor-gauge-transf-sec} we will see that these terms form the constraint that generates a part of the tensor gauge transformations.\\

If we insert the definition \eqref{can-EFT-original-canonical-momentum-A-redef} of the modified momenta $\mathcal{P}^m_N$ into the Hamiltonian \eqref{can-eq:can-Hamiltonian-EFT} we can see how much more cluttered the Hamiltonian is when expressed in terms of the canonical momenta $\Pi^m_N(A)$. Therefore $\mathcal{P}^m_N$ seem to be the best variables to use.  

\subsection{Fundamental Poisson brackets}\label{can-eft-sec-fundamental-poisson}
Before we can construct the complete set of canonical constraints we first need to define the fundamental Poisson brackets. In this section we use the notation $X_1 = (x_1, Y_1)$ to denote the tupel of spatial external and internal coordinates and define $X_1-X_2 = (x_1-x_2, Y_1-Y_2)$. The non-vanishing fundamental equal-time Poisson brackets are listed below.
\begin{align}
      \{ N(X_1) , \Pi(N)(X_2)\}  & =  \,\delta^{(4+27)}(X_1-X_2) \\
      \{ N^n(X_1) , \Pi_m(N^k)(X_2)\}   & =  \, \delta^n_m \delta^{(4+27)}(X_1-X_2) \\
      \{ e_n{}^a(X_1) , \Pi^m{}_b(e)(X_2)\}   & =  \, \delta_n^m \delta^a_b \delta^{(4+27)}(X_1-X_2) \\
      \{ A^M_t(X_1), \Pi_N(A^K_t)(X_2)\}   & =  \, \delta^M_N \delta^{(4+27)}(X_1-X_2) \\
      \{ A^M_m(X_1), \Pi^n_N(A^K_k)(X_2)\}  & =  \{ A^M_m(X_1), \mathcal{P}^n_N(X_2)\} \label{can-eft-funda-P-rel}\\
      & = \delta^M_N \delta^n_m \delta^{(4+27)}(X_1-X_2) \\
        \{ B_{tlR}(X_1), \Pi^{tnS}(B_{tqQ})(X_2)\}   & =  \delta^n_l \delta^S_R \,\delta^{(4+27)}(X_1-X_2)  \\
       \{ B_{klR}(X_1), \Pi^{mnS}(B_{pqQ})(X_2)\}  & =  \left(\delta^m_k \delta^n_l - \delta^m_l \delta^n_k\right)  \delta^S_R \,\delta^{(4+27)}(X_1-X_2) \\
    \{M_{MN}(X_1), \Pi^{PQ}(M)(X_2) \}  & =  \left(\delta^P_M \, \delta^Q_N + \delta^P_N \,\delta^Q_M \right) \delta^{(4+27)}(X_1-X_2) \label{sec-EFT-can-eqn:can-rel-Poisson-scalar}
\end{align}
The modified momenta $\mathcal{P}^n_N(A)$ do not affect the Poisson bracket \eqref{can-eft-funda-P-rel} with the one-forms. We need to be careful with the modified momenta however because the definition \eqref{can-EFT-original-canonical-momentum-A-redef} is not a canonical transformation and the momenta do not Poisson-commute among themselves $\{\mathcal{P}^l_L(A),\mathcal{P}^k_K(A) \} \neq 0$. Because of the B-field dependent term in \eqref{can-EFT-original-canonical-momentum-A-redef} we furthermore have to pay attention to the fact that $\{\mathcal{P}^l_L(A), \Pi^{mnS}(B)  \} \neq 0$.\\

As explained earlier we use the implicit formalism for the E$_{6(6)}/$USp(8) coset constraints (cf. reference \cite{KreutzerSUGRA}). Because of the implicit formalism the relation \eqref{sec-EFT-can-eqn:can-rel-Poisson-scalar} is the fundamental Poisson bracket of a generic scalar matrix and there is no coset projector term.

\subsection{Canonical constraints}\label{can-eft-can-constr-sec}
In this section we derive the secondary canonical constraints that arise as the consistency conditions of the primary constraints listed in section \ref{can-eft-sec-primary-constraints}. 

\subsubsection{Total Hamiltonian}\label{can-eft-sec-total-ham}
To verify the consistency of the primary constrains we need to make use of the total Hamiltonian \cite{Henneaux-Teitelboim}.
The total Hamiltonian is given by \eqref{can-eft-total-ham-prelim} and consists of the canonical Hamiltonian plus a generic phase space linear combination of the primary constraints.  
\begin{align}\label{can-eft-total-ham-prelim}
    \mathcal{H}_\text{ExFT-Total} :=&\, \mathcal{H}_\text{ExFT} + u_0\cdot\Pi(N) + (u_1)_a \cdot\Pi^a(N_a)+(u_2)^{ab}\cdot L_{ab} \\
    & +(u_3)^M\cdot \Pi^t_M(A)+(u_4)_{lN}\cdot(\mathcal{H}_\text{P1})^{lN}+ (u_5)_{slN}\cdot(\mathcal{H}_\text{P2})^{slN} \nonumber
\end{align}
With the fundamental Poisson brackets from section \ref{can-eft-sec-fundamental-poisson} we can verify that all primary constraints Poisson-commute --- that is with the exception of the Lorentz constraints among themselves, which form the Lorentz subalgebra --- this fact simplifies the consistency procedure. In particular this means that the primary constraints associated to the two-forms Poisson-commute, which is a result that we already know from the analogous model theory in section \ref{ham-can-B-sec-constr-consist}.
\begin{align}
    \{(\mathcal{H}_\text{P1})^{kK} ,(\mathcal{H}_\text{P1})^{lL}  \} &= 0 \\
    \{(\mathcal{H}_\text{P1})^{kK} ,(\mathcal{H}_\text{P2})^{mnM} \} &= 0\\
    \{(\mathcal{H}_\text{P2})^{klK}  ,(\mathcal{H}_\text{P2})^{mnM} \} & =0
\end{align}

\subsubsection{Secondary constraints}
For the formalism to be consistent the primary constraints have to be conserved in time under the time evolution generated by the total Hamiltonian \cite{Henneaux-Teitelboim}. The consistency of the shift-type primary constraints $\Pi(N)$, $\Pi^n(N_n)$, $\Pi_M(A^M_t)$ and $(\mathcal{H}_{P1})^{lM}(B)=\Pi^{tlM}(B)$ immediately leads to the Hamilton constraint \eqref{can-eft-Hamilton-constr}, the (external) diffeomorphism constraint \eqref{can-eft-diff-constr}, the (internal) generalised diffeomorphism constraint \eqref{can-eft-GD-constr} and the secondary B-field constraint \eqref{can-eft-S1-constr} respectively.
\begin{align}
    \mathcal{H}_{\text{Ham}} =& +\frac{1}{4 e} \Pi_{ab}(e)\,\,  \Pi_{ab}(e) - \frac{1}{12e} \Pi(e)^2 - e\,\,  \hat{R}  + e\, V_\text{HP} \label{can-eft-Hamilton-constr}\\
                              &  +\frac{3}{2e} \Pi^{MN}(M) \,\, \Pi^{RS}(M)\,\,  M_{MR}\,\,  M_{NS}   - \frac{e}{24} g^{kl} \,\, \mathcal{D}_k M_{MN} \,\, \mathcal{D}_l M^{MN}      \nonumber \\
                              &+\frac{e}{4} M_{MN}\,\,  g^{rm}\,\,  g^{sn} \,\, \mathcal{F}_{rs}^M\,\, \mathcal{F}_{mn}^N       +\frac{1}{2e} g_{lm}\,\,  M^{KL} \,\, \mathcal{P}^l_L\,\,  \mathcal{P}^m_K  \nonumber \\
(\mathcal{H}_{\text{Diff}})_n       =&  + 2\,\, \Pi^m{}_a(e)\,\, \mathcal{D}_{[n} e_{m]a} - e_{na}\,\, \mathcal{D}_m \Pi^m{}_a(e) \label{can-eft-diff-constr} \\
                              &+\frac{1}{2} \Pi^{MN}(M)\,\, \mathcal{D}_n M_{MN}     \nonumber \\
                              & + \mathcal{F}_{nl}^M \mathcal{P}^l_M + \partial_M\left(g_{mn}\, M^{MN}\, \mathcal{P}^m_N \right)    \nonumber \\
   (\mathcal{H}_{\text{GD}})^M      =& -\mathcal{D}_l \mathcal{P}^l_M -5 \,d^{NLS} \, d_{MNK}\, A^K_m \partial_S \mathcal{P}^m_L + (\mathcal{H}_{top})_M \label{can-eft-GD-constr}\\
                           &+\,\Pi^m{}_a(e)\, \partial_M e_{ma}   - \frac{1}{3} \partial_M \Pi(e) \nonumber \\
                          &+ \frac{1}{2}\,\Pi^{KL}(M) \, \partial_M M_{KL} - 6\, \mathbb{P}^R{}_K{}^S{}_M \, \partial_S\left(\Pi^{KL}(M)\, M_{RL} \right) \nonumber \\
 (\mathcal{H}_{\text{S1}})^{lM}      =& +10 \,d^{MKL} \partial_K \left( \mathcal{P}^l_L  - \kappa\, \epsilon^{lmnr}\, \mathcal{H}_{mnrL}\right) \label{can-eft-S1-constr} 
\end{align}
The Lorentz constraints do not lead to any secondary constraints.\\

The secondary constraints $\mathcal{H}_\text{S1}$ \eqref{can-eft-S1-constr} are the ExFT version of the constraints \eqref{hamilton-2forms-canonical-constr-S1-H} in the topological model theory, which themselves follow from the consistency condition \eqref{ham-can-B-P1-consistency-S1}. We should furthermore note the similarity of \eqref{can-eft-S1-constr} to the Lagrangian duality equations of motion of the two-forms \eqref{eft-eom-b}.\\

The only primary constraints whose consistency we have not yet considered are the two-form constraints $\mathcal{H}_\text{P2}$ \eqref{can-EFT-primary-canonical-constraint-B-nontrivial} which are not of shift type. In the model theory of section \ref{can-sec-twoforms-model} we have seen that the consistency of the analogous $\mathcal{H}_\text{P2}$ constraints leads to secondary constraints $\mathcal{H}_\text{S2}$ \eqref{hamilton-2forms-canonical-S2}, because of the non-vanishing bracket $\{\mathcal{H}_\text{P2}, \mathcal{H}_\text{S1}\} \neq 0$. Because the constraint $\mathcal{H}_\text{P2}$ Poisson-commutes with all other primary constraints its time evolution with respect to the total and the canonical Hamiltonian are identical and therefore the consistency conditions are too \eqref{can-eft-consist-P2-1}.
\begin{equation}\label{can-eft-consist-P2-1}
     0  \stackrel{!}{=}\{(\mathcal{H}_\text{P2})^{mnM} ,\mathcal{H}_\text{ExFT-Total}\} = \{(\mathcal{H}_\text{P2})^{mnM} ,\mathcal{H}_\text{ExFT}\}
\end{equation}
In contrast to the simple model of section \ref{can-sec-twoforms-model} the consistency condition \eqref{can-eft-consist-P2-1} is more complicated in ExFT because every secondary constraint in the Hamiltonian $\mathcal{H}_\text{ExFT}$ depends on the two-forms and these secondary constraints do not commute with $\mathcal{H}_\text{P2}$. At this point we are not allowed to apply primary constraints and thus there are actually no $\Pi^{klL}(B)$ terms in the canonical Hamiltonian --- as is expected of a field that appears with a single time derivative in the Lagrangian. Fortunately this means that we only need to care about the $\Pi^{mnM}(B)$ term of the constraint $\mathcal{H}_\text{P2}$ for the calculation of \eqref{can-eft-consist-P2-1}. This means we can express \eqref{can-eft-consist-P2-1} in terms of the transformation \eqref{can-eft-consist-P2-2}.
\begin{equation}\label{can-eft-consist-P2-2}
      \{(\mathcal{H}_\text{P2})^{mnM},\mathcal{H}_\text{ExFT}\}  =\{\Pi^{mnM}(B) ,\mathcal{H}_\text{ExFT}\}  \stackrel{!}{=} 0 
\end{equation}
The consistency condition \eqref{can-eft-consist-P2-2} is in direct analogy to the model theory where it leads to the secondary constraints \eqref{hamilton-2forms-canonical-S2}. Due to of the St\"uckelberg coupling of the two-forms to the field strength of the one-forms we get contributions to \eqref{can-eft-consist-P2-2} from every secondary constraint in ExFT. Because of the independence of the Lagrange multipliers of the secondary constraints in the Hamiltonian we can split up \eqref{can-eft-consist-P2-2} into the independent consistency conditions \eqref{can-eft-constr-missing-B-Ham}, \eqref{can-eft-constr-missing-B-Diff}, \eqref{can-eft-constr-missing-B-GD} and \eqref{can-eft-constr-missing-B-S1}. Each of these consistency conditions is really just a different transformation of $\Pi^{mnM}(B)$ with the parameter given by the relevant Lagrange multiplier.
\begin{align}
    \{(\mathcal{H}_\text{P2})^{mnM} ,\mathcal{H}_\text{Ham}[N]\} & = \{\Pi^{mnM}(B) ,\mathcal{H}_\text{Ham}[N]\} \stackrel{!}{=} 0 \label{can-eft-constr-missing-B-Ham}\\
    \{(\mathcal{H}_\text{P2})^{mnM} ,\mathcal{H}_\text{Diff}[N^l]\} & = \{\Pi^{mnM}(B) ,\mathcal{H}_\text{Diff}[N^l]\} \stackrel{!}{=} 0 \label{can-eft-constr-missing-B-Diff}\\
    \{(\mathcal{H}_\text{P2})^{mnM} ,\mathcal{H}_\text{GD}[A^L_t]\} & = \{\Pi^{mnM}(B) ,\mathcal{H}_\text{GD}[A^L_t]\} \stackrel{!}{=} 0 \label{can-eft-constr-missing-B-GD}\\
    \{(\mathcal{H}_\text{P2})^{mnM} ,\mathcal{H}_\text{S1}[B_{tlL}]\} & = \{\Pi^{mnM}(B) ,\mathcal{H}_\text{S1}[B_{tlL}]\} \stackrel{!}{=} 0 \label{can-eft-constr-missing-B-S1}
\end{align}
These constraints should exist due to the direct analogy with the $\mathcal{H}_\text{S2}$ constraints in the model from section \ref{can-sec-twoforms-model}. Furthermore we should expect the same structure of second class constraints as described in section \ref{ham-can-sec-gauge-dof}. In particular the second class system of constraints will again require the introduction of a Dirac bracket which then leads to the difficulties that we already addressed in section \ref{ham-can-sec-dirac-brackets}. We will come back to the two-form constraints in section \ref{can-eft-tensor-gauge-transf-sec} where we calculate the canonical transformations equivalent to  \eqref{can-eft-constr-missing-B-Ham}, \eqref{can-eft-constr-missing-B-Diff}, \eqref{can-eft-constr-missing-B-GD} and \eqref{can-eft-constr-missing-B-S1}.\\ 

We do not find any other secondary constraints and the consistency of the secondary constraints that we have found so far does not yield any further canonical constraints.

\section{Canonical (gauge) transformations in \texorpdfstring{E$_{6(6)}$}{E6(6)} ExFT}\label{can-eft-gauge-transformations}
In this section we investigate the gauge transformations that the canonical constraints generate via the Poisson brackets on the canonical coordinates. Schematically we can think of the canonical (gauge) transformations generated by a (first class) constraint $\mathcal{H}[\lambda]$ on a canonical coordinate $X$ as $\delta_{\mathcal{H}[\lambda]} X=\{X,\mathcal{H}[\lambda]\}$. We do not know which canonical constraints are first class functions without knowing the full constraint algebra. In the following we will intuitively use the term ``gauge transformation'' for the canonical transformations that we can identify with gauge transformation of the Lagrangian formulation. \\
In section \eqref{can-eft-gauge-transf-atk0} we briefly discuss how it can be computationally advantageous to consider the (gauge) transformations in the ``non-topological'' limit $\kappa=0$. Thereafter we analyse the canonical (gauge) transformations on a constraint by constraint basis. Throughout the following sections we make use of the smeared (or integrated) constraints in order to avoid writing derivatives of Dirac delta distributions.

\subsection{Gauge transformations at \texorpdfstring{$\kappa=0$}{k=0}}\label{can-eft-gauge-transf-atk0}

One of the main challenges in the canonical analysis of the E$_{6(6)}$ exceptional field theory is the complicated topological term $\mathcal{H}_{top}$ which inherits its complexity from the topological term in the Lagrangian \eqref{topological-term-E-six}. As a consequence of the existence of the topological term we have modified the canonical momenta of the one-forms, see section \ref{can-eft-momenta-oneform-sec}, in order to simplify the Hamiltonian. The modified momenta however do not Poisson-commute amongst themselves $\{\mathcal{P}^n_N(A),\mathcal{P}^m_M(A)\} \neq 0$. This situation is analogous to the case of the canonical formulation of five-dimensional E$_{6(6)}$ invariant supergravity \cite{KreutzerSUGRA}, but unfortunately the topological term in the E$_{6(6)}$ ExFT is much more complicated. To deal with this issue computationally we can nonetheless proceed in the same way. 
First we compute each expression at $\kappa=0$ and this result already contains much of the relevant information about the overall result because only the topological contribution is missing. We then want to proceed to compute the terms linear in the topological coefficient $\kappa$ to piece together the full result. The terms that are quadratic in $\kappa$ vanish because this only involves Poisson brackets of the one-forms and two-forms but no canonical momenta.\\

In some computations there is no difference between the result for $\kappa=0$ and the full result and in many cases the only difference is that the one-form momenta $\Pi^n_N(A)$ are replaced by the modified momenta $\mathcal{P}^n_N$. Some transformations are computationally very difficult to calculate in the case of $\kappa\neq0$. This is in particular the case for the transformations of $\mathcal{P}^n_N$ itself and for some of these transformations we only give the result at $\kappa=0$ because the computation of the remaining terms becomes too difficult. \\

What is completely missing at $\kappa=0$ is the topological dynamics of the B-field. We thus consider the case $\kappa=0$ only as a computational tool, because it removes one of the main difficulties of the canonical analysis and allows us to formulate partial results for some of the more difficult calculations. The limit $\kappa=0$ very likely does not correspond to any physically meaningful theory upon solution of the section condition. \\

If we do set $\kappa=0$ then by definition the modified one-form momenta reduce to the canonical momenta $\mathcal{P}^n_N = \Pi^n_N(A)$, the topological term vanishes $\mathcal{H}_{top}=0$ and so too does the $\partial_K\mathcal{H}_{klmN}$ term in $\mathcal{H}_\text{S1}$. The constraint $\mathcal{H}_\text{S1}$ does not completely vanish because the remaining $B_{tlM}$ term originates from the Stückelberg coupling terms in the one-form field strength $\mathcal{F}_{\mu\nu}^M$ in the Yang-Mills term. In the case $\kappa=0$ the canonical Hamiltonian becomes the much simpler expression $\mathcal{H}^{\,\kappa=0}_{\text{ExFT}}$ \eqref{can-eq:can-Hamiltonian-EFT-kappa0}.
\begin{align}\label{can-eq:can-Hamiltonian-EFT-kappa0}
    \mathcal{H}^{\,\kappa=0}_{\text{ExFT}}\,\, = &+ N \cdot \bigg{[} +\frac{1}{4 e} \Pi_{ab}(e)\,\,  \Pi_{ab}(e) - \frac{1}{12e} \Pi(e)^2 - e\,\,  \hat{R}  + e\, V_\text{HP} \\
                              & \hspace{1.3cm } +\frac{3}{2e} \Pi^{MN}(M) \,\, \Pi^{RS}(M)\,\,  M_{MR}\,\,  M_{NS}   - \frac{e}{24} g^{kl} \,\, \mathcal{D}_k M_{MN} \,\, \mathcal{D}_l M^{MN}      \nonumber \\
                              &\hspace{1.3cm }+\frac{e}{4} M_{MN}\,\,  g^{rm}\,\,  g^{sn} \,\, \mathcal{F}_{rs}^M\,\, \mathcal{F}_{mn}^N       +\frac{1}{2e} g_{lm}\,\,  M^{KL} \,\, \Pi^l_L(A)\,\,  \Pi^m_K(A) \bigg{]} \nonumber \\
                          & +N^n \cdot \bigg{[}   + 2\,\, \Pi^m{}_a(e)\,\, \mathcal{D}_{[n} e_{m]a} - e_{na}\,\, \mathcal{D}_m \Pi^m{}_a(e)     \nonumber \\
                              &\hspace{1.5cm } +\frac{1}{2} \Pi^{MN}(M)\,\, \mathcal{D}_n M_{MN}     \nonumber \\
                              & \hspace{1.5cm } + \mathcal{F}_{nl}^M\, \Pi^l_M(A) + \partial_M\left(g_{mn}\, M^{MN}\, \Pi^m_N(A) \right) \bigg{]}    \nonumber \\
                          & +A_t^M \cdot \bigg{[} -\mathcal{D}_l \Pi^l_M(A) -5 \,d^{NLS} \, d_{MNK}\, A^K_m \partial_S \Pi^m_L(A)  \nonumber\\
                           &\hspace{1.5cm} +\,\Pi^m{}_a(e)\, \partial_M e_{ma}   - \frac{1}{3} \partial_M \Pi(e) \nonumber \\
                          & \hspace{1.5cm}+ \frac{1}{2}\,\Pi^{KL}(M) \, \partial_M M_{KL} - 6\, \mathbb{P}^R{}_K{}^S{}_M \, \partial_S\left(\Pi^{KL}(M)\, M_{RL} \right)\bigg{]} \nonumber \\
                            & +B_{tlM} \cdot \bigg{[} +10 \,d^{MKL}\, \partial_K \Pi^l_L(A)  \bigg{]} \nonumber \\
                            &+ \dot{N} \cdot \Pi(N)  + \dot{N}_a\cdot \Pi^a(N_a) + \dot{A}_t^M \cdot\Pi_M(A_t) \nonumber
\end{align}

\subsection{Generalised exceptional diffeomorphisms}\label{can-eft-GD-sec}
When acting with the generalised diffeomorphism constraints $\mathcal{H}_\text{GD}$ \eqref{can-eft-GD-constr} on the canonical coordinates via the Poisson bracket we find that they (mainly) generate generalised exceptional diffeomorphisms in the form of the generalised Lie derivative, including the correct generalised weight terms. For the spatial vielbein, the scalar fields and their conjugate canonical momenta the generalised Lie derivative is the full result, which agrees with the Lagrangian formulation. This structure is already apparent in the constraints \eqref{can-eft-GD-constr}.  
    \begin{align}
         \{  e_{na} ,\mathcal{H}_{\text{GD}}[\zeta] \}  &= \mathbb{L}_\zeta e_{na}  \\
    \{ \Pi^n{}_a(e),\mathcal{H}_{\text{GD}}[\zeta]  \}  & =\mathbb{L}_\zeta \Pi^n{}_a(e) \\
      \{M_{MN} , \mathcal{H}_{\text{GD}}[\zeta]\}  & =\mathbb{L}_{\zeta} M_{MN} \\
    \{\Pi^{MN}(M) , \mathcal{H}_{\text{GD}}[\zeta]\}  &=  \mathbb{L}_\zeta \Pi^{MN}(M)
    \end{align}
The transformation of the differential forms is more complicated. For the one-forms the relevant part of the constraints are the momentum terms \eqref{can-eft-legendre-relation-gen-diff-gen2}. Equation \eqref{can-eft-GD-A-1} shows that $\mathcal{H}_\text{GD}$ canonically generates a generalised diffeomorphism, as represented by the generalised Lie derivative acting on the one-forms, but there are additional terms being generated.
\begin{align}
     \{ A^N_n, \mathcal{H}_{\text{GD}}[\zeta] \}    &= \mathbb{L}_{\zeta} A^N_n + \partial_n \zeta^N -5\,d^{NLR}\,\partial_L(d_{RMK}\,\zeta^M A^K_n) \label{can-eft-GD-A-1}\\
     &= \mathcal{D}_n\zeta^N +5\,d^{NLR}\,\partial_L(d_{RMK}\,\zeta^M A^K_n) \label{can-eft-GD-A-2}\\
      & = \mathcal{D}_n\zeta^N +\delta_{\mathcal{H}_{S1}[-\frac{1}{2}\,d_{\cdot MK}\,\zeta^M A^K_\cdot]}(A^N_n)  \label{can-eft-GD-A-3}
\end{align}
The second term in \eqref{can-eft-GD-A-1} is an abelian U$(1)^{27}$ gauge transformation which is the only part of the transformation that survives in the trivial solution of the section condition (cf. reference \cite{KreutzerSUGRA}). The last term in \eqref{can-eft-GD-A-1} originates from the extra momentum term in \eqref{can-eft-legendre-relation-gen-diff-gen2} which was found during the Legendre transformation. We can use the symmetrisation of the ExFT Dorfman bracket (see equation (2.19) of \cite{EFTI-E6}) to arrive at \eqref{can-eft-dorfmann-identity}.
\begin{equation}\label{can-eft-dorfmann-identity}
    \mathbb{L}_\zeta A^N_n = - \mathbb{L}_{A_n}\zeta^N +10\,d^{NLR}\,d_{RMK}\,\partial_L(\zeta^M A^K_n)
\end{equation}
Using the identity \eqref{can-eft-dorfmann-identity} and the definition of the covariant derivative we can rewrite \eqref{can-eft-GD-A-1} as \eqref{can-eft-GD-A-2}. The extra term of \eqref{can-eft-GD-A-1} does not cancel the term in \eqref{can-eft-dorfmann-identity} but instead the sign is switched. We will see that this extra term should be thought of as a tensor gauge transformation of the one-forms \eqref{can-eft-GD-A-3} coming from the $\mathcal{H}_{S1}$ constraints. Comparing \eqref{can-eft-GD-A-3} to the Lagrangian transformation \eqref{eft-cov-GD-A} we find that the expressions agree up to the tensor gauge transformation.\\

The transformation of the original canonical one-form momenta $\Pi^n_N(A)$ at $\kappa=0$ is given by equation \eqref{can-eft-GD-piA1}. We find the generalised Lie derivative term, but there is an extra term, which again originates from the additional term in \eqref{can-eft-legendre-relation-gen-diff-gen2}. Comparing the second term in \eqref{can-eft-GD-piA1} to the constraints at $\kappa=0$, see equation \eqref{can-eq:can-Hamiltonian-EFT-kappa0}, we can identify it as the $\kappa=0$ version of the secondary constraint $\mathcal{H}_\text{S1}$ and we arrive at \eqref{can-eft-GD-piA2}.
\begin{align}
     \{ \Pi^n_N(A), \mathcal{H}^{\kappa=0}_{\text{GD}}[\zeta] \}  &=\mathbb{L}_\zeta \Pi^n_N(A)-5\,d^{PKL}d_{NMP}\,\zeta^M\,\partial_K\Pi^n_L(A) \label{can-eft-GD-piA1}\\
     &=\mathbb{L}_\zeta \Pi^n_N(A)-\frac{1}{2}\,d_{NMP}\,\zeta^M\,(\mathcal{H}^{\kappa=0}_\text{S1})^{nP} \label{can-eft-GD-piA2}
\end{align}
The full calculation of the transformation of $\mathcal{P}^n_N$ is computationally exceedingly complicated because of the $\mathcal{P}^n_N$ terms in \eqref{can-eft-GD-constr} and because of the complexity of the topological term.
The full transformation should certainly contain the expression \eqref{can-eft-GD-PA3} where some of the topological contributions arrange into $\mathcal{P}^n_N(A)$ and into the full $\mathcal{H}_\text{S1}$ constraint, including the two-form field strength covariantisation terms. One may hope that the full transformation is indeed just given by equation \eqref{can-eft-GD-PA3} and that the remaining contributions cancel, however it is possible that this is too optimistic and that there are further more complicated transformations that need to be added to this transformation. Due to the complexity of the calculation the precise form of the transformation \eqref{can-eft-GD-PA3} remains to be determined. 
\begin{align}
      \{ \mathcal{P}^n_N(A),\mathcal{H}_{\text{GD}}[\zeta] \}  & \stackrel{?}{=} \mathbb{L}_\zeta \mathcal{P}^n_N(A)-\frac{1}{2}\,d_{NMP}\,\zeta^M\,(\mathcal{H}_\text{S1})^{nP} +\dots \label{can-eft-GD-PA3}
\end{align}
It may be interesting to observe the analogy between the role of the U$(1)^{27}$ one-form gauge transformations in the case of (external) diffeomorphisms in canonical ungauged maximal five-dimensional E$_{6(6)}$ invariant supergravity and the two-form gauge transformations, that we have seen above in the canonical (internal) generalised exceptional diffeomorphisms of ExFT. We can see this analogy by comparing the canonical transformations generated by the diffeomorphism constraint of five-dimensional supergravity on the one-forms and their momenta in \cite{KreutzerSUGRA} to the action of the generalised diffeomorphism constraint on the one-forms \eqref{can-eft-GD-A-3} and their momenta \eqref{can-eft-GD-piA2}. What we find is that where a U$(1)^{27}$ gauge transformation and the Gau\ss{} constraint (i.e. the constraint that generates the U$(1)^{27}$ transformation) appear for standard diffeomorphisms a tensor gauge transformation and the $\mathcal{H}_\text{S1}$ constraints appear respectively for the generalised diffeomorphisms of ExFT.\\

The Lagrangian kinetic term of the two-forms \eqref{can-eft-B-kinetic-langrangian} only has a single time derivative which causes this term to cancel in the Legendre transformation of section \ref{sec-can-legendre-2form-EFT}. Therefore the Hamiltonian cannot depend on the canonical momenta $\Pi^{klM}(B)$. The model Hamiltonian from section \ref{can-sec-twoforms-model} also illustrates this. In fact this absence of the canonical momenta in the Hamiltonian is a general feature of fields that are first order in the time derivative in the Lagrangian. As a direct consequence the two-forms do not transform under any of the secondary constraints which are all part of the Hamiltonian. The two-forms do however transform under the primary constraints $\mathcal{H}_\text{P2}$ \eqref{can-EFT-primary-canonical-constraint-B-nontrivial}. The constraints $\mathcal{H}_\text{P2}$ contain the two-form momenta by definition and relate the canonical momenta directly to the internal derivative of the two-forms themselves. From equation \eqref{can-eft-legendre-b-at-term-v1} we can see that we can use the primary constraints $\mathcal{H}_\text{P2}$ to insert the canonical momenta $\Pi^{klM}(B)$ into the B-field kinetic term that is part of $\mathcal{H}_{GD}$. We then find that \eqref{can-eft-legendre-b-at-term-v1} does indeed generate the generalised Lie derivative \eqref{can-eft-gauge-GD-B} if we apply the primary constraint $\mathcal{H}_\text{P2}=0$.
\begin{equation}
    \{ B_{mkZ}, \mathcal{H}_{\text{GD}}[\zeta]  \}  = \mathbb{L}_\zeta B_{mkZ} \,\,\,\text{ if $(\mathcal{H}_\text{P2})=0$ is used,$\,\,$ else 0} \label{can-eft-gauge-GD-B}
\end{equation}
The problem with \eqref{can-eft-gauge-GD-B} is that we are not allowed to make use of canonical constraints inside the Poisson bracket \cite{Henneaux-Teitelboim}. Furthermore we know from section \ref{can-sec-twoforms-model} that the two-form constraints are second class constraints and thus we should construct a Dirac bracket, which would then also allow us to apply the canonical constraints before evaluating the bracket. Nonetheless it seems likely that the result \eqref{can-eft-gauge-GD-B} would carry over to the Dirac bracket, possibly with further gauge transformation terms being generated.\\

The transformation of the two-form momenta $\Pi^{mkZ}(B)$ is relatively complicated and not easy to interpret. Because there are many two-form terms in the topological term of $\mathcal{H}_{GD}$ its Poisson bracket with the two-form momenta \eqref{can-eft-GD-PiB} has many topological contributions that do not have any obvious simplification. We know however from \eqref{can-eft-constr-missing-B-GD} that \eqref{can-eft-GD-PiB} with the parameter replaced by $\zeta^M =A_t^M$ should itself be a canonical constraint. This constraint is also analogous to the $\mathcal{H}_\text{S2}$ constraint from section \ref{can-sec-twoforms-model}, but without the proper Dirac bracket this constraint is difficult to interpret. 
\begin{align}\label{can-eft-GD-PiB}
      \{ \Pi^{mkZ}(B) ,\mathcal{H}_{\text{GD}}[\zeta]\}  &= -300\,\kappa\,\epsilon^{tmknr}\,d^{RNQ}\,d^{ZST}\,d_{MNT}\,\partial_S\left( \zeta^M\,\partial_R B_{nrQ} \right) \\
       &  -15\kappa\,\epsilon^{tmkls}\,d^{ZNR}\,d_{NKM}\,\partial_l\partial_R\left(\zeta^M\,A^K_s\right) \nonumber\\
       &  +30\kappa\,\epsilon^{tmklr}\,d^{ZNR}\,d_{NKT}\,\partial_R\left(\partial_L\zeta^T\,A^K_r\,A^L_l\right) \nonumber\\
        &  -10\kappa\,\epsilon^{tmklr}\,d^{ZNR}\,d_{NKT}\,\partial_R\left(\zeta^L\,A^K_r\,\partial_L A^T_l\right) \nonumber\\
         &  -150\kappa\,\epsilon^{tmklr}\,d^{ZNR}\,d_{NKT}\,d^{LTS}\,d_{PLQ}\,\partial_R\left(\partial_S\zeta^P\,A^K_r\,A^Q_l\right) \nonumber\\
            &  +50\kappa\,\epsilon^{tmklr}\,d^{ZNR}\,d_{NQL}\,d^{LKX}\,d_{MYX}\,\partial_R\left(\zeta^M\,A^Q_r\,\partial_K A^Y_l\right) \nonumber\\
     &  +5\kappa\,\epsilon^{tmklr}\,d^{ZNR}\,d_{NML}\,\partial_R\left(\zeta^M\,[A_l,A_r]^L_\text{E}\right) \nonumber
\end{align}
We do however find that the terms of equation \eqref{can-eft-legendre-b-at-term-v1} also generate the generalised Lie derivative of the two-form momenta $\mathbb{L}_{\zeta}\Pi^{mkZ}(B)$ if we were allowed to make use of the primary constraints $\mathcal{H}_\text{P2}$. Nonetheless we would see additional terms due to the topological (non-kinetic) two-form terms in the modified one-form momenta and $\mathcal{H}_\text{top}$.
    
\subsection{External diffeomorphisms}
\label{can-sec-External-diffeomorphisms}
Acting with the canonical external diffeomorphism constraints $\mathcal{H}_\text{Diff}$ \eqref{can-eft-diff-constr} on the canonical coordinates we find that they (mainly) generate covariantised external diffeomorphisms.\\

When acting with $\mathcal{H}_\text{Diff}$ on the spatial vielbein \eqref{can-eft-diff-e} and the scalar fields \eqref{can-eft-diff-M} we find that the resulting gauge transformations are the covariantised versions of standard diffeomorphisms, where all derivatives are replaced by covariant derivatives. These transformations are identical to the spatial part of the transformations of the Lagrangian gauge transformations \eqref{eft-diff-e} and \eqref{eft-diff-M}. The transformations of the canonical momenta of the vielbein \eqref{can-eft-diff-Pie} and the scalars \eqref{can-eft-diff-PiM} are also given by the $\mathcal{D}$-covariantised standard diffeomorphisms (with appropriate external diffeomorphism weight cf. reference \cite{KreutzerSUGRA}), however there are additional $\partial_N\xi^n$ terms.
\begin{align}
     \{  e_n{}^a,\mathcal{H}_\text{Diff}[\xi] \}= &+ \xi^l \, \mathcal{D}_l e_n{}^a + \mathcal{D}_n \xi^l \, e_l{}^a \label{can-eft-diff-e}\\
   \{  \Pi^n{}_a(e), \mathcal{H}_\text{Diff}[\xi] \}= &+  \xi^l \, \mathcal{D}_l   \Pi^n{}_a(e) - \mathcal{D}_l \xi^n \,   \Pi^l{}_a(e)  + \mathcal{D}_l \xi^l \,   \Pi^n{}_a(e) \label{can-eft-diff-Pie} \\
   &+2 \,e_{la}\,M^{MN}\,\partial_M\xi^{(l}\,\mathcal{P}^{n)}_N(A)  \nonumber\\
    \{M_{MN}, \mathcal{H}_\text{Diff}[\xi] \}= &  +\xi^n \,\mathcal{D}_n M_{MN} \label{can-eft-diff-M}\\
    \{\Pi^{MN}(M),  \mathcal{H}_\text{Diff}[\xi] \}= &+\xi^n \mathcal{D}_n \Pi^{MN}(M) + \mathcal{D}_n\xi^n \Pi^{MN}(M)\label{can-eft-diff-PiM} \\
    &- 2\,\partial_K \xi^m \,g_{mn}\,\mathcal{P}^n_L(A)\, M^{K(M} M^{N)L} \nonumber
\end{align}    
These additional terms originate from the $-g_{lm}\, M^{KL} \mathcal{P}_L^l(A) \, \partial_K N^m$ term in the Hamiltonian, which we identified in section \ref{sec-can-EFT-legendre-1forms} as coming from the interplay of the Einstein-Hilbert improvement and the Yang-Mills terms. This term depends on the spatial components of every field and thus leads to contributions in the transformations of all canonical momenta. Due to the dependence on the modified momenta it also contributes to the transformation of the one-forms. We find that the one-forms transform as \eqref{can-eft-diff-A}.
\begin{equation}\label{can-eft-diff-A}
      \{ A^N_n, \mathcal{H}_\text{Diff}[\xi] \}= +\xi^m\,\, \mathcal{F}^N_{mn} - g_{mn} \, M^{MN}\,\partial_M \xi^m
\end{equation}
The first term in the transformation \eqref{can-eft-diff-A} is the covariantised version of the standard diffeomorphism that one finds in five-dimensional supergravity \cite{KreutzerSUGRA}. This term is also the spatial version of the first term in the Lagrangian transformation \eqref{eft-diff-A}. The second term in \eqref{can-eft-diff-A} is identical to the spatial component of the $\partial_M\xi^m$ term found in \eqref{eft-diff-A}, but surprisingly it appears here with the opposite sign. We will come back to this fact at the end of this section where we discuss the origin of the $-g_{lm}\, M^{KL} \mathcal{P}_L^l(A) \, \partial_K N^m$ term in detail.\\

If we restrict the coordinate dependence of the gauge parameter of the external diffeomorphisms to only the external coordinates $\xi^n\left(x^\mu,Y^M\right)=\xi^n(x^\mu)$, equivalently $\partial_N\xi^n =0$, then the transformations generated by the canonical constraints $\mathcal{H}_\text{Diff}[\xi]$, that we have seen so far, are given exactly by the $\mathcal{D}$-covariantised version of standard diffeomorphisms. The individual terms of the E$_{6(6)}$ ExFT Lagrangian \eqref{eft-lagrangian-terms} are manifestly invariant under the action of external diffeomorphisms with parameter $\xi^n(x^\mu)$. The diffeomorphisms with parameter $\xi^n\left(x^\mu,Y^M\right)$ are not a manifest symmetry of \eqref{eft-lagrangian-terms} and instead connect different terms in the Lagrangian, thereby leading to the unique E$_{6(6)}$ ExFT action (up to the overall scaling) \cite{Hohm:2013pua,EFTI-E6}. 
It seems that the $\partial_M\xi^m$ terms in \eqref{can-eft-diff-Pie}, \eqref{can-eft-diff-PiM} and \eqref{can-eft-diff-A} serve a similar function. Because these terms depend on many different variables the transformations mix various fields and momenta. We may speculate that the canonical constraint algebra relations concerning the external diffeomorphism constraints depend on many cancellations of such mixing terms and that these cancellations depend on the precise coefficients of all terms in the Hamiltonian.\\

Due to the one-form dependent covariant derivatives in the diffeomorphism constraints \eqref{can-eft-diff-constr} the transformation of the conjugated original one-form momenta $\Pi^n_N(A)$ is more complicated than the above transformations. 
For the transformation of $\Pi^n_N(A)$ under the $\kappa=0$ version of the diffeomorphism constraints we find that the transformation is given by \eqref{can-EFT-k=0-ext-diff-gauge-transf-PiA}. The first line of \eqref{can-EFT-k=0-ext-diff-gauge-transf-PiA} is the covariantised standard Lie derivative of $\Pi^n_N(A)$. We can compare \eqref{can-EFT-k=0-ext-diff-gauge-transf-PiA} to the analogous transformation in five-dimensional supergravity \cite{KreutzerSUGRA} and find that the $\mathcal{H}_\text{GD}$ constraint term in \eqref{can-EFT-k=0-ext-diff-gauge-transf-PiA} is an extension of the U(1)$^{27}$ Gau\ss{} constraint term in the transformation of \cite{KreutzerSUGRA}. The constraints $\mathcal{H}_\text{S1}$ consist only of the $\Pi^n_N(A)$ term in \eqref{can-eq:can-Hamiltonian-EFT-kappa0} for the $\kappa=0$ case. This term and also the $\Pi^n_N(A)$ terms in $\mathcal{H}_\text{GD}$ originate from the transformation of the covariantised one-form field strength $\mathcal{F}^M_{mn}$ in the diffeomorphism constraint $\mathcal{H}_\text{Diff}$.
The remaining terms in \eqref{can-EFT-k=0-ext-diff-gauge-transf-PiA} are all of the $\partial_N\xi^n$ form.
In contrast to the transformations of the fields above the $\partial_N\xi^n$ terms in \eqref{can-EFT-k=0-ext-diff-gauge-transf-PiA} exist due to the covariant derivatives in $\mathcal{H}_\text{Diff}$ and in the case $\kappa=0$ the $-g_{lm}\, M^{KL} \Pi_L^l(A) \, \partial_K N^m$ term is irrelevant to the transformation of $\Pi^n_N(A)$. 
\begin{align}
      \{\Pi^n_N(A), \mathcal{H}^{\kappa=0}_\text{Diff}[\xi]  \}= & +\xi^k \,\mathcal{D}_k \Pi^n_N(A)- \mathcal{D}_k\xi^n \,\Pi^k_N(A)+\mathcal{D}_k\xi^k \, \Pi^n_N(A) \label{can-EFT-k=0-ext-diff-gauge-transf-PiA} \\
      & +\xi^n \left( \mathcal{H}^{\kappa=0}_\text{GD} \right)_N +\frac{1}{2}\,d_{NKL}\,\xi^k\,A^K_k (\mathcal{H}^{\kappa=0}_{S1})^{nL}   \nonumber\\
      & +\partial_N\xi^k \,\Pi^n{}_a(e)\,e_{k}{}^a -\frac{1}{3} \,\partial_N\xi^n\,\Pi^k{}_a(e)\,e_{k}{}^a\nonumber\\
      &-\frac{1}{3}\,\partial_N\xi^n\,\Pi^{KL}(M)\,M_{KL} -\partial_K\xi^n\,\Pi^{KL}(M)\,M_{LN}\nonumber\\
      &+10 \,d^{RSM}\,d_{KNR}\,\partial_S \xi^n \,\Pi^{KL}(M)\,M_{LM}\nonumber\\
      &+ 10\,d^{RSM}\,d_{KNR}\,\partial_S\xi^{[m}\,\Pi^{n]}_M(A)\,A^K_m\nonumber
\end{align}
For the calculation of the transformation of the modified momenta $\mathcal{P}^n_N(A)$ we need to remember the Poisson non-commutativity of this variable with itself which generates a large number of additional terms. We find that the transformation of $\mathcal{P}^n_N(A)$ can be expressed as \eqref{can-EFT-ext-diff-gauge-transf-P}. Most of the additional topological contributions coming from the two $\mathcal{P}^n_N(A)$ terms in $\mathcal{H}_\text{Diff}$ go into the complicated topological term \eqref{can-eft-Htop} that is inside $\mathcal{H}_\text{GD}$ and into $\mathcal{H}_\text{S1}$. To arrive at \eqref{can-EFT-ext-diff-gauge-transf-P} the Schouten identity (cf. appendix C of \cite{KreutzerSUGRA}) has to be applied many times in order to move the correct index to the gauge parameter. In order for the purely two-form dependent terms in the transformation \eqref{can-EFT-ext-diff-gauge-transf-P} to match the two-form kinetic term of $\mathcal{H}_\text{GD}$ we have to make use of the rewriting \eqref{can-eft-legendre-b-at-term-v2} which implies that the section condition has to be used in this calculation. This seems to be the only occurrence of the section condition before considering the canonical constraint algebra. This is further discussed in section \ref{section-conclusios}. Furthermore there are many topological $\partial_N\xi^n$ terms in \eqref{can-EFT-ext-diff-gauge-transf-P} for which there does not seem to be a simpler form.
Finally we are left with a rather large number of topological contributions that we have written in \eqref{can-EFT-ext-diff-gauge-transf-P} as $\kappa\,\Gamma(A,B)_N \, \xi^n$. Most of the terms hidden inside $\Gamma(A,B)$ only depend on the one-forms, but some also depend on the two-forms. There are no pure two-form terms in $\Gamma(A,B)$. Note that these terms are not of the $\partial_N\xi^n$ form and therefore there has to be another interpretation for them or else they may arrange to cancel in some non-trivial way to yield $\Gamma(A,B)=0$.
\begin{align}
      \{\mathcal{P}^n_N(A), \mathcal{H}_\text{Diff}[\xi]  \}= & +\xi^k \,\mathcal{D}_k \mathcal{P}^n_N(A)- \mathcal{D}_k\xi^n \,\mathcal{P}^k_N(A)+\mathcal{D}_k\xi^k \, \mathcal{P}^n_N(A) \label{can-EFT-ext-diff-gauge-transf-P}\\
      & +\xi^n \left( \mathcal{H}_\text{GD} \right)_N +\frac{1}{2}\,d_{NKL}\,\xi^k\,A^K_k (\mathcal{H}_{S1})^{nL}  \nonumber\\
      & +\partial_N\xi^k \,\Pi^n{}_a(e)\,e_{k}{}^a -\frac{1}{3} \,\partial_N\xi^n\,\Pi^k{}_a(e)\,e_{k}{}^a\nonumber\\
      &-\frac{1}{3}\,\partial_N\xi^n\,\Pi^{KL}(M)\,M_{KL} -\partial_K\xi^n\,\Pi^{KL}(M)\,M_{LN}\nonumber\\
      &+10 \,d^{RSM}\,d_{KNR}\,\partial_S \xi^n \,\Pi^{KL}(M)\,M_{LM}\nonumber\\
      &+ 10\,d^{RSM}\,d_{KNR}\,\partial_S\xi^{[m}\,\mathcal{P}^{n]}_M(A)\,A^K_m\nonumber\\
          &  -\frac{9}{2}\,\kappa\,\epsilon^{lmqn}\, d_{NQT}\,\partial_K A^Q_m\,A^K_q\,\partial_W\xi^k\, g_{lk} \,M^{TW}  \nonumber\\
          &  -\frac{3}{2}\,\kappa\,\epsilon^{lmqn}\, d_{NQT}\,\partial_K A^K_q\,A^Q_m\,\partial_W\xi^k\, g_{lk} \,M^{TW}  \nonumber\\
           &  +3 \,\kappa\,\epsilon^{lmnr}\, d_{MQT}\,\partial_N A^M_r\,A^Q_m\,\partial_W\xi^k\, g_{lk} \,M^{TW}  \nonumber\\
            &  +\frac{30}{4}\,\kappa\,\epsilon^{lmqn}\, d_{MQT}\,d^{MRK}\,d_{RSN}\,\partial_K A^Q_m\,A^S_q\,\partial_W\xi^k\, g_{lk} \,M^{TW}  \nonumber\\
            &  +\frac{45}{2}\,\kappa\,\epsilon^{lmqn}\, d_{MQT}\,d^{MRK}\,d_{RSN}\,\partial_K A^S_q\,A^Q_m\,\partial_W\xi^k\, g_{lk} \,M^{TW}  \nonumber\\
              &  -15\,\kappa\,\epsilon^{lnqr}\, d_{MNT}\,d^{MRK}\,d_{RSL}\,\partial_K A^L_r\,A^S_q\,\partial_W\xi^k\, g_{lk} \,M^{TW}  \nonumber\\
               &  +6\,\kappa\,\epsilon^{lmqn}\, d_{MNT}\,\partial_q A^M_m\,\partial_W\xi^k\, g_{lk} \,M^{TW}  \nonumber\\
                & -30\,\kappa\,\epsilon^{lmqn}\, d^{MQR}\,d_{QNT}\,\partial_R B_{mqM} \,\partial_W\xi^k\, g_{lk} \,M^{TW}  \nonumber\\
      & + \kappa\,\Gamma(A,B)_N\,\xi^n \nonumber
\end{align} 
In section \ref{can-eft-GD-sec} we argued that because the B-field kinetic term of the Lagrangian only has one time derivative the secondary constraints do not contain any two-form momenta and therefore the two-forms do not transform under any of the secondary constraints. In the case of the generalised diffeomorphism constraint we found that we could make use of the primary constraints $\mathcal{H}_\text{P2}$ to rewrite the B-field kinetic term in a way that would make the B-fields transform properly by inserting the canonical momenta $\Pi^{klM}(B)$. The same procedure does not work for the diffeomorphism constraint $\mathcal{H}_\text{Diff}$ however because here the two-forms only appear inside $\mathcal{F}^M_{mn}$ and $\mathcal{P}^n_N(A)$. In the canonical formalism it is therefore not possible to make the topological two-forms transform under external diffeomorphisms --- even if we use the primary constraints. Furthermore this is consistent with what we found in the model theory of the topological two-forms in section \ref{can-sec-twoforms-model}. In the canonical formalism this behaviour should be expected for any field whose kinetic terms are located inside a topological term --- such as \eqref{topological-term-E-six} --- which by definition does not depend on the metric $G_{\mu\nu}$ and as a consequence thereof not on the shift vector $N^n$. The two-forms therefore do not ``see'' the external diffeomorphisms and canonically they transform trivially as \eqref{can-eft-diff-B}.
\begin{equation}
    \{ B_{klM}, \mathcal{H}_\text{Diff}[\xi] \}=  0 \label{can-eft-diff-B}
\end{equation}
By the same argument we do not find any Lie derivative terms in the transformation of the conjugated two-form momenta $\Pi^{pvS}(B)$. Due to the two-form dependent $\mathcal{F}^M_{mn}$ and $\mathcal{P}^n_N(A)$ terms in $\mathcal{H}_\text{Diff}$ we do nonetheless get a non-vanish transformation \eqref{can-eft-diff-PiB} for the momenta $\Pi^{pvS}(B)$. The vanishing of the transformation \eqref{can-eft-diff-PiB} with the replacement $\xi^n=N^n$ is then equivalent to the consistency condition \eqref{can-eft-constr-missing-B-Diff}. We will come back to this when discussing the tensor gauge transformations in section \ref{can-eft-tensor-gauge-transf-sec}.
\begin{align}
    \{\Pi^{pvS}(B), \mathcal{H}_\text{Diff}[\xi]  \}= &  +20 \, d^{TKS}\,\partial_K\left(\xi^{[p}\,\mathcal{P}^{v]}_T(A)\right) \label{can-eft-diff-PiB} \\
                                      & +30\kappa\, \epsilon^{lpvr}d^{SNR}d_{NKT}\,\partial_R\left(\partial_L\xi^k\,g_{kl}\,M^{LT}\,A^K_r\right) \nonumber\\
                                      &  -30\kappa\, \epsilon^{lpvr}d^{SNR}d_{NKT}\partial_R\left(\xi^k\,\mathcal{F}^T_{kl}\,A^K_r\right) \nonumber
\end{align}

\subsubsection*{The origin and sign of the \texorpdfstring{$-g_{lm}\, M^{KL} \mathcal{P}_L^l(A) \, \partial_K N^m$}{-gMPdN} term in the Hamiltonian}
The transformation of the one-forms \eqref{can-eft-diff-A} is the only diffeomorphism transformation with a $\partial_N\xi^n$ term that we can compare to an analogous Lagrangian gauge transformation \eqref{eft-diff-A}. What we have found is that the sign of the $\partial_N\xi^n$ term is different when compared to the Lagrangian transformation. The minus sign in \eqref{can-eft-diff-A} is a direct consequence of the sign of the term $-g_{lm}\, M^{KL} \mathcal{P}_L^l(A) \, \partial_K N^m$ in the Hamiltonian. In this paragraph we want to explain the origin of this term in the Hamiltonian.\\

In the ADM decomposition of the Einstein-Hilbert improvement term \eqref{can-eft-einstein-improve-adm} we find the term $+\frac{e}{N}\,\mathcal{F}^M_{tn}   \,\partial_M N^n$. This term comes from taking the {$\mu=t, \, \nu=n,$}{$ \, \alpha=0, \, \beta=b,$}{$ \, \rho=\{t,r\}$} parts of the five-dimensional indices and using the identity $\partial_M(e_b{}^n\,e_r{}^b)=0$ to unify the {$ \, \rho=\{t,r\}$} contributions. When computing the canonical momenta $\Pi^l_T(A)$ \eqref{can-EFT-original-canonical-momentum-A} the term $+\frac{e}{N}\,\mathcal{F}^M_{tn}   \,\partial_M N^n$ in the Lagrangian leads to the contribution $+ \frac{e}{N} \partial_T N^l $ to $\Pi^l_T(A)$ and consequently to $\mathcal{P}^l_T(A)$. Because the term $+\frac{e}{N}\,\mathcal{F}^M_{tn}   \,\partial_M N^n$ is linear in time derivatives it cancels in the Legendre transformation in the step from \eqref{can-eft-legendre-a-2} to \eqref{can-eft-legendre-a-3} against the Legendre transformation term of the one-forms $\dot A_{n}^N\cdot \Pi_N^n(A_n^N)$. And in equation \eqref{can-eft-legendre-a-3} the only term with time derivatives in the Legendre transformation is the quadratic Yang-Mills term which contains in particular the term $  + \frac{e}{2\,N}\,g^{sn}  \, M_{MN}\, \dot{A}^M_s  \,\dot{A}^N_n$. From this $  + \frac{e}{2\,N}\,g^{sn}  \, M_{MN}\, \dot{A}^M_s  \,\dot{A}^N_n$ term in \eqref{can-eft-legendre-a-3} we get many other terms by inserting the expression \eqref{can-eft-legendre-a-5-P-inverse} for $\dot{A}^N_n $. The terms where one $\dot{A} \sim \mathcal{P}(A)$ and the other $\dot{A} \sim -\frac{e}{N} \partial_T N^l$ then lead to the term $-g_{lm}\, M^{KL} \mathcal{P}_L^l(A) \, \partial_K N^m$ in the Hamiltonian. The minus sign of this term originates in the inversion of the momenta for $\dot{A}(\mathcal{P})$ \eqref{can-eft-legendre-a-5-P-inverse}.\\
When we then act with the term $-g_{lm}\, M^{KL} \mathcal{P}_L^l(A) \, \partial_K N^m$ on the one-forms, as we do in \eqref{can-eft-diff-A}, we immediately arrive at the transformation \eqref{can-eft-diff-A-v2}.
\begin{equation}\label{can-eft-diff-A-v2}
      \{ A^N_n, -g_{lm}\, M^{ML} \,\mathcal{P}_L^l(A) \, \partial_M N^m \}=  - g_{mn} \, M^{MN}\,\partial_M N^m
\end{equation}
In section \ref{sec-eft-e6-lagrangian} we have explained the origin of the analogous term in the Lagrangian gauge transformation \eqref{eft-diff-A} as coming from a compensating Lorentz transformation (cf. references \cite{EFTI-E6,Berman-review-eft}) when considering the Kaluza-Klein-like rewriting of eleven-dimensional supergravity. In the derivation of this sign from the gauge transformations of eleven-dimensional supergravity there does not seem to be any choice for changing this sign.\\
Another possible origin of this difference in sign could be a diverging convention, however the conventions chosen in this work, in particular the signature of the Minkowski metric, seem to agree with the conventions used in \cite{Hohm:2013pua,EFTI-E6}. \\
When viewed purely from the canonical perspective the sign in \eqref{can-eft-diff-A} is not immediately problematic, however it remains to be checked if the sign might affect the closure of the canonical constraint algebra.\\
Considering the above factors we do not have an explanation for the difference in sign of the $\partial_N\xi^n$ term in \eqref{can-eft-diff-A} when compared to the Lagrangian formulation.

\subsection{Time evolution}\label{can-eft-sec-hamilton-time-evolution}
The Hamilton constraint $\mathcal{H}_\text{Ham}$ \eqref{can-eft-Hamilton-constr} acting on the canonical coordinates generates time evolution. 
The time evolution of the spatial vielbein \eqref{can-eft-ham-e}, the scalar fields \eqref{can-eft-ham-M} and the one-forms \eqref{can-eft-ham-A} generated by the Hamilton constraint are in form identical to the analogous transformations in five-dimensional E$_{6(6)}$ invariant supergravity \cite{KreutzerSUGRA}. These results are expected because the canonical momenta terms in the Hamilton constraint are of the same form as in the five-dimensional E$_{6(6)}$ invariant supergravity Hamiltonian. We should note that the time evolution of the one-forms \eqref{can-eft-ham-A} is most concisely written in terms of the modified momenta $\mathcal{P}_N^n(A)$ which may be seen as another argument in favour of the modified momenta.
\begin{align}
   \{  e_{na} ,\mathcal{H}_\text{Ham}[\phi] \} = & +\frac{\phi}{2e} \,\, g_{mn} \,\,\Pi^m{}_a(e) - \frac{\phi}{6e}\,\,\Pi(e)\,\,e_{na} \label{can-eft-ham-e} \\
    \{ M_{MN},  \mathcal{H}_\text{Ham}[\phi]  \} = & +\frac{6}{e} \phi\,\, \Pi^{QP}(x)\,\, M_{MQ}\,\,M_{NP} \label{can-eft-ham-M} \\
      \{ A^N_n, \mathcal{H}_\text{Ham}[\phi]  \} = & + \frac{\phi}{e} \,\, g_{nl} \,\, M^{NL} \,\, \mathcal{P}^l_L(A)  \label{can-eft-ham-A}
\end{align}
While the time evolution of the fields in E$_{6(6)}$ ExFT agrees in form with that of five-dimensional E$_{6(6)}$ invariant supergravity \cite{KreutzerSUGRA} the time evolution of the canonical momenta is significantly more complicated in ExFT. Part of the complexity is expected because the analogous transformations are already relatively complicated in five-dimensional supergravity but the scalar potential and covariant derivatives lead to many additional terms.\\

The canonical time evolution of the vielbein momenta $\Pi^n{}_a(e)$ is in five-dimensional E$_{6(6)}$ invariant supergravity \cite{KreutzerSUGRA} given by the spatial Einstein equation in vielbein form and contributions from all other terms of the Hamilton constraint because all fields couple to the metric. In ExFT we find the analogous covariantised time evolution \eqref{eqn:pi-of-e-time-evolution} which contains a number of additional terms, in particular due to the scalar potential \eqref{can-Hamiltonian-eq:EFT-potential} which does not exist in five-dimensional E$_{6(6)}$ invariant supergravity. The covariant derivative $\nabla_n$ contains only the Levi-Civita connection.
\begin{align}
      \{ \Pi^n{}_a(e), \mathcal{H}_\text{Ham}[\phi]  \} = & + \frac{\phi}{4e}\,\,\Pi_{bc}(e)\,\,\Pi_{bc}(e)\,\, e_a{}^n -\frac{\phi}{2e}\Pi^k{}_b(e)\,\,\Pi^n{}_b(e)\,\,e_{ka} \label{eqn:pi-of-e-time-evolution}\\
   & -\frac{\phi}{12e}\,\,\Pi^2(e)\,\,e_a{}^n  + \frac{\phi}{6e}\,\,\Pi(e)\,\,\Pi^n{}_a(e) \nonumber\\
    &-2\phi e \left( \hat{R}^{nk} \,\,e_{ka} - \frac{1}{2}\hat{R}\,\,e_a{}^n\right) \nonumber\\
    &  +2e \left( \nabla_a \nabla^n \phi - \nabla^k \nabla_k \phi  \,\, e_a{}^n \right)   \nonumber \\
     & +\frac{3\phi}{2e} \Pi^{MN}(M)\,\,\Pi^{RS}(M)\,\,M_{MR}\,\,M_{NS}\,\,e_a{}^n  \nonumber\\
      &  + \frac{\phi e}{24}\partial_k M_{MN}\,\,\partial_l M^{MN} \,\,g^{kl}\,\,e_a{}^n -   \frac{\phi e}{12} \partial_k M_{MN}\,\,\partial_l M^{MN}\,\,g^{ln}\,\,e_a{}^k  \nonumber\\
       &- \frac{\phi e}{4} M_{MN}\,\,\mathcal{F}^M_{rs}\,\,\mathcal{F}^N_{km}\,\,g^{rk}\,\,g^{sm}\,\,e_a{}^n +    \phi e \,\,M_{MN}\,\,\mathcal{F}^M_{rs}\,\,\mathcal{F}^N_{km}\,\,g^{rk}\,\,g^{mn}\,\,e_a{}^s  \nonumber\\
        &+ \frac{\phi}{2e} M^{KL}\,\,\mathcal{P}^l_L(A)\,\,\mathcal{P}^k_K(A)\,\,g_{lk}\,\,e_a{}^n -     \frac{\phi}{e}M^{KL}\,\,\mathcal{P}^n_K(A)\,\,\mathcal{P}^l_L(A)\,\,e_{la}  \nonumber\\
    & +\frac{e\phi}{4}M^{MN}\,\partial_Mg^{kl}\,\partial_Ng_{kl}\,e_a{}^n +\partial_M\left(\frac{e\phi}{2} M^{MN}\,\partial_N g_{kl} \right)g^{kn}\,e_a{}^l \nonumber\\
    &  -\partial_N\left(\frac{e\phi}{2} M^{MN} \partial_Mg^{mn}\right)e_{ma} \nonumber\\
    &  -\frac{\phi}{e}\,M^{MN}\,\partial_M e\, \partial_N e\,e_a{}^n-2\, \partial_N\left(\frac{\phi}{e} M^{MN}\,\partial_M e \right)\,e\,e_a{}^n \nonumber\\
    & +\frac{e\,\phi}{24}M^{MN}\partial_MM^{KL}\partial_NM_{KL}\, e_a{}^n  -\frac{e\,\phi}{2}M^{MN}\partial_MM^{KL}\partial_LM_{NK}\, e_a{}^n \nonumber\\
    &  -e\,\phi\,\partial_M\partial_N M^{MN}\, e_a{}^n  -2e\,\partial_M\partial_N\phi\,M^{MN}\, e_a{}^n -2e\,\partial_N\phi\,\partial_MM^{MN}\, e_a{}^n\nonumber\\
    & -2\,e\,\phi\,\mathcal{F}^M_{mk}\,g^{mn}\,\partial_Me^k_a - \partial_M\left(e\,\phi\,\mathcal{F}^M_{mk}\,g^{mn}\,e^k_a\right) \nonumber\\
    &  -e\,\phi\,\mathcal{F}^M_{mk}\,\partial_Me^b_r \,\left( e^n_b\,e^k_a\,g^{mr}+e^k_b\,e^m_a\,g^{rn}+e^k_b\,e^r_a\,g^{mn}\right) \nonumber 
\end{align}
The canonical time evolution of the scalar momenta $\Pi^{KL}(M)$ in five-dimensional E$_{6(6)}$ invariant supergravity \cite{KreutzerSUGRA} is relatively simple. It consists of contributions from the scalar kinetic terms and is only slightly complicated by the fact that the scalar fields $M_{MN}$ are used as a ``generalised metric'' to contract the E$_{6(6)}$ indices in the one-form kinetic terms. In ExFT we find the corresponding expressions, but due to the covariant derivatives in the scalar kinetic term and in particular due to the scalar potential \eqref{can-Hamiltonian-eq:EFT-potential} the full ExFT result \eqref{can-eft-ham-PiM} is much more complicated. There may exist a slightly simpler and more covariant form to write the additional internal derivative terms in \eqref{can-eft-ham-PiM} coming from the scalar kinetic term. For the scalar potential contributions to \eqref{can-eft-ham-PiM} there seems to be little hope of significant simplification.
\begin{align}
  &\{\Pi^{KL}(M) , \mathcal{H}_\text{Ham}[\phi] \} =  - \frac{6 \phi }{e} \,\,  \Pi^{PK}(M)\,\, \Pi^{LR}(M)\,\,M_{PR}  \label{can-eft-ham-PiM} \\
  &- \partial_l\left( \frac{\phi e}{6}\,\,g^{kl}\,\,\partial_kM^{KL}\right)- \frac{\phi e}{6} \,\, g^{kl} \,\,\partial_kM_{MN}\,\,\partial_lM^{KM}\,\,M^{LN}   \nonumber \\
   & - \frac{\phi e}{2} \,\, g^{rm}\,\, g^{sn}\,\,\mathcal{F}^{K}_{rs}\,\,\mathcal{F}^{L}_{mn} + \frac{\phi}{e}  g_{lm} \,\, \mathcal{P}^l_M(A) \,\,\mathcal{P}^m_N(A) \,\, M^{KM}\,\, M^{LN} \nonumber \\
   &  -\partial_n\left( \frac{e\,\phi}{12} \,g^{mn}\,\mathbb{L}_{A_m}M_{MN}\right)\,M^{KM}\,M^{LN}+ \partial_n\left( \frac{e\,\phi}{12}\,g^{mn}\,\mathbb{L}_{A_m}M^{KL}\right) \nonumber\\
    & +\partial_R\left( \frac{e\,\phi}{12}\,g^{mn}\,\mathcal{D}_nM^{KL}\,A^R_m  \right) - \frac{e\,\phi}{18}g^{mn}\,\mathcal{D}_nM^{KL}\,\partial_R A^R_m  \nonumber\\
    &   -\frac{e\,\phi}{6}g^{mn}\,\mathcal{D}_nM^{M(K}\partial_M A^{L)}_m + \frac{5\,e\,\phi}{3}g^{mn}\,\mathcal{D}_nM^{M(K}\,d^{L)TX}\,d_{UXM}\,\partial_T A^U_m \nonumber\\
    &  -\partial_R \left( \frac{e\,\phi}{12} g^{mn} \,\mathcal{D}_n M_{MN} \, A^R_m \right) \, M^{KM}\,M^{LN} -\frac{e\,\phi}{18} g^{mn}\,\mathcal{D}_n M_{MN}\,\partial_R A^R_m \, M^{KM} \, M^{LN}  \nonumber\\
    &  -\frac{e\,\phi}{6} g^{mn} \,\mathcal{D}_n M_{MN} \, \partial_R A^M_m\, M^{N(K}\,M^{L)R} + \frac{5\,e\,\phi}{3} g^{mn}\,\mathcal{D}_n M_{MN} \, d^{XTM}\,d_{RUX}\, \partial_T A^U_m \, M^{N(K} \, M^{L)R}  \nonumber\\
    &  -\frac{e\,\phi}{2} \partial_M g^{mn}\,\partial_N g_{mn}\,M^{M(K}\,M^{L)N} - \frac{2\phi}{e} \partial_M e \, \partial_N e\, M^{KM} \, M^{LN} \nonumber\\
    &  +4 \phi\, \partial_M\partial_N e \, M^{KM}\,M^{LN} + 2 \partial_N\partial_M(e\phi) \, M^{KM} \, M^{LN}  -4 \partial_M(\phi \partial_N e) \, M^{M(K}\,M^{L)N} \nonumber\\
    & -\frac{e\,\phi}{12} \partial_M M^{RS}\,\partial_N M_{RS}\,M^{M(K}\,M^{L)N} +\partial_M \left( \frac{e\,\phi}{12} M^{MN}\,\partial_N M_{RS} \right) \, M^{KR}\,M^{LS}   \nonumber\\
    &  - \partial_N\left( \frac{e\,\phi}{12}\partial_M M^{KL}\, M^{MN} \right) + \partial_S \left( e\,\phi\, M^{M(K}\,\partial_M M^{L)S} \right) \nonumber\\
    &  +e\,\phi\, \partial_M M^{RS}\,\partial_S M_{NR}\, M^{M(K}\,M^{L)N} - \partial_M \left( e\,\phi\, M^{MN}\, \partial_S M_{NR} \right) \, M^{R(K}\, M^{L)S} \nonumber
   \end{align}
The canonical time evolution of the modified one-form momenta $ P^k_K(A)$ in ungauged five-dimensional E$_{6(6)}$ invariant supergravity \cite{KreutzerSUGRA} is quite simple and consists of only two terms. One contribution from the abelian Maxwell-like term and one topological contribution from the Poisson non-commutativity of the modified momenta $P^k_K(A)$ --- which are much simpler in \cite{KreutzerSUGRA} due to the simpler topological term. Because the one-forms are used in ExFT to gauge the generalised diffeomorphism symmetry the canonical time evolution of their conjugate modified momenta $\mathcal{P}^k_K(A)$ is very complicated. There are many contributions coming from the covariant non-abelian field strength terms, the covariant derivatives of other fields and in particular there is a very large number of topological contributions from the Poisson non-commutativity of the modified momenta with the $\mathcal{P}^2$ term. We do not give an expression for the full result of the transformation of $\mathcal{P}^k_K(A)$, but without the topological contributions, at $\kappa=0$, we find that the canonical time evolution of $ \Pi^k_K(A)$ is given by \eqref{can-eft-ham-PiA-k=0}. The first line of \eqref{can-eft-ham-PiA-k=0} is the covariantised version of the analogous result in \cite{KreutzerSUGRA}, but without the topological contribution due to $\kappa=0$. The remaining terms in \eqref{can-eft-ham-PiA-k=0} come from the covariantised field strengths in the generalised Yang-Mills and Einstein-Hilbert terms, as well as from the covariant derivatives in the scalar kinetic term.
\begin{align}
      \{  \Pi^k_K(A) , \mathcal{H}_\text{Ham}[\phi]\} = &+ \partial_m \left( e\,\phi\,\, M_{MK} \,\,g^{rm} \,\,g^{ks} \,\,\mathcal{F}^M_{rs}   \right) \label{can-eft-ham-PiA-k=0} \\
       & - \partial_R \left( e\,\phi\,\, M_{MK} \,\,g^{rm} \,\,g^{ks} \,\,\mathcal{F}^M_{rs} \,A^R_m  \right) \nonumber \\
   & +  e\,\phi\,\, M_{MN} \,\,g^{rk} \,\,g^{sn} \,\,\mathcal{F}^M_{rs}\,\partial_KA^N_n \nonumber \\
    & -5 \,d^{NRS}\,d_{RKL}\,e\,\phi\,\, M_{MN} \,\,g^{rk} \,\,g^{sn} \,\,\mathcal{F}^M_{rs}\,\partial_S A^L_n \nonumber \\
    & + 5 \,d^{NRS}\,d_{RKL}\,\partial_S \left( e\,\phi\,\, M_{MN} \,\,g^{rm} \,\,g^{ks} \,\,\mathcal{F}^M_{rs} \,A^L_m  \right)\nonumber \\
    & -\frac{e\phi}{12} \,g^{kn}\,\mathcal{D}_n M^{MN}\,\partial_K M_{MN}\nonumber \\
    & + \mathbb{P}^R{}_M{}^L{}_K\, \partial_L\left(e\,\phi\,g^{kn}\,\mathcal{D}_n M^{MN}\,M_{RN} \right)\nonumber \\
    & -2 \,\partial_m\left(e\,\phi\,e_a{}^{[m}\,e_b{}^{k]}(e^{ar}\partial_Ke_r{}^b) \right) \nonumber \\
    &  +2 \,\partial_R\left(e\,\phi\,A^R_m\,e_a{}^{[m}\,e_b{}^{k]}(e^{ar}\partial_Ke_r{}^b) \right)\nonumber \\
    &-2\,e\,\phi\,\partial_KA^M_n\,e_a{}^{[k}\,e_b{}^{n]}(e^{ar}\partial_Me_r{}^b)  \nonumber \\
    & +10\,d^{MRS}\,d_{RLK}\,e\,\phi\,\partial_S A^L_n\,e_a{}^{[k}\,e_b{}^{n]}(e^{ar}\partial_Me_r{}^b)\nonumber \\
    &  -10\,d^{MRS}\,d_{RLK} \,\partial_S\left(e\,\phi\,A^L_m\,e_a{}^{[m}\,e_b{}^{k]}(e^{ar}\partial_Ke_r{}^b) \right)\nonumber 
\end{align}
As was discussed for the external and internal diffeomorphism constraints the two-forms cannot transform under secondary constraints because their Lagrangian kinetic term is topological and linear in the time derivative. Similarly we find that their time evolution generated by the Hamilton constraint is vanishing \eqref{can-eft-ham-B}. What this means is that the overall canonical time evolution of the two-forms is given purely in terms of the gauge transformations that do lead to transformations of the two-forms. In the topological model theory in section \ref{can-sec-twoforms-model} we find that there are no propagating degrees of freedom for the topological two-forms of the model. Similarly we should not expect the two-forms in ExFT to have a canonical time evolution given in terms of their canonical momenta $\Pi^{klM}(B)$, as would be the case for propagating fields (cf. \eqref{can-eft-ham-e}, \eqref{can-eft-ham-M} and \eqref{can-eft-ham-A}). Moreover it is not possible to make use of the primary constraints to construct a $\left(\Pi(B)\right)^2$ term because the Stückelberg coupling term in the one-form field strength is not of the same form as the primary constraints $\mathcal{H}_\text{P2}$. 
\begin{equation}
       \{  B_{klM} , \mathcal{H}_\text{Ham}[\phi]\} = 0    \label{can-eft-ham-B}
\end{equation}
In contrast the canonical two-form momenta $\Pi^{pvW}(B)$ do transform under the secondary constraints, including the Hamilton constraint, because of the Stückelberg coupling in $\mathcal{F}_{mn}^M$ and due to the two-form term inside the modified one-form momenta $\mathcal{P}_M^m$. The transformation of $\Pi^{pvW}(B)$ generated by the Hamilton constraint is given by \eqref{can-eft-ham-PiB}. In analogy to the above argument and to the model from section \ref{can-sec-twoforms-model} we should not interpret \eqref{can-eft-ham-PiB} as a normal time evolution. Furthermore replacing the parameter with the lapse function $\phi=N$ in \eqref{can-eft-ham-PiB} we find the explicit form of the consistency condition \eqref{can-eft-constr-missing-B-Ham}.
\begin{align}
     \{  \Pi^{pvW}(B) , \mathcal{H}_\text{Ham}[\phi]\} = &\,  -\partial_N\left(20\,d^{WMN}\phi \,e\,e_a{}^{[p}\,e_b{}^{v]}\,(e^{ra}\,\partial_M e_r{}^b)  \right) \label{can-eft-ham-PiB}\\
     &+\partial_R\left(100\,\phi \,e\,d^{MKL} \,d^{NRW}\,M_{MN}\,g^{rp}\,g^{sv}\,\partial_K B_{rsL} \right) \nonumber\\
      &-\partial_K\left(\frac{30\kappa\phi}{e}\,g_{mn}\,M^{MN}\,\mathcal{P}^n_N\,A^S_q\,\epsilon^{tmpvq}\,d^{RKW}\,d_{MRS} \right) \nonumber
\end{align}    

\subsection[Tensor gauge transformations]{\texorpdfstring{$B_{\mu\nu M}$}{B-field} tensor gauge transformations}\label{can-eft-tensor-gauge-transf-sec}
In this section we discuss the transformations generated by the constraints $\mathcal{H}_\text{P1}$ \eqref{can-EFT-primary-canonical-constraint-B}, $\mathcal{H}_\text{P2}$ \eqref{can-EFT-primary-canonical-constraint-B-nontrivial} and $\mathcal{H}_\text{S1}$ \eqref{can-eft-S1-constr} and identify some of the transformations with the tensor gauge transformations of the two-forms. We also comment on the consistency conditions \eqref{can-eft-constr-missing-B-Ham}, \eqref{can-eft-constr-missing-B-Diff}, \eqref{can-eft-constr-missing-B-GD} and \eqref{can-eft-constr-missing-B-S1} analogous to the $\mathcal{H}_\text{S2}$ constraints of the model theory from section \ref{can-sec-twoforms-model}.\\

The primary constraints $\mathcal{H}_\text{P1}$ and $\mathcal{H}_\text{P2}$ are in form identical to those of the model theory in section \ref{can-sec-twoforms-model} and therefore they lead to the same transformations. The shift type primary constraints  $\mathcal{H}_\text{P1}$ \eqref{can-EFT-primary-canonical-constraint-B} generate shift transformations \eqref{can-eft-P1-B} on the time components of the two-forms $B_{tmN}$.  
\begin{equation}\label{can-eft-P1-B}
    \{ B_{tmN}, \mathcal{H}_\text{P1}[\lambda] \} = 2\,\lambda_{mN} 
\end{equation}
The primary constraints $\mathcal{H}_\text{P2}$ similarly generate shifts \eqref{can-eft-P2-B} of the spatial B-fields $B_{mnS}$. The general shift transformations \eqref{can-eft-P1-B} and \eqref{can-eft-P2-B} include the restricted $\mathcal{O}_{\mu\nu M}$ shift transformations \eqref{eft-GD-B}, as well as the more specific tensor gauge transformations $\Xi_{\mu M}$ \eqref{eft-GD-B}, which is illustrated by equation \eqref{hamilton-2forms-canonical-gauge-transf-shift2-example}. As was discussed in section \ref{can-sec-twoforms-model} in principle there should be a way of explicitly bringing the canonical shift transformations into the Lagrangian form but it is not clear how this can be done for the tensor gauge transformations.\\
The momenta $\Pi^{mnS}(B)$ transform under $\mathcal{H}_\text{P2}$ as \eqref{can-eft-P2-PiB}, which is identical to the transformation \eqref{ham-two-forms-gauge-P2-on-PiB}.\\
Due to the two-form term inside the modified momenta $\mathcal{P}^m_M(A)$ they transform as \eqref{can-eft-P2-PA}. This is not actually a new transformation but a transformation induced by \eqref{can-eft-P2-B} in the composite modified momenta $\mathcal{P}^m_M(A)$ which are not fundamental canonical coordinates.
\begin{align}
   \{ B_{mnS} , \mathcal{H}_\text{P2}[\rho] \} &= 2\,\rho_{mnS} \label{can-eft-P2-B}\\
    \{ \Pi^{mnS}(B), \mathcal{H}_\text{P2}[\rho] \}&=  30\kappa\,\epsilon^{tmnkl}\,d^{SRN}\,\partial_R \rho_{klN}  \label{can-eft-P2-PiB}\\
     \{ \mathcal{P}^m_M(A), \mathcal{H}_\text{P2}[\rho] \}&=  -30\,\kappa\,\epsilon^{tklmn} \,d^{KST}\,d_{TNM}\,A^N_n\,\partial_S \rho_{klK}  \label{can-eft-P2-PA}
\end{align} 
The one-forms $A^N_n$ transform under the secondary constraints $\mathcal{H}_\text{S1}$ as \eqref{can-eft-S1-A}. Comparing the canonical $\mathcal{H}_\text{S1}$ transformation \eqref{can-eft-S1-A} to the Lagrangian $\Xi_{\mu M}$ transformation in \eqref{eft-GD-A} we can identify \eqref{can-eft-S1-A} precisely as the tensor gauge transformations of $A^N_n$ induced by the Stückelberg coupling. The transformation \eqref{can-eft-S1-A} of $A^N_n$ therefore exists even at $\kappa=0$ because the one-form momentum term in \eqref{can-eq:can-Hamiltonian-EFT-kappa0} comes from the Stückelberg coupling and is not dependent on the topological term.
\begin{equation}
  \{ A^N_n, \mathcal{H}_\text{S1}[\Xi] \}=    -10\,d^{LMN}\,\partial_L \Xi_{nM} \label{can-eft-S1-A}
\end{equation}
At $\kappa=0$ the canonical momenta $ \Pi^n_N(A)$ do not transform  \eqref{can-eft-S1-PiA} under $\mathcal{H}_\text{S1}$. The modified momenta $\mathcal{P}^n_N$ do transform under $\mathcal{H}_\text{S1}$ in a very complicated way because a large number of terms are generated by the Poisson non-commutativity of the $\mathcal{P}^n_N$ and due to the covariantisation terms in $\mathcal{H}_{klmN}$. The explicit transformation of $\mathcal{P}^n_N$ remains to be calculated.
\begin{equation}\label{can-eft-S1-PiA}
 \{ \Pi^n_N(A),  \mathcal{H}^{\kappa=0}_\text{S1}[\Xi] \} = 0  
\end{equation}
The transformation of the two-forms \eqref{can-eft-S1-B} under the secondary constraints $\mathcal{H}_\text{S1}$ vanishes, which agrees with what was found in the model in section \ref{can-sec-twoforms-model}. The two-form momenta $\Pi^{qsR}(B) $ transform as \eqref{can-eft-S1-PiB1}, which can be expand as \eqref{can-eft-S1-PiB2}. We can see the covariantised version of \eqref{ham-two-forms-gauge-S1-on-PiB} plus an additional one-form dependent term in \eqref{can-eft-S1-PiB1}. From \eqref{can-eft-S1-PiB2} we can see that it is not possible to simplify this expression in a meaningful way because of the different structures of the terms.
\begin{align}
 \{ B_{qsR}, \mathcal{H}_\text{S1}[\Xi]  \} = & \,0 \label{can-eft-S1-B}\\
 \{ \Pi^{qsR}(B) ,  \mathcal{H}_\text{S1}[\Xi]\} = & +300\,\kappa\,\epsilon^{lqsr}\,d^{MKL}\,d^{SNR}\,d_{QSL}\,\partial_N\left(\partial_K\Xi_{lM}\,A^Q_r\right) \label{can-eft-S1-PiB1} \\
 & + 60\,\kappa\,\epsilon^{lmqs} \,d^{MKR} \,\mathcal{D}_m\partial_K\Xi_{lM}\nonumber\\
 =&\,+300\,\kappa\,\epsilon^{lqsr}\,d^{MKL}\,d^{SNR}\,d_{QSL}\,\partial_N\left(\partial_K\Xi_{lM}\,A^Q_r\right)\label{can-eft-S1-PiB2} \\
 & + 60\,\kappa\,\epsilon^{lmqs} \,d^{MKR} \,\partial_m\partial_K\Xi_{lM}\nonumber\\
 & -60\,\kappa\,\epsilon^{lqsr}\,d^{MKR}\,A^N_r\,\partial_N\partial_K\Xi_{lM}\nonumber\\
 &+60\,\kappa\,\epsilon^{lmqs}\,d^{MKN}\,\partial_K\Xi_{lM}\,\partial_N A^R_m\nonumber\\
 &-600\,\kappa\,\epsilon^{lmqs}\,d^{MKL}\,d^{RPS}\,d_{LTP}\,\partial_s A^T_m\,\partial_K\Xi_{lM}\nonumber
\end{align}
When replacing $\Xi_{nM}=B_{tnM}$ in \eqref{can-eft-S1-PiB2} we find the explicit form of the consistency condition \eqref{can-eft-constr-missing-B-S1}. This is in direct analogy to the constraints $\mathcal{H}_\text{S2}$ defined by equation \eqref{hamilton-2forms-canonical-S2} of the simpler model in section \ref{can-sec-twoforms-model}. In the model the constraints $\mathcal{H}_\text{S2}$ consisted only of the second term in \eqref{can-eft-S1-PiB2}. \\

Now that we have calculated the explicit form of \eqref{can-eft-constr-missing-B-S1} we have found all of the consistency conditions \eqref{can-eft-constr-missing-B-Ham}, \eqref{can-eft-constr-missing-B-Diff}, \eqref{can-eft-constr-missing-B-GD} and \eqref{can-eft-constr-missing-B-S1} that follow from \eqref{can-eft-consist-P2-2}. All of these consistency conditions are non-vanishing and since they are independent we should think of all of them as canonical constraints, in analogy with the constraints $\mathcal{H}_\text{S2}$ of section \ref{can-sec-twoforms-model}. It is the form of the topological term \eqref{topological-term-E-six} combined with the Stückelberg coupling in \eqref{eft-GD-F-covariant-stuckelberg} that leads to these constraints. \\
What is unusual about these constraints is that they depend on the Lagrange multipliers $N$, $N^n$, $A_t^M$ and $B_{tnN}$. This is already the case for the $\mathcal{H}_\text{S2}$ constraints in section \ref{can-sec-twoforms-model} which depend on $B_{tnN}$. In the model this leads to the constraint algebra relation \eqref{hamilton-2forms-canonical-S1-P2-Poisson-alg} which makes $\mathcal{H}_\text{S2}$ and the primary constraints $\mathcal{H}_\text{P1}$ into second class constraints. The $\mathcal{H}_\text{S2}$ constraints in the model are themselves of the same form as the term in the \eqref{hamilton-2forms-canonical-S2-P1-Poisson-alg} algebra relation that makes the other constraints into second class constraints, which suggests some relation between these unusual constraints and the need for Dirac brackets. It is possible that a similar relation exists in ExFT for the constraints \eqref{can-eft-constr-missing-B-Ham}, \eqref{can-eft-constr-missing-B-Diff}, \eqref{can-eft-constr-missing-B-GD} and \eqref{can-eft-constr-missing-B-S1}, which turn the primary constraints that are the canonical momenta conjugate to the Lagrange multipliers into second class constraints. To make sense of these constraints a better understanding of the model in section \ref{can-sec-twoforms-model} and in particular of the $\mathcal{H}_\text{S2}$ constraints is needed. 

\subsection{Shifts and Lorentz transformations}
Finally we can briefly describe the transformations generated by the remaining primary constraints from section \ref{can-eft-sec-primary-constraints} which are of the usual form and appear identically in five-dimensional E$_{6(6)}$ invariant supergravity \cite{KreutzerSUGRA}.\\

The primary constraints of shift type generate shift transformations on the conjugate canonical variables. 
\begin{align}
   \{ N , \Pi(N)[\lambda_1] \} &= \lambda_1 \\
    \{ N_a, \Pi(N_b)[\lambda_2] \} &=(\lambda_2)_a \\
     \{ A_t^N, \Pi(A_t^M)[\lambda_3] \}  &=(\lambda_3)^N 
\end{align} 
The Lorentz constraints \eqref{can-eft-Lorentz-constraints} generate spatial Lorentz transformations on the spatial vielbein and their conjugate momenta. The equivalent Lorentz transformations in the five-dimensional E$_{6(6)}$ invariant supergravity have been discussed in detail in \cite{KreutzerSUGRA}.
\begin{align}
      \{   e_n{}^a, L[\gamma] \}= &+ e_{nb}\,\, \gamma^{ba} \\
       \{    \Pi^n{}_a(e),  L[\gamma] \}= & +\Pi^n{}_c(e)   \,\, \gamma^{cb}\,\,\delta_{ba}
\end{align}

\section{Canonical constraint algebra}\label{sec-canonical-algebra}
In this section we discuss part of the algebra that the canonical constraints form under the Poisson bracket. Some of the relations of the canonical constraint algebra are very difficult to compute and because not all of the transformations of the modified one-form momenta $\mathcal{P}^m_M$ have been fully computed we can only give speculative results for some relations of the algebra.\\

The primary constraints all Poisson-commute --- with the exception of the Lorentz constraints which form the Lorentz subalgebra \eqref{eqn:alg-LL}. For the primary two-form constraints $\mathcal{H}_\text{P1}$ and $\mathcal{H}_\text{P2}$ their Poisson-commutativity was already verified in section \ref{can-sec-twoforms-model}.\\

We can also look at the algebra relations between the Lorentz constraints and the secondary constraints. In the canonical formulation of five-dimensional E$_{6(6)}$ invariant supergravity \cite{KreutzerSUGRA} one finds that the Lorentz constraints Poisson-commute with the Hamilton constraint. We find the same result for ExFT \eqref{eqn:alg-L-Ham} because the scalar potential can be written entirely in terms of the metric and is therefore Lorentz invariant. For the bracket of the external diffeomorphism constraints with the Lorentz constraints we find the relation \eqref{eqn:alg-diff-L}, where the Lorentz parameters are transformed by an external diffeomorphism. The relation \eqref{eqn:alg-diff-L} is the covariantisation of the equivalent expression in five-dimensional E$_{6(6)}$ invariant supergravity \cite{KreutzerSUGRA}. Irrespective of its sign the contribution from the $-g_{lm}\, M^{KL} \mathcal{P}_L^l(A) \, \partial_K N^m$ term in $\mathcal{H}_\text{Diff}$ to the bracket \eqref{eqn:alg-diff-L} vanishes due to antisymmetry of the Lorentz constraints.
The Poisson bracket of the Lorentz constraints with the generalised diffeomorphism constraints is given by \eqref{eqn:alg-L-GD}, where the Lorentz parameters are transformed by a generalised diffeomorphism. The analogous relation in five-dimensional E$_{6(6)}$ invariant supergravity \cite{KreutzerSUGRA} vanishes, which is consistent because the trivial solution to the section condition leads to the vanishing of the generalised derivative in the gauge parameter. The bracket \eqref{eqn:alg-L-S1} vanishes trivially due to the form of the constraints.
\begin{align}
\{L[\gamma] ,L[\kappa] \}      &=   L[-2 \gamma^{c[a}\,\,\kappa^{b]c}]  \label{eqn:alg-LL}\\ 
\{ \mathcal{H}_\text{Ham}[\phi] ,  L[\gamma]  \}   &=  0  \label{eqn:alg-L-Ham}\\ 
\{ \mathcal{H}_\text{Diff}[\lambda] ,  L[\gamma]  \}   &=     L[\lambda^m\,\mathcal{D}_m\gamma^{ab}] \label{eqn:alg-diff-L}\\  
\{ \mathcal{H}_\text{GD}[\Lambda] , L[\gamma]  \}   &=   L[ \mathbb{L}_\Lambda \gamma^{ab}]   \label{eqn:alg-L-GD}\\
\{ \mathcal{H}_\text{S1}[\Xi] , L[\gamma]  \}   &=  0 \label{eqn:alg-L-S1}
\end{align}
So far we have not needed to make use of the section condition \eqref{eft-e6-section-condition-1}. The section condition is however needed many times in the computation of the algebra relation \eqref{can-eft-algebra-GD-GD} concerning the generalised diffeomorphism constraints. The computation of the relation \eqref{can-eft-algebra-GD-GD} is more complicated than in the Lagrangian formalism because the constraints $\mathcal{H}_\text{GD}$ do not just generate generalised diffeomorphisms but also contain some information about the tensor gauge transformations, cf. equations \eqref{can-eft-GD-A-3}  and \eqref{can-eft-GD-piA2}. Because of the Poisson non-commutativity of $\mathcal{P}^m_M$ and because of the topological term in $\mathcal{H}_\text{GD}$ a very large number of terms is being generated in the computation. The relation \eqref{can-eft-algebra-GD-GD} was verified at $\kappa=0$. There are additional terms that may rearrange into further constraints, possibly related to the tensor gauge transformations, however this remains to be computed. The cubic $d$-symbol relations \eqref{eft-cubic-d-1} and \eqref{eft-cubic-d-2} need to be applied repeatedly in the computation of \eqref{can-eft-algebra-GD-GD} to move E$_{6(6)}$ indices between objects.
\begin{equation}\label{can-eft-algebra-GD-GD}
\{\mathcal{H}_\text{GDC}[\Lambda] ,\mathcal{H}_\text{GDC}[\zeta]\} =\mathcal{H}_\text{GDC}\left[  [\Lambda,\zeta]_E \right] + \dots
\end{equation}
The seeming difference in sign between \eqref{eft-alg-ii-lambda} and \eqref{can-eft-algebra-GD-GD} is due to the fact that the constraints act in the Poisson brackets from the right onto the fields. This can be verified explicitly by using the relation $\delta_{\mathcal{H}_\text{GDC}[\Lambda]} =\{\,\cdot\,,\mathcal{H}_\text{GDC}[\Lambda]\}$ to translate \eqref{eft-alg-ii} into the canonical formalism and using the Jacobi identity and antisymmetry of the Poisson bracket.\\

The algebra of the canonical constraints of E$_{6(6)}$ ExFT has to be consistent with that of five-dimensional E$_{6(6)}$ invariant supergravity \cite{KreutzerSUGRA}. The ExFT algebra relations need to reduce to those of \cite{KreutzerSUGRA} for the trivial solution of the section condition. Combining this fact with the information about the Lagrangian gauge algebra described in \cite{Baguet:2015xha} and summarised in section \ref{sec-EFT-lagrangian} we can make some conjectures about the form of the remaining canonical constraint algebra relations. In all of the following algebra relations it is possible that additional canonical constraints, in particular those related to the two-forms may appear. In general the form of the gauge algebra in the Lagrangian formalism does not have to be identical to that of the constraint algebra in the Hamiltonian formalism because one can always choose a different basis for the algebra.\\ 

From the relations \eqref{eft-alg-ee} and \eqref{eft-alg-ee-xi} one might conjecture the algebra relations \eqref{eqn:alg-diff-diff} and \eqref{eqn:alg-diff-GD}, but these relations have not been computationally verified. Regarding the consistency of \eqref{eqn:alg-diff-diff} and \eqref{eqn:alg-diff-GD} with five-dimensional E$_{6(6)}$ invariant supergravity the field-strength term in the parameter of \eqref{eqn:alg-diff-diff} seems to be problematic as it does not seem to lead to a vanishing term in the trivial solution of the section condition --- therefore it seems that this term should not appear canonically, possibly due to the different parametrisations of the algebra. Furthermore the sign of the second term in the transformation \eqref{can-eft-diff-A} appears in two of the gauge parameters, which adds further uncertainty about these relations. In contrast the $\mathcal{H}_\text{Diff}$ terms on the right hand sides seem more likely to be correct. It seems probable that the section condition may play a role in the computation of these relations because the generalised diffeomorphism constraints are involved.  
\begin{align}
\{ \mathcal{H}_\text{Diff}[\lambda] ,  \mathcal{H}_\text{Diff}[\xi] \}   &\stackrel{?}{=} \mathcal{H}_\text{Diff}\left[ \lambda^\mu \,\mathcal{D}_{m  }\xi^n   - \xi^m   \,\mathcal{D}_m    \lambda^n    \right] \nonumber\\
&\quad+ \mathcal{H}_\text{GD}[ \cancel{\lambda^m\,\xi^n\,\mathcal{F}^M_{mn}}+  M^{MN}\,g_{mn}\left(\lambda^{m}\partial_M\xi^n-\xi^{m}\partial_M\lambda^n \right)]+\dots \label{eqn:alg-diff-diff}\\  
\{ \mathcal{H}_\text{GD}[\Lambda]  , \mathcal{H}_\text{Diff}[\xi] \} &\stackrel{?}{=} \mathcal{H}_\text{Diff}[\mathbb{L}_{\Lambda}\xi^n] \nonumber\\
&\quad+ \mathcal{H}_\text{S1}[d_{MNK}\,\Lambda^K\left(\xi^m\,\mathcal{F}^N_{mn}- M^{KL}\,g_{mn}\,\partial_L \xi^n  \right)] + \dots   \label{eqn:alg-diff-GD}
\end{align}
The relation $\{\mathcal{H}_\text{Ham}[ \theta] ,  \mathcal{H}_\text{GD}[\xi]  \}$ is particularly difficult to compute but since this relation vanishes in five-dimensional E$_{6(6)}$ invariant supergravity we cannot extrapolate the result to ExFT. Finally we may conjecture \eqref{eft-alg-cf-sugra-eq:alg-ham-ham} and  \eqref{eft-alg-cf-sugra-ALG-Diff-Ham} as relations of this form seem to be required by comparison to the algebra of canonical five-dimensional E$_{6(6)}$ invariant supergravity (the appearance of the spin connection $\omega_{nab}$ in \eqref{eft-alg-cf-sugra-eq:alg-ham-ham} has been discussed in \cite{KreutzerSUGRA}). There may be additional constraints appearing on the right hand sides of \eqref{eft-alg-cf-sugra-eq:alg-ham-ham} and  \eqref{eft-alg-cf-sugra-ALG-Diff-Ham}.
\begin{align}
\{ \mathcal{H}_\text{Ham}[ \theta] , \mathcal{H}_\text{Ham}[ \tau]\}        &\stackrel{?}{=}   \mathcal{H}_\text{Diff}[(\theta\,\nabla_m\tau-\tau\,\nabla_m\theta) \, g^{mn}]  \nonumber\\ 
&\quad- L\left[ (\theta\,\nabla_m\tau-\tau\,\nabla_m\theta) \, g^{mn}\, \omega_{nab} \right]  +\dots  \label{eft-alg-cf-sugra-eq:alg-ham-ham}\\
\{ \mathcal{H}_\text{Diff}[\lambda],\mathcal{H}_\text{Ham}[ \theta] \}     &\stackrel{?}{=}  \mathcal{H}_\text{Ham}[\lambda^m\,\mathcal{D}_m\theta] + \mathcal{H}_\text{GD}\left[\frac{\theta}{e}\lambda^p\,\,g_{pk}\,\,\mathcal{P}^k_L\,\,M^{LM}\right]+ \dots \label{eft-alg-cf-sugra-ALG-Diff-Ham}  
\end{align}
From five-dimensional E$_{6(6)}$ invariant supergravity we cannot get any information about the algebra relations concerning the two-form constraints as there are no two-forms in the theory. From the results of the model theory in section \ref{can-sec-twoforms-model} we should expect that the two-forms do indeed not contribute any propagating degrees of freedom to the theory. Moreover we do not yet have a good understanding of the relations that follow from the $\mathcal{H}_\text{P2}$ consistency condition \eqref{can-eft-consist-P2-2}.\\

To confirm that the number of physical degrees of freedom in field space is indeed 128 the full canonical constraint algebra needs to be known as otherwise we cannot know which canonical constraints are first class and which ones are second class.\\
From the canonical analysis of the five-dimensional ungauged maximal E$_{6(6)}$ invariant supergravity \cite{KreutzerSUGRA} we should expect that the (bosonic) E$_{6(6)}$ ExFT, without the two-forms, does have 128 physical degrees of freedom. To these 128 physical degrees of freedom the external metric $G_{\mu\nu}$ contributes 5, while 42 come from the scalar fields $M_{MN}$ and 81 come from the generalised one-forms $A_\mu^M$. And indeed the canonical analysis of the topological two-form model in section \ref{can-sec-twoforms-model} suggests that the two-forms $B_{\mu\nu M}$ should not contribute any propagating physical degrees of freedom. This may lead us to naively suspect that (with the implicit treatment of the scalar coset constraints) only the canonical constraints coming from the two-forms are second class, in which case the counting of the physical degrees of freedom would work out to be 128, but this remains to be verified by the computation of the full constraint algebra.

\section{The generalised vielbein and USp(8)}\label{sec-vielbein-usp8}
The canonical formulation of E$_{6(6)}$ exceptional field theory, written in terms of the generalised metric $M_{MN}$, was constructed in section \ref{section-can-eft-chapter}. The description in terms of the generalised metric is sufficient for the bosonic sector of ExFT, however it can be useful for some applications (e.g. coupling to fermions or manifesting the USp(8) symmetry) to consider the formulation in terms of the generalised USp(8) vielbein $\mathcal{V}^{AB}_M$. In this section we discuss how the canonical formulation of E$_{6(6)}$ ExFT can be reformulated in terms of the generalised vielbein.\\

The generalised USp(8) vielbein is essential in the supersymmetric formulation of E$_{6(6)}$ ExFT and we use the same conventions for the USp(8) invariant form as \cite{Musaev2015,Baguet:2015xha}. The unitary symplectic Lie group USp(8) is 36 dimensional and has an 8 dimensional fundamental representation. The indices $A,B,\dots,F = 1,\dots,8$ are used here to denote the fundamental representation of USp(8).\\

The internal generalised metric $M_{MN}$ of E$_{6(6)}$ ExFT is an {E$_{6(6)}/$USp(8)} coset representative. There is a direct analogy of $M_{MN}$ as a coset element to the external metric $G_{\mu\nu}$ of general relativity, which is a GL$(d)$/SO$(1,d-1)$ coset element. In analogy to the vielbein (or frame field) $E_\mu{}^\alpha$ of the external metric \eqref{eft-vielbein-def} we can introduce a generalised internal vielbein $\mathcal{V}_M^{AB}=\mathcal{V}_M^{[AB]}$ as in \eqref{usp8-gen-vielbein-def} by making use of the local USp(8) invariance. In the definition \eqref{usp8-gen-vielbein-def} the USp(8) symplectic form $\Omega_{AB}$ takes the place of the Minkowski metric in \eqref{eft-vielbein-def}.
\begin{equation}\label{usp8-gen-vielbein-def}
    M_{MN} =: \mathcal{V}_M^{AB}\, \mathcal{V}_N^{CD}\, \Omega_{AC}\, \Omega_{BD} = \mathcal{V}_M^{AB}\, \mathcal{V}_{NAB}
\end{equation}
In \eqref{usp8-gen-vielbein-def} the $\mathcal{V}_{NAB}$ is defined by $\mathcal{V}_{NAB} := \mathcal{V}_N^{CD}\Omega_{AC} \Omega_{BD}$. We define the inverse symplectic form by $\Omega_{AB}\,\Omega^{CB} := \delta_A^C$, which is equivalent to $\Omega_{AB}\,\Omega^{BC} = -\delta_A^C$.
The generalised vielbein furthermore satisfies the identity \eqref{usp8-vielbein-omega-prop}. With the condition \eqref{usp8-vielbein-omega-prop} the index pair $[AB]$ has $27$ components, which agrees with the dimension of the fundamental E$_{6(6)}$ representation.
\begin{equation}\label{usp8-vielbein-omega-prop}
     \mathcal{V}_M^{AB}\, \Omega_{AB} =0
\end{equation}
The inverse generalised vielbein is defined by \eqref{usp8-inverse-vielbein1} and \eqref{usp8-inverse-vielbein2} \cite{Musaev2015}.
\begin{align}
    \mathcal{V}_M^{AB}\, \mathcal{V}_{AB}^N &:= \delta_M^N \label{usp8-inverse-vielbein1}\\
        \mathcal{V}_M^{AB}\,  \mathcal{V}_{CD}^M &:= \frac{1}{2}(\delta^A_C \delta^B_D - \delta^A_D \delta^B_C ) - \frac{1}{8} \Omega^{AB}\,  \Omega_{CD} \label{usp8-inverse-vielbein2}
\end{align}
The canonical momenta of the generalised vielbein that follow from the kinetic term of the scalar fields \eqref{can-eft-adm-scalar} are given by \eqref{usp8-canonical-momentum-V}.
\begin{align}
\label{usp8-canonical-momentum-V}
 \Pi^M_{AB}(\mathcal{V}) =  \frac{e}{3N} &\bigg[  \dot{\mathcal{V}}_N^{CD}  \mathcal{V}^M_{CD}  \mathcal{V}^N_{AB} +  
 \dot{\mathcal{V}}_K^{EF}  \mathcal{V}^K_{CD}  \mathcal{V}^N_{EF}   \mathcal{V}^{M CD}   \mathcal{V}_{N AB}  \\
 & + \mathbb{L}_{A_t}  \mathcal{V}^M_{AB} \quad+   \mathbb{L}_{A_t}  \mathcal{V}^N_{CD} \mathcal{V}_{N AB} \mathcal{V}^{M CD} \nonumber\\
 & + N^n \mathcal{D}_n  \mathcal{V}^M_{AB} +  N^n \mathcal{D}_n  \mathcal{V}^N_{CD} \mathcal{V}_{N AB} \mathcal{V}^{M CD} \bigg] \nonumber
\end{align}
We can relate the canonical momenta \eqref{usp8-canonical-momentum-V} to the rescaled canonical momenta of the generalised metric by \eqref{usp8-PiVtoPiM}. The inverse relation \eqref{usp8-PiMtoPiV} is directly analogous to the relation (2.9) in reference \cite{KreutzerSUGRA} about the canonical momenta of the metric and the standard vielbein in general relativity.
\begin{align}
 \Pi^M_{AB}(\mathcal{V}) & = 2\, {\Pi}^{MN}(M) \, \mathcal{V}_{N AB} \label{usp8-PiVtoPiM}\\
    {\Pi}^{MN}(M)   & =\frac{1}{2} \Pi^{(M}_{AB}({\mathcal{V}}) \, \mathcal{V}^{N) AB} \label{usp8-PiMtoPiV}
\end{align}
In analogy to the Lorentz constraints in general relativity \eqref{can-eft-Lorentz-constraints} there are the primary canonical USp(8)-constraints $\mathcal{H}_\text{USp(8)}$ \eqref{usp8-k-constraint}.
\begin{equation}
\label{usp8-k-constraint}
 (\mathcal{H}_\text{USp(8)})^{AD} := \mathcal{V}_M^{AB}\, \Omega_{BC}\, \Pi^{M  CD} + \mathcal{V}_M^{DB}\, \Omega_{BC}\, \Pi^{M  CA} = 0
\end{equation}
In contrast to the Lorentz constraints the contraction of the canonical momenta with the generalised vielbein in \eqref{usp8-k-constraint} is symmetrised. The constraints $(\mathcal{H}_\text{USp(8)})^{AD}$ thus have 36 components, which is equal to the dimension of USp(8).\\

The fundamental Poisson bracket of the generalised vielbein can be defined by \eqref{usp8-Poisson}.
\begin{equation}
      \{ \mathcal{V}_M^{AB}, \Pi_{CD}^M(\mathcal{V}) \} := \frac{1}{2}(\delta^A_C \delta^B_D - \delta^A_D \delta^B_C ) - \frac{1}{8} \Omega^{AB}\,  \Omega_{CD} \label{usp8-Poisson}
\end{equation}
In order to rewrite the ExFT Hamiltonian \eqref{can-eq:can-Hamiltonian-EFT} in terms of the generalised vielbein $\mathcal{V}_M^{AB}$ we first need to verify that the Legendre transformation is invariant under this change of canonical coordinates. An additional symplectic term coming from the identity \eqref{usp8-inverse-vielbein2} vanishes due to the projection \eqref{usp8-vielbein-omega-prop} and we find that the Legendre transformation is indeed invariant \eqref{usp8-legendre-equiv}.
\begin{equation}\label{usp8-legendre-equiv}
    \frac{1}{2}\sum_{\substack{R,S=1,\dots,27}}\dot M_{RS}\cdot \tilde{\Pi}^{RS}(M)= \frac{1}{2}\sum_{\substack{M=1,\dots,27\\A,B=1,\dots,8}} \dot{\mathcal{V}}_M^{AB}\cdot {\Pi}^M_{AB}(\mathcal{V})
\end{equation}
Due to the identity \eqref{usp8-legendre-equiv} we can now insert the definitions of the generalised vielbein \eqref{usp8-gen-vielbein-def} and their canonical momenta \eqref{usp8-PiMtoPiV} into the canonical Hamiltonian \eqref{can-eq:can-Hamiltonian-EFT} thus replacing the generalised metric and its momenta. We find that the part of the Hamiltonian coming from the scalar kinetic term \eqref{can-eft-scalar-legendre} can be written in terms of the generalised vielbein as \eqref{usp8-legendre-SK-term-vielbein}. In \eqref{usp8-legendre-SK-term-vielbein} there are always two differently contracted versions of each term of \eqref{can-eft-scalar-legendre}, this is a general feature for the generalised USp(8) vielbein.
\begin{align}\label{usp8-legendre-SK-term-vielbein}
&\frac{1}{2}\sum_{\substack{M=1,\dots,27\\A,B=1,\dots,8}} \dot{\mathcal{V}}_M^{AB}\cdot {\Pi}^M_{AB}(\mathcal{V})- \mathcal{L}_\text{sc}   \\
= N \cdot &\bigg[ \frac{3}{16e} \Pi^M_{AB} \Pi^S_{CD} \mathcal{V}_M^{CD} \mathcal{V}_S^{AB}  
+\frac{3}{16e} \Pi^M_{AB} \Pi^R_{CD} \mathcal{V}^{S CD} \mathcal{V}_S^{AB} \mathcal{V}_M^{EF} \mathcal{V}_{R EF}  \nonumber\\
& -\frac{e}{12} g^{kl} \mathcal{D}_k \mathcal{V}_M^{AB} \mathcal{D}_l \mathcal{V}^M_{AB}
-\frac{e}{12} g^{kl} \mathcal{D}_k \mathcal{V}_M^{AB} \mathcal{D}_l \mathcal{V}^N_{CD}  \mathcal{V}_{N AB}  \mathcal{V}^{M CD} \bigg]\nonumber\\
+ N^l \cdot & \bigg[ \frac{1}{4} \Pi^M_{AB} \mathcal{D}_l \mathcal{V}_{M}^{AB}
+  \frac{1}{4} \Pi^M_{AB}  \mathcal{D}_l \mathcal{V}_{N CD} \mathcal{V}^{N AB} \mathcal{V}_M^{CD} \bigg]\nonumber\\
+ A_t^K \cdot & \bigg[ \frac{1}{4} \Pi^M_{AB} (\partial_K  \mathcal{V}_M^{AB} -6 \mathbb{P}^{P\,\, L}_{\,\,M\,\, K} \partial_L  \mathcal{V}_P^{AB})  \nonumber\\
& + \frac{1}{4} \Pi^M_{AB}   (\partial_K  \mathcal{V}_N^{CD} -6 \mathbb{P}^{P\,\, L}_{\,\,N\,\, K} \partial_L  \mathcal{V}_P^{CD}) \mathcal{V}^{N AB}  \mathcal{V}_{M CD} \bigg]   \nonumber
\end{align}

\section{Conclusions and outlook}\label{section-conclusios}
In this work we have constructed the canonical Hamiltonian of (bosonic) E$_{6(6)}$ exceptional field theory, computed most of the canonical gauge transformations and discussed some relations of the canonical constraint algebra. \\

In order to carry out the Legendre transformation of the E$_{6(6)}$ ExFT Lagrangian \eqref{eft-lagrangian-terms} we had to construct the explicit non-integral (non-manifestly gauge invariant) topological term of E$_{6(6)}$ ExFT \eqref{topological-term-E-six}. The topological term \eqref{topological-term-E-six} was found by making an ansatz inspired by the topological term of the five-dimensional maximal gauged supergravity \cite{deWit:2004nw} and then comparing its general variation to the variation \eqref{can-top-lag-general-var-EFTI} to fix all coefficients of the ansatz.\\
The resulting topological term \eqref{topological-term-E-six} is very intricate and consists of many terms. This complexity is one of the major computational challenges in the canonical formulation of E$_{6(6)}$ ExFT because many calculations involve a very large number of terms. As a consequence some of the canonical transformations concerning the modified momenta $\mathcal{P}^m_M$, but in particular many of the relations of the canonical constraint algebra have not been calculated. As this is a purely computational issue it should be feasible to fully perform these calculations with the assistance of a suitable computer algebra program that is able to handle the multitude of mathematical structures involved in canonical ExFT simultaneously. Alternatively one might instead want to consider the E$_{8(8)}$ ExFT \cite{Hohm:2014fxa}, which has a somewhat simpler topological term, although there are other complications that arise in the E$_{8(8)}$ theory, such as e.g. constrained compensator fields.\\

The dynamics of the two-forms $B_{\mu\nu M}$ in the E$_{6(6)}$ ExFT is purely topological and governed by \eqref{topological-term-E-six}. In analogy to gauged supergravity the two-forms are needed to absorb the non-covariance in the transformation \eqref{eft-GD-F-naive} of the one-forms. In section \ref{can-sec-twoforms-model} we have investigated the canonical formulation of the model consisting of the two-form kinetic term in isolation and confirmed canonically that there are no propagating degrees of freedom. We have furthermore found that due to its topological nature the two-forms do not see the external diffeomorphisms canonically. Because the two-forms in ExFT only couple to the external metric through the Stückelberg coupling terms this is also true for the full ExFT. From the model theory we have furthermore learned that the tensor gauge transformations appear canonically in a very different form. This can be compared to the gauge transformations of Maxwell theory which canonically also appear in a different form (cf. \cite{KreutzerSUGRA}) but which can be brought into the usual Lagrangian form $\delta_\lambda A_\mu = \partial_\mu \lambda$ by means of the extended Hamiltonian formalism, see chapter 19 of \cite{Henneaux-Teitelboim}. For our topological model this method does not work because all constraints are second class and the extended Hamiltonian is identical to the total Hamiltonian. Because the canonical structure of the topological two-form model of section \ref{can-sec-twoforms-model} is similar to the canonical structure of three-dimensional Chern-Simons theory it may be possible to find an analogy between these theories that allows us to find a way of canonically bringing the tensor gauge transformations into the standard form.\\

To deal with the second class constraints of the two-form model in section \ref{can-sec-twoforms-model} we wrote down a possible definition \eqref{ham-can-B-model-dirac-bracket} for the Dirac bracket in the ExFT geometry. Because the non-constraint terms in the constraint algebra of the model theory are not constant but instead contain both internal and external derivatives we found that we would need to solve equations of the form \eqref{ham-can-stammfunktion} for a primitive function of the $5+27$-dimensional Dirac delta distribution in order to find an explicit expression for the Dirac bracket of the two-form model. It is not clear which functions solve the equation \eqref{ham-can-stammfunktion}. If one could solve \eqref{ham-can-stammfunktion} we should be able to find the explicit form of \eqref{ham-can-B-model-dirac-bracket} in the case of the model --- which may also shed some light on the structure of the tensor gauge transformations.\\

In section \ref{section-can-eft-chapter} we have calculated the ADM decomposition of the E$_{6(6)}$ ExFT Lagrangian, the canonical momenta of all fields and carried out the Legendre transformation to arrive at the canonical Hamiltonian of E$_{6(6)}$ ExFT \eqref{can-eq:can-Hamiltonian-EFT}. In analogy to \cite{KreutzerSUGRA} we found that the most concise description of the canonical theory is given in terms of the modified one-form momenta-like variables $\mathcal{P}^m_M(A)$ \eqref{can-EFT-original-canonical-momentum-A-redef} where all topological contributions have been subtracted from the original canonical momenta $\Pi^m_M(A)$. While these modified momenta give the simplest Hamiltonian and have nice gauge transformation properties the redefinition \eqref{can-EFT-original-canonical-momentum-A-redef} is not a canonical transformation and the new variables do not Poisson commute $\{\mathcal{P}^m_M(A),\mathcal{P}^n_N(A) \}\neq0$. This Poisson non-commutativity, which in turn again stems from the complexity of the topological term \eqref{topological-term-E-six}, further complicates the canonical calculations.\\
With the exception of the purely internal terms we found that the canonical Hamiltonian \eqref{can-eq:can-Hamiltonian-EFT} is given by the covariantisation of the canonical Hamiltonian of five-dimensional ungauged maximal E$_{6(6)}$ invariant supergravity \cite{KreutzerSUGRA}. The ExFT Hamiltonian \eqref{can-eq:can-Hamiltonian-EFT} (mainly) consists of the secondary Hamilton constraint $\mathcal{H}_\text{Ham}$, the (external) diffeomorphism constraints $\mathcal{H}_\text{Diff}$, the generalised diffeomorphism constraints $\mathcal{H}_\text{GD}$ and the two-form tensor gauge constraints $\mathcal{H}_\text{S1}$. In the Legendre transformation we found that the ExFT scalar potential remains largely unchanged and only a single potential term cancels \eqref{can-eft-potantial-relation}. The scalar potential is the main addition to the Hamilton constraint when compared to five-dimensional supergravity \cite{KreutzerSUGRA}. The generalised diffeomorphism constraint, which is associated to the Lagrange multiplier $A^M_t$, is an extension of the abelian U(1)$^{27}$ constraint of the five-dimensional supergravity because the vector fields are used to gauge the generalised diffeomorphism symmetry. The generalised diffeomorphism constraint furthermore contains the extensive Hamiltonian topological term \eqref{can-eft-Htop}. The secondary two-form tensor gauge constraints $\mathcal{H}_\text{S1}$ are in form very similar to the Lagrangian on-shell duality relation \eqref{eft-eom-b}. It may be suspected that the on-shell duality of the one- and two-forms is canonically implied by the transformations generated by the two-form constraints, though the details of this duality are not yet fully understood canonically.\\
Verifying the consistency of the primary constraints of shift type (i.e. vanishing canonical momenta) we confirmed that the above secondary constraints are required. The two-form primary constraints $\mathcal{H}_\text{P2}$ are special because they are not of shift type, instead they relate the spatial two-form canonical momenta directly to the spatial two-form components --- these constraints are a direct consequence of the fact that the two-forms appear with only a single time derivative in the Lagrangian. The consistency condition \eqref{can-eft-consist-P2-2} that is required by $\mathcal{H}_\text{P2}$ implies the existence of secondary constraints that depend on the Lagrange multipliers. In the two-form model of section \ref{can-sec-twoforms-model} these are the $\mathcal{H}_\text{S2}$ constraints that depend on the Lagrange multiplier $B_{tnM}$. In ExFT the consistency condition \eqref{can-eft-consist-P2-2} seems to yield several independent constraints as they depend on all of the independent Lagrange multipliers $N,\,N^n,\,A^M_t$ and $B_{tnM}$. These constraints are of the form of the transformations of $\Pi^{klM}(B)$ but with the parameters given by the Lagrange multipliers. The transformations generated by these constraints in the Poisson bracket are not very illuminating, but in analogy with the model in section \ref{can-sec-twoforms-model} we should expect that these transformations are indeed related to the tensor gauge transformations of $B_{\mu\nu M}$. Ultimately these constraints exist because of the linearity in the time derivative of the two-forms in the Lagrangian and because the two-forms couple to the other secondary constraints via the Stückelberg coupling terms and the topological term couplings. To better understand the constraints analogous to $\mathcal{H}_\text{S2}$ and the transformations generated by them we first need a better understanding of the model two-form theory of section \ref{can-sec-twoforms-model}, including the construction of the Dirac bracket.\\

In section \ref{can-eft-gauge-transformations} we have calculated most of the transformations generated by the canonical constraints. In particular the full transformations of the modified momenta $\mathcal{P}^m_M$ are very complicated due to the number of topological contributions. For some of the transformations of $\mathcal{P}^m_M$ we have thus only given the result at $\kappa=0$, thereby dropping all topological contributions, which we considered as a purely computational tool.\\
We found that all fields transform under the generalised diffeomorphism constraint as the generalised Lie derivative. For the one-forms and their momenta there are additional tensor gauge transformation terms that appear. The tensor gauge constraints $\mathcal{H}_\text{S1}$ and transformations that appear canonically in the transformations generated by $\mathcal{H}_\text{GD}$ are in analogy to the one-form gauge transformations that appear in the canonical transformations of the $\mathcal{H}_\text{Diff}$ constraints in five-dimensional supergravity \cite{KreutzerSUGRA} --- or analogously in ExFT to the $\mathcal{H}_\text{GD}$ terms that appear in the transformations of the $\mathcal{H}_\text{Diff}$ constraints, because in ExFT the $\mathcal{H}_\text{GD}$ constraints also take the role of the one-form gauge transformations. For the two-forms and their momenta one only finds that their transformations under $\mathcal{H}_\text{GD}$ are given by the generalised Lie derivative when applying the $\mathcal{H}_\text{P2}$ constraints and for the momenta there are additional terms that are likely related to the tensor gauge transformations. In general the two-forms do not transform under any secondary constraints because the Lagrangian is linear in their time derivative.\\
For the external diffeomorphism constraints we found that the external vielbein and the scalar fields transform as the covariantised external Lie derivative, which agrees with the covariantised external diffeomorphisms described in the Lagrangian formulation \cite{EFTI-E6}. The transformation of the one-forms \eqref{can-eft-diff-A} agrees in form with the transformation given in \cite{EFTI-E6}, however the $\partial_M \xi^n$ term, that is responsible for fixing the relative coefficients in the Lagrangian, has the opposite sign in the canonical transformation. As was shown in section \ref{can-sec-External-diffeomorphisms} the sign in the canonical formulation follows directly from the ADM decomposition of the Einstein-Hilbert improvement term in \eqref{eft-lagrangian-EH}. In section \ref{sec-EFT-lagrangian} we showed that the term in the Lagrangian formulation followed directly from a compensating Lorentz transformation, see also \cite{EFTI-E6, Berman-review-eft}. In both formulations there does not seem to be any choice involved in creating this term, in particular with respect to Lorentz transformations. The conventions in both formulations also seem to agree and therefore we do not have any explanation for this difference in sign. The canonical momenta of the vielbein, the scalars and the one-forms also transform as the covariantised external Lie derivative, however in these transformations there also appear $\partial_M \xi^n$ terms, which we should expect to play a similar role in determining the relative coefficients in the Lagrangian or canonically leading to precise cancellations in the constraint algebra. For the one-form momenta there furthermore appeared additional $\mathcal{H}_\text{GD}$ and $\mathcal{H}_\text{S1}$ constraint terms.\\
For the canonical time evolution generated by the Hamilton constraint $\mathcal{H}_\text{Ham}$ we found that the transformation of the vielbein, scalar fields and one-forms is identical in form to that found in five-dimensional E$_{6(6)}$ invariant supergravity \cite{KreutzerSUGRA}. The two-forms do not transform under this constraint because it is a secondary constraint --- the time evolution of the two-forms comes purely from the other constraints, similarly to the model in section \ref{can-sec-twoforms-model}. The time evolution of the canonical momenta that we found in ExFT is more complicated than in five-dimensional E$_{6(6)}$ invariant supergravity, mainly due to the scalar potential, the covariant derivatives and to some extend also to the topological term. The expressions we found for the time evolution reduce to those found in \cite{KreutzerSUGRA} when applying the trivial solution of the section condition, but the expressions given here are likely not the simplest and most covariant form of the equations of motion.\\ 
The $\mathcal{H}_\text{S1}$ constraints generate the transformations also found in the model theory, but we also identified the transformation generated by them on the one-forms with the tensor gauge transformation of the one-forms given in \cite{EFTI-E6}.\\

In section \ref{sec-canonical-algebra} we computed the canonical algebra relations of the Lorentz constraints with the secondary constraints. The second class algebra relations of the two-form constraints were computed in section \ref{can-sec-twoforms-model} for the two-form model. For the Poisson bracket of the generalised diffeomorphism constraints with themselves we found that at least at $\kappa=0$ the correct terms appear, but we cannot rule out that further constraints appear in this relation canonically. For the remaining algebra relations we could for now only give speculative results based on \cite{KreutzerSUGRA,EFTI-E6,Baguet:2015xha} because not all transformations of $\mathcal{P}^m_M$ are fully known and because of the above-described computational difficulties.\\

Concerning the section condition \eqref{eft-e6-section-condition-1} we found that we do not need to use the section condition to arrive at the canonical Hamiltonian. In the calculation of the canonical gauge transformations done in this work the section condition was only needed once in the transformation of $\mathcal{P}^m_M$ under the external diffeomorphism constraint \eqref{can-EFT-ext-diff-gauge-transf-P} in order to match the two-form terms of the $\mathcal{H}_\text{GD}$ constraints that appear. It may be that there is some ambiguity in the definition of the topological term and one might be able to add a term that vanishes under the section condition without affecting the variation of the overall term, in which case one may not need to use the section condition before computing the constraint algebra --- but such a modification of the topological term remains to be determined and as it is the section condition has to be used in the computation of \eqref{can-EFT-ext-diff-gauge-transf-P}. In the canonical constraint algebra computations that were done in this work we could only confirm that the section condition is needed many times in the computation of the closure of the generalised diffeomorphisms. We have not been able to find any natural interpretation of the section condition in the canonical formalism and the section condition has to be applied ad hoc. It is not possible to view the section condition as a canonical constraint because we cannot add the condition to the Lagrangian with Lagrange multipliers as it is not a condition on the canonical variables but instead on the derivatives and coordinates themselves.\\
 
What remains to be done to complete the canonical analysis of E$_{6(6)}$ ExFT is to compute the full transformations of $\mathcal{P}^m_M$, to better understand the role of the constraints $\mathcal{}{H}_\text{S2}$, to compute the full algebra of canonical constraints, to understand the difference in sign in \eqref{can-eft-diff-A} compared to the Lagrangian formulation and finally to explicitly construct appropriate Dirac brackets.\\

As an outlook it may be interesting to look for a generalisation of the Ashtekar canonical variables of general relativity. First described in 1986 the Ashtekar variables (or Ashtekar connection) $ A_{ma}$ are alternative phase space coordinates in four-dimensional (or three-dimensional) general relativity that lead to canonical constraints that are of polynomial form \cite{Ashtekar:1986-new-coords,Ashtekar:1987-coords-discussion}. In general relativity the Asthekar variables can be written as \eqref{ashtekar-eqn}, where $\omega_{mab}$ is the spin connection, $\Pi^m{}_a(e)$ are the canonical momenta of the spatial dreibein and $\gamma$ is a constant \cite{Nicolai:LQG-outside-view}. The coordinates that are canonically conjugate to the Asthekar variables are the inverse densitised dreibein $\tilde{E}_a{}^m:=e\,e_a{}^m$ \cite{Nicolai:LQG-outside-view}.
\begin{equation}\label{ashtekar-eqn}
    A_{ma} := -\frac{1}{2} \,\epsilon_{abc}\,\omega_{mbc}+\frac{\gamma}{e}(\Pi_{ma}(e)-\frac{1}{2}e_{ma}\,\Pi(e)) 
\end{equation}
In reference \cite{Melosch:1997wm} it was found that the internal vielbein (which combines metric and three-form degrees of freedom) of eleven-dimensional supergravity written in an external-internal SO(16) invariant form, behaves in a similar way as the $\tilde{E}_a{}^m$ variables and in particular renders the supersymmetry constraints and transformation in polynomial form. It was thus argued in \cite{Melosch:1997wm,Nicolai:1999yvash,Wit_2001} that there might exist such generalised Ashtekar variables in the context of eleven-dimensional supergravity. Because one should be able to see the internal vielbein of \cite{Melosch:1997wm} as part of the internal vielbein of E$_{8(8)}$ ExFT one might hope to see a similar structure in canonical ExFT. \\
In section \ref{sec-vielbein-usp8} we have investigated the rewriting of the canonical E$_{6(6)}$ ExFT in terms of the generalised USp(8) vielbein $\mathcal{V}_M^{AB}$. While the contractions of the canonical generalised vielbein momenta terms in the Hamilton constraint \eqref{usp8-legendre-SK-term-vielbein}, combined with the ``internal Ricci scalar'' of the scalar potential, look somewhat like the ADM general relativity Hamilton constraint we have not found that the generalised vielbein in this formulation has the desired properties of the Ashtekar connection --- in particular they do not seem to render the canonical constraints in polynomial form as is. Nonetheless it may be interesting to further look into the possible existence of a generalised Ashtekar connection. As such a construction may depend on the dimensions involved it may be advantageous to look at the E$_{8(8)}$ ExFT, as is suggested by the SO(16) covariant results of \cite{Melosch:1997wm}.\\
If such a generalised Ashtekar connection existed for the extended exceptional geometry one could attempt non-perturbative and background independent quantisation approaches to ExFT along the lines of loop quantum gravity \cite{Thiemann:moderncanquantGR,Nicolai:LQG-outside-view}. Similarly it was attempted in \cite{Bodendorfer:2011xe,Bodendorfer:2011hs,Bodendorfer:2011nv} to quantise eleven-dimensional supergravity in a background independent way by borrowing methods from loop quantum gravity. It may also be possible to find that the quantised ExFT is no longer equivalent to supergravity --- in particular when quantised in terms of generalised Ashtekar variables. Alternatively one could attempt the canonical quantisation of ExFT in the present coordinates. Some loop calculations in ExFT have already been carried out in \cite{Bossard:2017kfv,Bossard:2015foa} for special geometries. Double field theory has recently been discussed in the context of geometric quantisation in \cite{Alfonsi:2021bot} and the extension of the results to ExFT has also been commented on.
 
\appendix

\acknowledgments

This project has received funding from the European Research Council 
(ERC) under the European Union's Horizon 2020 research and innovation 
programme ("Exceptional Quantum Gravity", grant agreement No 740209).\\

I am supported by the  International  Max  Planck  Research  School  for  Mathematical and Physical Aspects of Gravitation, Cosmology and Quantum Field Theory. \\

I am particularly grateful to Axel Kleinschmidt and Hermann Nicolai for their support, help and many useful discussions. Furthermore I would like to thank Matteo Broccoli, Franz Ciceri, Jan Gerken, Amaury Leonard, Emanuel Malek and Henning Samtleben for comments and useful discussions. 


\providecommand{\href}[2]{#2}\begingroup\raggedright\endgroup


\end{document}